\documentclass[11pt]{article}

\usepackage[utf8]{inputenc}
\usepackage[T1]{fontenc}
\usepackage{lmodern}
\usepackage{amsmath, amssymb, amsthm, amsfonts}
\usepackage{bm}
\usepackage{graphicx}
\usepackage{booktabs}
\usepackage{enumitem}
\usepackage{geometry}
\usepackage{natbib}
\usepackage{algorithm}
\usepackage{algpseudocode}

\usepackage{hyperref}

\usepackage{cleveref}
\usepackage{algpseudocode}
\usepackage{mathtools}
\usepackage{setspace}
\algrenewcommand\algorithmicrequire{\textbf{Input:}}
\algrenewcommand\algorithmicensure{\textbf{Output:}}
\usepackage{xcolor}
\usepackage{colortbl}
\usepackage{booktabs}

\newcommand{\covariatecell}[1]{%
  \shortstack[l]{\strut\\#1\\\strut}%
}

\newcommand{\lightrowrule}{%
  \arrayrulecolor{black!18}%
  \specialrule{0.35pt}{1.3pt}{1.3pt}%
  \arrayrulecolor{black}%
}
\newtheorem{definition}{Definition}
\newtheorem{proposition}{Proposition}
\newtheorem{theorem}{Theorem}
\newtheorem{lemma}{Lemma}
\newtheorem{remark}{Remark}
\newtheorem{corollary}{Corollary}
\newtheorem{assumption}{Assumption}
\newcommand{\Var}{\mathrm{Var}}

\newcommand{\ind}{\mathbf{1}}
\title{Faithful Generative Modeling for Censored Time-to-Event Data: From Distributional Fidelity to Statistical Inference}
\author{Yin Liu$^1$, Shanshan Zhao$^2$, Didong Li$^1$\\
$^1$Department of Biostatistics, University of North Carolina at Chapel Hill\\
$^2$Biostatistics and Computational Biology Branch,\\ National Institute of Environmental Health Sciences}
\date{}

\begin{document}

\maketitle

\begin{abstract}
Synthetic data has emerged as a powerful tool for data sharing, clinical trial simulation, and digital twin development, with growing applications in biomedical research and drug development. However, limited attention has been devoted to synthetic time-to-event data, which are central to clinical trials and survival analysis. In particular, theoretical guarantees for the distributional fidelity of synthetic survival data remain largely unexplored. More importantly, even when a generator achieves distributional fidelity, it remains unclear whether the downstream statistical analysis is reliable. To address these gaps, we propose a general two-stage generative framework for survival data (TSGSurv) that jointly generates observed times and covariates conditional on event status, and establish its distributional convergence rate. Focusing on the widely used Cox proportional hazards model, we establish conditions under which generator fidelity guarantees consistency of model coefficient estimates. We further characterize the limitations of naive inference based on synthetic data, and develop a variance correction and a distribution calibration procedure to recover valid inference of original data. We demonstrate the effectiveness of the proposed generative and inferential framework with extensive simulations and two real data applications.

\end{abstract}
\tableofcontents
\section{Introduction}
\label{sec:intro}

Recent years have witnessed rapid advances in generative modeling, with increasingly powerful models capable of learning complex, high-dimensional data distributions and generating realistic synthetic records, supported by both remarkable empirical performance and a growing body of theoretical research. Generative adversarial networks (GANs,~\citealp{goodfellow2014generative,arjovsky2017wasserstein}), variational autoencoders (VAEs,~\citealp{kingma2014auto}), diffusion models~\citep{ho2020denoising,song2019generative}, and large language models (LLMs,~\citealp{brown2020language}) have substantially expanded the scope of modern generative methods. Beyond generating unstructured content such as text and images, these developments have stimulated growing interest in generative modeling across diverse data modalities and scientific domains, including clinical and biomedical research \citep{jordon2022synthetic,goncalves2020generation}. Synthetic data may facilitate data sharing, clinical-trial simulation, reproducible methodological research, and the development and validation of statistical workflows when access to original patient-level data is restricted \citep{jordon2022synthetic,chen2021synthetic}. 

Time-to-event outcomes play a central role in clinical research and drug development. Clinical trials routinely use endpoints such as overall survival, progression-free survival, time to disease, recurrence, or treatment discontinuation to evaluate treatment efficacy and safety~\citep{fleming2013counting}. These endpoints directly inform about  benefit--risk assessment, regulatory decisions, and clinical practice. A key challenge in survival analysis is censoring, where event time may not be fully observed because participants are lost to follow-up or remain event free when study ends ~\citep{klein2003survival}. This partial observation feature of survival data makes both its modeling and interpretation fundamentally different from those of ordinary tabular data~\citep{andersen1993statistical}. Survival data present a particularly challenging setting for synthetic data: a complete synthetic record should preserve not only the covariate and observed follow-up-time distributions, but also event status and the dependence among these components.

Several specialized models have been proposed for generating censored survival data, including SurvivalGAN~\citep{norcliffe2023survivalgan}, the outcome-conditioned method \citep{ashhad2025generating}, and SurvDiff \citep{brockschmidt2025survdiff}. These methods differ primarily in how they factorize the joint distribution of covariates, observed times, and event indicators. SurvivalGAN first generates covariates and then assigns an observed time through a deterministic regression model. In contrast, the outcome-conditioned method generates times and event indicators first and then generate covariates conditional on them. SurvDiff jointly generates all components through a diffusion model augmented with a weighted Cox partial log-likelihood loss, which introduces a trade-off between distributional fidelity and a model-specific survival objective. Although these methods have advanced synthetic survival data generation, theoretical understanding of their distributional fidelity remains limited.  More importantly, even if a generator achieves distributional fidelity, it remains unclear whether this guarantees reliable downstream statistical analysis, including accurate parameter estimation and valid statistical inference.

To address these gaps, we first investigate the theoretical limitations of existing approaches. We show that the deterministic regression-based time generation employed by SurvivalGAN incurs an irreducible approximation error, preventing recovery of the joint observed data distribution. We further propose a general two-stage generative framework, TSGSurv, which first generates the event indicator and then jointly generates observed follow-up time and covariates conditional on event status using a suitable existing generative model. We further establish its distributional convergence rate, showing that the two-stage factorization introduces no asymptotic penalty beyond the convergence rates of its component generators.

Next, we investigate whether distributional fidelity translates into reliable downstream inference, focusing on the Cox proportional hazards model. In particular, we show that under mild regularity conditions, distributional proximity measured by Wasserstein distance controls the discrepancy between the Cox coefficients obtained from the original and synthetic distributions. We further establish the convergence rate of synthetic data Cox model parameter estimators and the stability of their asymptotic covariance matrices. We further demonstrate that accurate Cox coefficients and covariance do not necessarily guarantee valid statistical inference. Naive inference based on synthetic data may substantially inflate type I error because it fails to account for uncertainty inherited from the original sample. Moreover, p-values from synthetic data do not necessarily reproduce those from the original data. To overcome these challenges, we develop a variance correction that accounts for both original and synthetic sampling uncertainty, and a distribution calibration procedure that aligns Cox parameter estimates from synthetic and original data, under a set of practical conditions. 

Finally, we conduct extensive simulation studies and real data experiments using data from National Health and Nutrition Examination Survey (NHANES) and Study to Understand Prognoses and Preferences for Outcomes and Risks of Treatments (SUPPORT) to evaluate the proposed two-stage framework against SurvivalGAN, SurvDiff, and the outcome-conditioned method. We investigate their empirical distributional convergence and their preservation of Cox regression coefficients, baseline cumulative hazards, predictive performance, and inferential decisions.

The main contributions of this paper are summarized as follows:\\
\textbf{Faithful generation of censored survival data.} We identify an irreducible approximation error in deterministic regression-based time generation and propose a general TSGSurv framework that accommodates existing generative architectures. We establish its distributional convergence rate and demonstrate its effectiveness through simulations and real-data applications.\\
\textbf{Theoretical guarantees for downstream Cox estimation.} We establish the local stability of Cox regression coefficients under total variation and Wasserstein perturbations, derive convergence rates for synthetic-data Cox estimators, and establish the stability of their asymptotic covariance matrices. These results connect distributional fidelity with downstream statistical estimation.\\
\textbf{Valid inference and inferential preservation.} We demonstrate that coefficient and covariance preservation do not necessarily guarantee valid synthetic-data inference and distinguish inferential validity from agreement with original-data conclusions. We develop variance correction and distribution calibration procedures and establish theoretical guarantees for valid hypothesis testing and preservation of original-data inferential conclusions.

The remainder of this paper is organized as follows. \Cref{sec:related_work} introduces the problem setup and reviews existing methods for synthetic survival-data generation and evaluation. \Cref{sec:method2} presents the proposed TSGSurv framework and its distributional convergence theory. \Cref{sec:theory} develops the theoretical analysis of downstream Cox regression, including estimation stability, inferential validity, and distribution calibration. \Cref{sec:sim,sec:real} present simulation studies and real-data applications, respectively. \Cref{sec:dis} discusses the limitations of our findings and future work. \Cref{app:data-source} provides code and data sources, \Cref{app:proof} contains proofs of the theoretical results, and \Cref{app:experiment-details} provides additional experimental details.

\section{Background and Related Work}
\label{sec:related_work}

\subsection{Problem Setup and Notation}
\label{subsec:notation}

Let $T$ denote the event time, $C$ the censoring time, and $X\in\mathcal{X}\subseteq\mathbb{R}^{p}$ a vector of baseline covariates. The corresponding observed follow-up time and event indicator are
$$
Y=\min(T,C),
\qquad
\Delta=\mathbf{1}(T\leq C),
$$
respectively. We assume that follow-up is limited to a finite interval $[0, \tau]$, where $\tau<\infty$, therefore
$
0\leq Y\leq C \leq\tau.
$
The true distribution and all fitted generative distributions considered below are assumed to share this time support. We observe $$
\mathcal{D}_{n}=\{D_i\}_{i=1}^{n},\qquad D_i=(X_i,Y_i,\Delta_i)\overset{\mathrm{i.i.d.}}{\sim}P_0,
$$
where $n$ is the sample size, $P_0$ is their true joint distribution on
$\mathcal{X}\times[0,\tau]\times\{0,1\}$. A synthetic data generator first estimates $P_0$ by $\widehat P_n$ and then produces a synthetic dataset with $m$ samples: $$
\widetilde{\mathcal{D}}_{m}
=
\{\widetilde D_j\}_{j=1}^{m},
\qquad
\widetilde D_j=(\widetilde{X}_i,\widetilde{Y}_i,\widetilde{\Delta}_i)\overset{\mathrm{i.i.d.}}{\sim}\widehat P_n.
$$
Our goal is two-fold: to generate synthetic data that resemble the original data, and to ensure that analyses of $\widetilde{\mathcal{D}}_m$ yield valid statistical inference and reproduce the corresponding results from $\mathcal{D}_n$.


\subsection{Related Work}
\label{subsec:existing_methods}

Existing approaches to synthetic survival data generation differ primarily in how they represent and generate the joint distribution of covariates $X$, observed follow-up time $Y$, and event indicator $\Delta$. Broadly, existing methods either adopt a sequential factorization that generates one component conditional on the others or model all components jointly. We briefly review three representative approaches that illustrate these strategies and motivate the framework developed in this section.

\noindent\textbf{SurvivalGAN (deterministic time generation).}
SurvivalGAN~\citep{norcliffe2023survivalgan} generates synthetic survival data through a sequence of conditional generation and deterministic prediction steps. First, a condition vector and event indicator are sampled from their empirical distribution or specified by the user. A conditional Wasserstein generative adversarial network then generates covariates given these quantities. Next, a pretrained survival model estimates survival probabilities over a set of pre-defined time points. A time regressor takes these probabilities, covariates and the event indicator to generate the observed follow-up time. SurvivalGAN uses DeepHit for estimating survival probabilities and XGBoost with squared-error loss for predicting the log-transformed time.

This construction explicitly incorporates censoring status and accommodates imbalances across event types and follow-up horizons. However, its final time-generation step is based on a deterministic regression function  $f_n (X,\Delta)$, trained to approximate $f_*(X, \Delta) = \exp\{E(\log(Y) \mid X, \Delta)\}$. Consequently, even if $f_n (X,\Delta)$ consistently estimates $f_*(X, \Delta) $, the generated times cannot recover the full conditional variation of  $\log(Y)|(X,\Delta)$ when the conditional distribution is nondegenerate. We formalize this limitation mathematically in \Cref{subsec:deterministic_error2}.

\noindent\textbf{Outcome-conditioned generation (Ashhad).}
\citet{ashhad2025generating} propose an alternative sequential construction based on the factorization
$$
p(X,Y,\Delta)
=
p(\Delta)\,
p(Y\mid\Delta)\,
p(X\mid Y,\Delta).
$$
The event indicator is first generated according to its empirical frequency, followed by the observed follow-up time conditional on event status using a Dirichlet process mixture model. Finally, a conditional tabular generator produces covariates given the generated time and event indicator. This framework can be combined with different tabular generative architectures, including conditional GANs, variational autoencoders, and diffusion models.

By generating the observed time before the covariates, this approach avoids deterministic regression-based time generation. It instead requires the final generator to learn the potentially high-dimensional conditional distribution $p(X\mid Y,\Delta)$. Because $Y$ is continuous, records may be sparse around particular conditioning values, especially in less densely observed regions of the follow-up distribution. Its performance therefore depends on how effectively the conditional generator captures variation in the covariate distribution across follow-up times and censoring statuses.

\noindent\textbf{SurvDiff.}
SurvDiff takes an end-to-end approach that jointly generates covariates, observed follow-up times, and event indicators through a diffusion model \citep{brockschmidt2025survdiff}. Continuous covariates and follow-up times are perturbed through a Gaussian diffusion process, whereas categorical covariates and the event indicator are handled through a discrete masking process. A shared denoising network reconstructs all components simultaneously during the reverse process.

In addition to the standard diffusion objective, SurvDiff incorporates a survival-specific loss derived from the Cox partial log-likelihood to encourage preservation of event-time ordering and censoring-dependent risk sets. The reverse diffusion process then jointly generates complete synthetic records $(X,Y,\Delta)$. This formulation avoids an explicit sequential factorization, but requires a single model to learn a mixed-type joint distribution containing high-dimensional covariates, a continuous follow-up time, and a binary event indicator.

Together, these approaches illustrate different tradeoffs in modeling the observed-data distribution: deterministic time generation may restrict conditional variability, outcome-conditioned generation places a potentially demanding continuous-conditioning problem on the covariate generator, and fully joint generation requires a single model to simultaneously represent heterogeneous data types. These considerations motivate a factorization that retains joint generation of follow-up times and covariates while conditioning only on the discrete event indicator.

\subsection{Evaluation of Synthetic Survival Data}
\label{subsec:existing_evaluation}

Existing studies assess synthetic survival data using distributional and survival-specific criteria. Covariate distributions are commonly compared through Jensen--Shannon divergence (JS), Maximum Mean Discrepancy (MMD), Wasserstein$-p$ distances ($W_p$), marginal histograms, and low-dimensional visualizations. Survival-specific measures include Kaplan--Meier divergence, optimism, short-sightedness, and event-time divergence~\citep{norcliffe2023survivalgan}. These measures examine selected features of the event and censoring distributions, follow-up times, and covariate patterns across event times.

Predictive utility is often evaluated using a train-on-synthetic, test-on-real design. A survival model fitted to synthetic data is evaluated on held-out real data using measures such as the concordance index and integrated Brier score~\citep{harrell1982evaluating,graf1999assessment}. This design assesses whether synthetic data support the development of a predictive model. It does not directly establish that analyses fitted to synthetic and original data yield similar effect estimates or inferential conclusions.

Recent studies have also compared Kaplan--Meier curves, Cox hazard ratios, confidence intervals, and the significance of effects between real and synthetic data~\citep{kuo2024ck4gen,thangaraj2026novel}. These comparisons assess important aspects of downstream fidelity. We extend this evaluation by jointly examining the distribution of $(X,Y,\Delta)$, censor-aware dependence inverse-probability-of-censoring-weighted Kendall's tau (IPCW Kendall,~\citealp{wang2020interaction}) and  distance correlation for right-censored survival outcomes (censored dCor,~\citealp{edelmann2022consistent}), Cox coefficient and baseline-hazard discrepancies, and agreement in model-based inferential results. Together, these reference-based criteria distinguish joint-distributional, survival-dependence, estimation, and inferential fidelity.

\section{Two-stage Generative (TSGSurv) Framework}
\label{sec:method2}
In this section, we first discuss a theoretical limitation of SurvivalGAN in \Cref{subsec:deterministic_error2}, which motivated us to propose a two-stage generative framework (TSGSurv) described in \Cref{subsec:factorization_generation2}. We then establish the convergence rate of the proposed TSGSurv framework in \Cref{subsec:two_stage_rate2}.

\subsection{Approximation Error of SurvivalGAN (deterministic time generation)}
\label{subsec:deterministic_error2}
We first investigate the distributional fidelity of SurvivalGAN, which generates $(X,\Delta)$ and subsequently assigns $Y$ through a deterministic regression function.

Let $P_{X,\Delta}$ denote the true marginal distribution of $(X,\Delta)$, and  $\widehat P_{X,\Delta,n}$ denote the fitted generative distribution of $(X,\Delta)$. Let $f_n(\widetilde X,\widetilde\Delta)$ be the fitted time regression function, with population target $f_*(X,\Delta)=\exp\{E(\log(Y) \mid X, \Delta)\}$. The resulting synthetic distribution $\widehat P_n^{\mathrm{reg}}$ is defined by 
$$\widehat P_n^{\mathrm{reg}} =
\mathcal{L}
\left\{
\widetilde X,\widetilde Y,\widetilde \Delta
\right\}, \qquad
(\widetilde X,\widetilde\Delta)
\sim
\widehat P_{X,\Delta,n} ,
\qquad
\widetilde Y
=
f_n(\widetilde X,\widetilde\Delta).
$$
Under this framework, the limiting joint distribution is $$
P^\dagger
=
\mathcal{L}
\left\{
X,f_*(X,\Delta),\Delta
\right\}.
$$
This framework assigns a deterministic follow-up time to each $(X,\Delta)$, eliminating the conditional variation in $Y$ given $X,\Delta$. Now we show that SurvivalGAN admits an irreducible error.

\begin{lemma}[Irreducible error in SurvivalGAN]
\label{lem:deterministic_error2}
Assume the marginal generator $\widehat P_{X,\Delta,n}$ has rate $\gamma_g$ and regressor $f_n(x,\delta)$ has rate $\gamma_r$, i.e.,
\[
W_1\left(
\widehat P_{X,\Delta,n},
P_{X,\Delta}
\right)
=
O_p(n^{-\gamma_g}),\quad\int
\left|
f_n(x,\delta)-f_*(x,\delta)
\right|
\,d\widehat P_{X,\Delta,n}(x,\delta)
=
O_p(n^{-\gamma_r}).
\]
Further assume that the mapping
$
(x,\delta)
\mapsto
\{x,f_*(x,\delta),\delta\}
$ is Lipschitz, let $b\coloneqq W_1(P^\dagger,P_0)$, then $$
W_1(\widehat P_n^{\mathrm{reg}},P_0)
=
b+O_p(n^{-\gamma_g}+n^{-\gamma_r})\xrightarrow[]{n\to \infty}b.$$
\end{lemma}
In practice, $Y|X,\Delta$ is generally random with a positive variance, so that $b>0$. Thus, even when both the generator and the regressor are consistent, SurvivalGAN retains a strictly positive irreducible error and cannot consistently recover the true joint distribution. This result suggests a basic design principle for survival data generation: observed time should be modeled probabilistically rather than deterministically, which motivates us to propose a  two-stage generative framework in the following section.



\subsection{TSGSurv}
\label{subsec:factorization_generation2}

Now we propose TSGSurv that directly models the joint conditional distribution of $(X,Y)|\Delta$, thereby avoiding the irreducible error resulting from the deterministic prediction. Observe that $$
P_0(X,Y,\Delta) = P_\Delta(\Delta)\, P_{X,Y|\Delta}(X,Y|\Delta).
$$
Let $ \pi=P_\Delta(\Delta=1), \widehat\pi_n = n^{-1}\sum_{i=1}^{n}\Delta_i $. A synthetic event indicator can be generated according to $
\widetilde{\Delta}_j \sim \operatorname{Bernoulli}(\widehat\pi_n), j=1,\ldots,m. $

Conditional on $\widetilde{\Delta}_j=\delta\in\{0,1\}$, the second stage then estimates $ P_{X,Y|\Delta=\delta}(X,Y|\Delta=\delta)$ by conditional generative distributions $\widehat G_{\delta,n_\delta},$ such that $
(\widetilde X_j,\widetilde Y_j  | \widetilde{\Delta}_j=\delta) \sim \widehat G_{\delta,n_\delta}, $
where $ n_\delta = \sum_{i=1}^{n}\mathbf{1}(\Delta_i=\delta). $ 
Hence, we have the resulting synthetic observation $ \widetilde D_j = (\widetilde X_j,\widetilde Y_j,\widetilde\Delta_j) $, and the corresponding induced joint generative distribution 
\begin{equation}\label{eqn:2-stage}
\widehat P_n = \widehat\pi_n\widehat G_{1,n_1} + (1-\widehat\pi_n)\widehat G_{0,n_0}. 
\end{equation}
This TSGsurv framework is summarized in
Algorithm~\ref{alg:two_stage_generation}.
\begin{algorithm}[htbp]
\caption{TSGSurv algorithm }
\label{alg:two_stage_generation}
\begin{algorithmic}[1]
\Require Original data
$\mathcal D_n=\{(X_i,Y_i,\Delta_i)\}_{i=1}^n$
and target synthetic sample size $m$.

\State Estimate the event probability
$\widehat\pi_n=n^{-1}\sum_{i=1}^n\Delta_i$
and subgroup sizes
$n_\delta=\sum_{i=1}^n\mathbf 1(\Delta_i=\delta)$.

\State Fit conditional generators
$\widehat G_{\delta,n_\delta}$
for $\{(X_i,Y_i)\}_{\Delta_i=\delta}$.

\For{$j=1,\ldots,m$}
    \State Draw
    $\widetilde\Delta_j
    \sim\operatorname{Bernoulli}(\widehat\pi_n)$.

    \State Given $\widetilde\Delta_j=\delta$, draw
    $(\widetilde X_j,\widetilde Y_j)
    \sim\widehat G_{\delta,n_\delta}$.
\EndFor


\Ensure Synthetic survival data
$\widetilde{\mathcal D}_m=\{(\widetilde X_j,\widetilde Y_j,\widetilde\Delta_j)\}_{j=1}^m$.
\end{algorithmic}
\end{algorithm}

This two-stage factorization offers three advantages. First, $Y$ is generated jointly with $X$ as a random variable, allowing the generator to represent conditional time variability. Second, conditioning only on the binary event indicator avoids estimating the potentially complex distribution $P(X\mid Y,\Delta)$ across continuously varying values of $Y$. Third, the framework is compatible with different generative architectures, including GAN and diffusion models, and its convergence rate is determined by the rates of the component generators, as established in \Cref{thm:rate_2_stage} in the next section.

\subsection{Convergence Rate of TSGSurv}
\label{subsec:two_stage_rate2}

We establish the convergence rate of TSGSurv under the $W_1$ distance. 

\begin{theorem}[Convergence rate of TSGSurv]
\label{thm:rate_2_stage}
Assume that the conditional generator has rate $\gamma$, i.e.,$
W_1(\widehat G_{\delta,n_\delta},P_{X,Y|\Delta=\delta})
=
O_p(n_\delta^{-\gamma}),~\delta\in\{0,1\}.
$
Then the induced joint generator defined in \Cref{eqn:2-stage}  has rate
$$
W_1(\widehat P_n,P_0)
=
O_p\left(
n^{-\min\{\gamma,\frac{1}{2}\}}
\right).
$$
\end{theorem}

The result shows that the two-stage framework achieves the same convergence rates as its two conditional generators, without introducing any additional asymptotic penalty. Throughout the paper, we assume that all generative and regression estimators considered have convergence rates of order $n^{-\gamma}$ with $0<\gamma\leq 1/2$, where $\gamma=1/2$ corresponds to the parametric rate.

\begin{remark}
The result in \Cref{thm:rate_2_stage} extends to any metric $d$ that satisfies, 
\begin{align*}
\text{Joint convexity:~}&d\left(\lambda P+(1-\lambda)Q,\,
         \lambda R+(1-\lambda)S\right)
\leq \lambda d(P,R)+(1-\lambda)d(Q,S),\\
\text{Homogeneity:~}&d\!\left(\lambda P+(1-\lambda)Q,\,
         \lambda'P+(1-\lambda')Q\right)
=|\lambda-\lambda'|\,d(P,Q),\quad \forall P,Q,R,S,~\forall\lambda,\lambda'\in[0,1].
\end{align*}
This class of metrics includes commonly used metrics such as total variation, the bounded-Lipschitz distance, and MMD with a bounded characteristic kernel. 
\end{remark}



\section{Theory for Cox Regression on Synthetic Data}
\label{sec:theory}

This section studies the theoretical properties of downstream Cox regression using synthetic datagenerated by TSGSurv. We first introduce the Cox model notations in \Cref{subsec:cox_notation}, and then establishes the local stability of the estimated parameters in Cox model under total variation and Wasserstein perturbations, and derives rates for synthetic sample estimation and preservation of coefficient directions in \Cref{subsec:cox_stability} . We then establishcovariance stability, characterize the limitations of naive synthetic data inference, and develop corrected inference together with a constructive calibration procedure in \Cref{sec:cox_inference}.

\subsection{Cox Model Setup}\label{subsec:cox_notation}
For a Cox proportional hazards model $
\lambda(t\mid X) = \lambda_0(t)\exp(\beta^\top X),$
we are interested in estimating the Cox coefficient $\beta\in \mathbb{R}^p$. Let $ N(t)\coloneqq \mathbf{1}(Y\leq t,\Delta=1)$ be the counting process, and $R(t)\coloneqq \mathbf{1}(Y\geq t)$ be the at risk process. For any distribution $P$, define
$$S_P^{(k)}(\beta,t) \coloneqq \mathbb{E}_P \left[ R(t)\exp(\beta^\top X)X^{\otimes k} \right], k=0,1,2, \qquad\overline X_P(\beta,t) \coloneqq S_P^{(1)}(\beta,t)/S_P^{(0)}(\beta,t). $$
Here, $S_P^{(k)}(\beta,t)$ represents the $k$th weighted moment for those at risk at $t$, and $\overline X_P(\beta,t)$ is the corresponding weighted mean of the covariates.

The population Cox partial log-likelihood $\ell_P(\beta)$, score $U_P(\beta)$, and information matrix $J_P(\beta)$ are given by 
\begin{align*}
\ell_P(\beta) &\coloneqq \mathbb E_P\!\left[ \Delta\{\beta^\top X-\log S_P^{(0)}(\beta,Y)\} \right],\\
U_P(\beta)&\coloneqq \nabla_\beta \ell_P(\beta)= \mathbb{E}_P \left[ \Delta\left\{ X-\overline X_P(\beta,Y) \right\} \right]
=
\mathbb{E}_P\left[ \int_0^\tau \left\{ X-\overline X_P(\beta,t) \right\} \,dN(t) \right],\\
J_{P}(\beta)&\coloneqq-\nabla_\beta U_{P}(\beta).
\end{align*}
\begin{definition}\label{def:U}
   We denote the unique solution of  $U_P(\beta)=0$  as $\beta(P)$ whenever it exists.
\end{definition}

To characterize the uncertainty of coefficient estimates, we introduce the influence function and its asymptotic covariance matrix~\citep{lin1989robust}. Because the Cox score depends on the underlying at risk process, its influence function must account for the corresponding estimations. Specifically, we define 
\begin{equation}
\xi_P(D)
\coloneqq
\int_0^\tau \left\{ X-\overline X_P\bigl(\beta(P),t\bigr) \right\} \,dN(t)
- e^{\beta(P)^\top X} \int_0^\tau R(t)
\frac{X-\overline X_P\bigl(\beta(P),t\bigr)}{ S_P^{(0)}\bigl(\beta(P),t\bigr) } \,\mathbb E_P\{dN(t)\}.
\label{eq:cox_xi}
\end{equation}
The corresponding influence function $\psi_P(D)$ and asymptotic sandwich covariance matrix $\Sigma(P)$ are 
\begin{equation}
\psi_P(D)
\coloneqq
J_P\{\beta(P)\}^{-1}\xi_P(D),
\qquad
\Sigma(P)
\coloneqq
\mathbb E_P
\left[
\psi_P(D)\psi_P(D)^\top
\right].
\label{eq:cox_if}
\end{equation}
We focus on the asymptotic sandwich covariance matrix \(\Sigma(P)\) instead of the model-based information covariance matrix \(J_P\{\beta(P)\}^{-1}\) because it does not require the information identity, i.e., $\mathbb E_P\!\left[ \xi_P(D)\xi_P(D)^\top \right] = J_P\{\beta(P)\}.$ Under a correctly specified Cox proportional hazards model and the standard independent censoring assumption, this identity holds, and \(\Sigma(P)\) reduces to \(J_P\{\beta(P)\}^{-1}\).

\subsection{Local Stability of the $ \beta$ Estimation}
\label{subsec:cox_stability}

We impose the following regularity conditions to establish the stability of the Cox parameter under distributional perturbations.

\begin{assumption}[Compact parameter space]
\label{ass:cox_parameter}
The parameter space for $\beta$, $B\subset\mathbb{R}^p$, is compact, with
$\beta(P_0)\in\operatorname{int}(B)$.
\end{assumption}

\begin{assumption}[Uniformly bounded covariates]
\label{ass:cox_covariates}
There exists a constant $K<\infty$ such that $\|X\|\leq K$ almost surely for all distributions in a neighborhood
$\mathcal P_0$ of $P_0$.
\end{assumption}

\begin{assumption}[Nondegenerate risk sets]
\label{ass:cox_risksets}
The analysis horizon satisfies $0<t\leq\tau<\infty$, and there exists
$a>0$ such that
\[
\inf_{P\in\mathcal P_0}
\inf_{\beta\in B}
\inf_{t\in[0,\tau]}
S_P^{(0)}(\beta,t)
\geq a.
\]
\end{assumption}

\begin{assumption}[Local identification]
\label{ass:cox_identification}
The population score function satisfies $U_{P_0}\{\beta(P_0)\}=0$, and
$\beta(P_0)$ is the unique solution in its neighborhood
$\mathcal N_{\beta(P_0)}\subset B$.
\end{assumption}

\begin{assumption}[Uniformly nonsingular information]
\label{ass:cox_information}
The map $\beta\mapsto U_{P_0}(\beta)$ is continuously differentiable on $\mathcal N_{\beta(P_0)}$, and there exists $\kappa>0$ such that
\[
\inf_{\beta\in\mathcal N_{\beta(P_0)}}
\lambda_{\min}\!\left\{
J_{P_0}(\beta)
\right\}
\geq \kappa.
\]
\end{assumption}

Assumptions~\ref{ass:cox_parameter}--\ref{ass:cox_information} are standard or slightly strengthened versions of regularity conditions commonly used in Cox regression theory~\citep{andersen1982cox,andersen1993statistical}. 
In particular, Assumption~\ref{ass:cox_risksets} requires a nonnegligible proportion of the population to remain at risk throughout the analysis period, including at the fixed follow-up horizon $\tau$.
Together with Assumption~\ref{ass:cox_parameter}--\ref{ass:cox_covariates}, they ensure that the at risk moments and their ratios are uniformly bounded and stable under distributional perturbations. Assumptions \ref{ass:cox_identification} and \ref{ass:cox_information} ensure local identification and sufficient curvature of the population partial likelihood, allowing perturbations of the score function to be translated into perturbations of its solution. Under these  mild assumptions, we establish the following stability result.

\begin{theorem}[Local stability of $\beta$ under $\mathrm{TV}$]
\label{thm:cox_population_stability}
Under Assumptions~\ref{ass:cox_parameter}--\ref{ass:cox_information}, there exist a constant
$\varepsilon>0$ such that, for every
$P\in\mathcal P_0$ satisfying $d_{\mathrm{TV}}(P,P_0)\leq\varepsilon$, the Cox parameter $\beta(P)$ is well defined in
$\mathcal N_{\beta(P_0)}$ and satisfies
\[
\left\|
\beta(P)-\beta(P_0)
\right\|
\leq
C\,d_{\mathrm{TV}}(P,P_0),
\]
where $C<\infty$ is a constant.
\end{theorem}

Next, we extend the stability result to the Wasserstein-1 distance, which is widely used in generative modeling and for which convergence rates are available for various classes of generators. Unlike total variation, $W_1$ controls differences in expectations through Lipschitz functions. The score function, however, involves the discontinuous at risk indicator $R(t)=\mathbf{1}(Y\geq t)$. We therefore impose an additional regularity condition on the distribution of the observed time over the interior of the follow-up interval, while allowing administrative censoring at the end of follow-up time.

\begin{assumption}[Regularity of observed follow-up times]
\label{ass:cox_w1_times}
For every $P\in\mathcal P_0$, the distribution of $Y$ over $[0,\tau)$ is absolutely continuous with respect to Lebesgue measure, with density $f_P$ satisfying $\sup_{P\in\mathcal P_0}\sup_{0\leq y<\tau}f_P(y)\leq M$ for some $M<\infty$. In addition, $P(Y=\tau,\Delta=1)=0$. Positive probability of administrative censoring at $(Y,\Delta)=(\tau,0)$ is permitted.
\end{assumption}
\Cref{ass:cox_w1_times} controls the effect of the discontinuous risk set indicator under Wasserstein perturbations. The uniform density bound limits the probability mass that can lie in a short interval around an interior time point, allowing differences in risk-set expectations to be controlled through Lipschitz approximations. The restriction $P(Y=\tau,\Delta=1)=0$ excludes the possibility that an event happens at the end of follow-up horizon $\tau$, while still allowing censoring at $\tau$.

\begin{theorem}[Local stability of $\beta$ under $W_1$]
\label{thm:cox_wasserstein_stability}
Under Assumptions~\ref{ass:cox_parameter}--\ref{ass:cox_w1_times}, there exist a constant $\varepsilon>0$ such that, for every $P\in\mathcal P_0$ satisfying $W_1(P,P_0)\leq\varepsilon$, the population Cox parameter $\beta(P)$ is well defined in $\mathcal N_{\beta(P_0)}$ and satisfies
\[\left\| \beta(P)-\beta(P_0) \right\| \leq C\,W_1(P,P_0),\]
where $C<\infty$ is a  constant.
\end{theorem}

\Cref{thm:cox_wasserstein_stability} characterizes how Wasserstein approximation error in the learned distribution propagates to the Cox parameter by setting $P = \widehat P_n$. In practice, an additional source of variability arises because downstream analyses are performed on a finite sample drawn from the learned generative distribution $\widehat P_n$. Let $\widehat P_{n,m}$ denote its empirical distribution based on the i.i.d.\ synthetic sample 
$\widetilde{\mathcal{D}}_m=\{(\widetilde X_j,\widetilde Y_j,\widetilde\Delta_j)\}_{j=1}^{m}$ from $\widehat P_n$. 
Let $P_n$ denote the empirical distribution of the original samples 
$\mathcal{D}_n=\{(X_j, Y_j,\Delta_j)\}_{j=1}^{n}$ from $ P_0$. The following result integrates the generative and synthetic sampling errors and gives a direct comparison with the Cox estimator fitted to the original sample.

\begin{corollary}[Estimation rate and preservation of coefficient signs]
\label{cor:cox_synthetic}

Assume that the conditions of either \Cref{thm:cox_population_stability} or \Cref{thm:cox_wasserstein_stability} hold, and that $\widehat P_n\in\mathcal P_0$. That is, respectively, $d_{\mathrm{TV}}(\widehat P_n,P_0)=O_p(r_n)$ or $W_1(\widehat P_n,P_0)=O_p(r_n)$, where $r_n\to0$. Let $m\to\infty$, further assume that the standard regularity conditions for the empirical Cox estimator hold uniformly over the fitted distribution $\widehat P_n$. Then:

\begin{enumerate}[label=(\roman*),leftmargin=*]

\item \textbf{Estimation rates.}
The Cox parameter under $P_0$, $\widehat{P}_n$, and $\widehat{P}_{n,m}$ satisfies 
\begin{align*}
\|\beta(\widehat P_n)-\beta(P_0)\|&=O_p(r_n),\\
\|\beta(\widehat P_{n,m})-\beta(P_0)\|&=O_p(r_n+m^{-1/2}),\\
\|\beta(\widehat P_{n,m})-\beta(P_n)\|
&=O_p(r_n+m^{-1/2}+n^{-1/2}).
\end{align*}
\item \textbf{Preservation of coefficient signs.}
Let $\mathcal J\coloneqq \{j:\beta_j(P_0)\neq 0\}$, then
\[
\Pr\left\{
\operatorname{sign}
\left(\beta_j(\widehat P_{n,m})\right)
=
\operatorname{sign}
\left(\beta_j(P_0)\right),~\forall j\in\mathcal J
\right\}
\longrightarrow 1.
\]

\end{enumerate}
\end{corollary}

Part (i) separates three sources of discrepancy between the synthetic and original Cox analyses as approximation error in the learned distribution, finite synthetic sample variability, and original sample variability. Increasing $m$ reduces the synthetic sampling term $m^{-1/2}$, leaving the discrepancy increasingly dominated by the generative approximation error and the uncertainty in the original sample. Part (ii) shows that consistency of the synthetic Cox coefficients guarantees preservation of their signs for nonzero effects, ensuring that risk and protective factors identified from the original data retain their directions of association in the synthetic data.

\begin{remark}[Empirical Cox regularity] 
The standard regularity conditions for the empirical Cox estimator in \Cref{cor:cox_synthetic} is the usual regularity condition for the empirical Cox estimator, imposed uniformly over $\widehat P_n$. Specifically, conditional on  $\widehat P_n$, the synthetic records are i.i.d.\ from $\widehat P_n$, the empirical risk set momonets converge uniformly to their population counterparts, and the empirical score function satisfies the corresponding conditional central limit theorem. Moreover, the risk-set denominator $\widehat S_{n,m}^{(0)}(\beta,t)$ remains uniformly bounded away from zero, and the empirical information matrix remains uniformly nonsingular. These standard conditions yield
\[
\left\|
\beta(\widehat P_{n,m})-\beta(\widehat P_n)
\right\|=O_p(m^{-1/2}).
\]
\end{remark}


\subsection{Stability of the Inference}
\label{sec:cox_inference}
Moving from coefficient fidelity to inferential fidelity also requires control of uncertainty. We first establish the local stability of the Cox covariance matrix $\Sigma$, which provides the basis for comparing standard errors and $p$-values between the original and synthetic analyses.


\begin{theorem}[Local stability of $\Sigma$ under $W_1$]
\label{thm:cox_covariance_stability}
Under Assumptions~\ref{ass:cox_parameter}--\ref{ass:cox_w1_times} hold, there exist a constant $\varepsilon>0$ such that, for every $P\in\mathcal P_0$ satisfying $W_1(P,P_0)\leq\varepsilon$,
\[
\left\|
\Sigma(P)-\Sigma(P_0)
\right\|
\leq
C\,W_1(P,P_0),
\]
where $C<\infty$ is a constant.
\end{theorem}
Consequently, if $W_1(\widehat P_n,P_0)=o_p(1)$ and $m\to\infty$, then, under the standard regularity conditions for consistent estimation of the Cox sandwich covariance, $\Sigma(\widehat P_{n,m}) \xrightarrow{p}\Sigma(P_0).$ An analogous conclusion holds for the original sample: $\Sigma(P_n)\xrightarrow{p}\Sigma(P_0).$ We assume that $\Sigma_{jj}(P_0)>0$ for each tested coordinate.
\begin{remark}The proof in \Cref{proof:cox_covariance_stability} also establishes the stability of the information matrix:
\[ \left\| J_P\{\beta(P)\}^{-1} - J_{P_0}\{\beta(P_0)\}^{-1} \right\| \leq C_{J^{-1}}W_1(P,P_0), \]
The sandwich covariance is more general than the inverse information matrix because it remains applicable under Cox model misspecification, when the information identity may fail.
\end{remark}
\noindent\textbf{Limits of coefficient and covariance preservation.}
However, preservation of $\beta$ and $\Sigma$ does not by itself guarantee valid inference from a finite synthetic sample. Observe that the bias of the synthetic coefficient estimator $\beta(\widehat P_{n,m})$ can be decomposed as

\begin{equation*}
\beta(\widehat P_{n,m})-\beta(P_0)
=
\underbrace{\{\beta(P_n)-\beta(P_0)\}}_{\text{Original sampling error}}
+
\underbrace{\{\beta(\widehat P_n)-\beta(P_n)\}}_{\text{Generative error}}
+
\underbrace{\{\beta(\widehat P_{n,m})-\beta(\widehat P_n)\}}_{\text{Synthetic sampling error}}.
\end{equation*}
Even if the generator preserves the original data Cox coefficient, eliminating the generator error, the synthetic estimator still inherits the original sampling error. However, standard errors computed from the synthetic sample alone account only for synthetic sampling variability, potentially underestimating the total uncertainty and leading to invalid inference. 

Consider the two-sided test $H_{0,j}:\beta_j(P_0)=b_j^0$ at level $\alpha\in(0,1)$. Define
\begin{equation*}
Z_{\mathrm{real},j}
=
\frac{
\beta_j(P_n)-b_j^0
}{
\sqrt{\Sigma_{jj}( P_{n})/n}
}, \qquad
Z_{\mathrm{syn},j}
=
\frac{
\beta_j(\widehat P_{n,m})-b_j^0
}{
\sqrt{\Sigma_{jj}(\widehat P_{n,m})/m}
}.
\end{equation*}
For any Wald statistic $Z$, define the two-sided $p$-value by $p(Z)=2\{1-\Phi(|Z|)\}$, and denote $p_{\mathrm{real},j}=p(Z_{\mathrm{real},j})$ and $p_{\mathrm{syn},j}=p(Z_{\mathrm{syn},j})$. The following proposition shows that naive inference is asymptotically invalid even when the generator achieves a root-$n$ distributional convergence rate.and satisfies the stated regularity conditions.

\begin{proposition}
[A coefficient-preserved generator with invalid naive inference]
\label{prop:cox_naive_invalid} 
Assume Assumptions~\ref{ass:cox_parameter}-- \ref{ass:cox_w1_times} hold with $\Sigma(P_0)>0$. Under the synthetic sampling scheme defined in \Cref{subsec:notation}, suppose $\beta(\widehat P_n)=\beta(P_n), W_1(\widehat P_n,P_0)=O_p(n^{-1/2}).$ Under $H_{0}$, if $m/n\longrightarrow r\in(0,\infty)$, then $Z_{\mathrm{syn}}\xrightarrow{d}N(0,1+r)$. Consequently, its limiting rejection probability is
\[
2\left[
1-\Phi\left\{
\frac{z_{1-\alpha/2}}{\sqrt{1+r}}
\right\}
\right]>\alpha.
\]
\end{proposition}
For $m=n$ and $\alpha=0.05$, the actual type I error approaches $0.166$. More strikingly, as $m/n\to\infty$, type I error tends to one, indicating that generating more synthetic data can make naive inference increasingly unreliable. To achieve valid naive inference in this setting, one would need $m/n\to 0$, a counterintuitive restriction that sacrifices statistical precision. This motivates inference procedures that account for the uncertainty inherited from the original data rather than restricting the synthetic sample size. 


\Cref{prop:cox_naive_invalid} establishes that naive synthetic-data inference can be invalid even when the original-data Cox coefficient is exactly preserved. A separate question is whether valid synthetic-data inference necessarily reproduces the original-data inferential conclusions. The following proposition shows that this is not guaranteed, even for an oracle generator that samples directly from the true distribution.

\begin{proposition}
[Valid $p$-values need not reproduce the original $p$-value]
\label{prop:cox_oracle}
Under $H_0$, suppose the oracle generator is $\widehat{P}_n=P_0$, then both the original and synthetic $p$-values are asymptotically uniform on $[0,1]$. Nevertheless, their level-$\alpha$ testing decisions disagree with limiting probability $2\alpha(1-\alpha)>0$.
\end{proposition}
Together, \Cref{prop:cox_naive_invalid,prop:cox_oracle} identify two distinct challenges: accounting for uncertainty inherited from the original sample to ensure valid inference, and preserving the original-data analysis to achieve inferential agreement. These challenges motivate our variance correction and distribution calibration procedures, respectively.


\noindent\textbf{Corrected variance and distribution calibration.} To account for both original and synthetic sampling uncertainty, we first define the combined sampling scale
\[
s_{n,m}\coloneqq\sqrt{n^{-1}+m^{-1}}.
\]
Let $\widehat s_{\mathrm{syn},j} \coloneqq \{\Sigma_{jj}(\widehat P_{n,m})/m\}^{1/2}$ be the uncorrected synthetic standard error involved in $Z_{\mathrm{syn},j}$, then the corrected synthetic standard error, and corresponding Wald statistic and $p$-value are
\begin{equation}
\widehat s_{\mathrm{corrected},j}
\coloneqq
s_{n,m}\left\{\Sigma_{jj}(\widehat P_{n,m})\right\}^{1/2},~~
Z_{\mathrm{corrected},j}
\coloneqq
\frac{
\beta_j(\widehat P_{n,m})-b_j^0
}{
\widehat s_{\mathrm{corrected},j}
},
~~
p_{\mathrm{corrected},j}
\coloneqq
p(Z_{\mathrm{corrected},j}).
\label{eq:cox_corrected_z}
\end{equation}
The corrected standard error is $\sqrt{1+m/n}$ times the naive synthetic standard error, accounting for uncertainty inherited from the original sample. However, variance correction alone is insufficient if the fitted generator does not adequately preserve the original Cox coefficient. To characterize this discrepancy, we define the score residual as
\begin{definition}[Score residual] The score residual
    \[c_n\coloneqq\left\|U_{\widehat P_n}\{\beta(P_n)\}\right\|\]
measures how closely the fitted generator distribution preserves the original Cox estimating equation.
\end{definition} We impose the following condition on $c_n$ to establish the validity and convergence rates of corrected inference.




\begin{assumption}[Score calibration]
\label{ass:cox_weak_calibration}
As $n,m\to\infty$, for any relative growth of $m$ and $n$ under consideration, $
c_n=o_p(s_{n,m}).$
\end{assumption}


\begin{remark}\Cref{ass:cox_weak_calibration} does not follow automatically from distributional fidelity, but it can be satisfied exactly by a generator designed to preserve the original-data Cox estimate. For example, suppose $X\sim\operatorname{Bernoulli}(1/2)$ and $T\mid X=x\sim\operatorname{Exp}(e^{\beta_0x})$, with administrative censoring at $\tau$. A generator that replaces $\beta_0$ with $\beta(P_n)$ and retains the same covariate distribution and censoring mechanism satisfies $\beta(\widehat P_n)=\beta(P_n)$. In this case, $c_n=0$, hence \Cref{ass:cox_weak_calibration} holds.
\end{remark}



For a general generator such as GAN or diffusion, \Cref{ass:cox_weak_calibration} is not guaranteed. In practice, $c_n$ can be approximated using a large independent sample from the fitted generator to access calibration accuracy, see also \Cref{subsec:real-convergence}. 

\begin{theorem}[Validity and fidelity of corrected Cox inference] \label{thm:cox_pvalue_preservation} 
Suppose Assumptions~\ref{ass:cox_parameter}--\ref{ass:cox_weak_calibration} hold and let $n,m\to\infty$. Then for every $\widehat P_n \in\mathcal P_0$ satisfying $W_1(\widehat P_n,P_0)=O_p(r_n), r_n\to0$, under the null, we have:
\begin{enumerate}[label=(\roman*),leftmargin=*]

\item \textbf{Consistency under correction.} 
\begin{align*}
&\frac{
\beta_j(\widehat P_{n,m})-\beta_j(P_0)
}{
\widehat s_{\mathrm{corrected},j}
}
\xlongrightarrow{d} N(0,1),\qquad\sup_{u\in[0,1]} \left| \Pr(p_{\mathrm{corrected},j}\le u)-u \right| \longrightarrow0.
\end{align*}
In addition, if $m/n\to\infty$, then 
for every fixed $\alpha\in(0,1)$,
\begin{align*}
&Z_{\mathrm{corrected},j}-Z_{\mathrm{real},j} =o_p(1),\qquad p_{\mathrm{corrected},j}-p_{\mathrm{real},j} =o_p(1),\\
&\Pr\left\{ \mathbf 1(p_{\mathrm{corrected},j}\le\alpha) \neq \mathbf 1(p_{\mathrm{real},j}\le\alpha) \right\} \longrightarrow0.
\end{align*}

\item \textbf{Rates under correction.}
Assume $m/n\to\infty$ and $c_n=O_p(\epsilon_n)$ for some deterministic sequence $\epsilon_n=o(n^{-1/2})$ such that
\begin{align*}
Z_{\mathrm{corrected},j}-Z_{\mathrm{real},j}
&=
O_p\left(
\sqrt n\,\epsilon_n
+r_n+n^{-1/2}
+\sqrt{n/m}
\right),\\
p_{\mathrm{corrected},j}-p_{\mathrm{real},j}
&=
O_p\left(
\sqrt n\,\epsilon_n
+r_n+n^{-1/2}
+\sqrt{n/m}
\right).
\end{align*}
\end{enumerate}
\end{theorem}
In Part (ii), the terms $\sqrt n$ and $\epsilon_n$ quantify the contribution of imperfect score calibration to the discrepancy between corrected synthetic and original data inference. In particular, if $r_n=n^{-1/2}$, $c_n=O_p(r_n/\sqrt{n})=O(n^{-1})$, and $m=\Omega(n^2)$ so that $\sqrt{n/m}=O(r_n)$, then \[Z_{\mathrm{corrected},j}-Z_{\mathrm{real},j}=O_p(r_n),\qquad
p_{\mathrm{corrected},j}-p_{\mathrm{real},j}=O_p(r_n).\]
Thus, under sufficiently accurate score calibration and a sufficiently large synthetic sample size, the distributional convergence rate $r_n$ of the generator is inherited by the downstream Wald statistics and $p$-values.

When the \Cref{ass:cox_weak_calibration} 
is not satisfied, standard-error correction alone does not ensure valid inference. We then seek to calibrate $\widehat P_n$ so that its score $c_n$ is sufficiently small.

\begin{theorem}[Inferential fidelity under distribution calibration]
\label{thm:cox_pvalue_preservation_cali}
Suppose Assumptions~\ref{ass:cox_parameter}--\ref{ass:cox_w1_times} hold. Further assume that there exists a calibrated generative distribution $\overline P_n\in\mathcal P_0$ such that 
\[\|U_{\overline P_n}\{\beta(P_n)\}\|=O_p(r_n/\sqrt n),\qquad W_1(\overline P_n,P_0)=O_p(r_n).\] Then 
under the null and $m/n\to\infty$,
\begin{align*}
Z_{\mathrm{corrected},j}-Z_{\mathrm{real},j}
&=
O_p\left(
r_n+n^{-1/2}
+\sqrt{n/m}
\right),\\
p_{\mathrm{corrected},j}-p_{\mathrm{real},j}
&=
O_p\left(
r_n+n^{-1/2}
+\sqrt{n/m}
\right).
\end{align*}
\end{theorem}

If $r_n=n^{-1/2}$ and $m=\Omega(n^2)$, both discrepancies above are $O_p(r_n)$. Thus, sufficiently accurate calibration allows the distributional convergence rate $r_n$ of the generator to transfer to downstream Wald statistics and $p$-values. 

Next, we investigate whether a calibrated distribution $\overline P_n$ satisfying the conditions of \Cref{thm:cox_pvalue_preservation_cali} can be constructed from $\widehat P_n$ without substantially compromising its distributional fidelity. We show that this is indeed possible through a local adjustment of the fitted distribution using the original-data Cox estimate \(\beta(P_n)\) as the calibration target.


\noindent\textbf{Calibration of the generative distribution.}\label{sec:calibration} We calibrate $\widehat P_n$
through the exponential tilt
\[
\frac{d\overline P_{n,\lambda}}{d\widehat P_n}(D)
\coloneqq
\exp\left\{
\lambda^\top\xi_{\widehat P_n}(D)-\zeta_{\widehat P_n}(\lambda)
\right\},
\qquad
\zeta_{\widehat P_n}(\lambda)
\coloneqq
\log\mathbb E_{\widehat P_n}
\left[
e^{\lambda^\top\xi_{\widehat P_n}(D)}
\right].
\]
Here, $\xi_{\widehat P_n}(D)$ is defined in Equation~\eqref{eq:cox_xi}, evaluated at $\beta(\widehat P_n)$. We choose $\lambda$ so that $U_{\overline P_{n,\lambda}}\{\beta(P_n)\}=0$ and denote the resulting distribution by $\overline P_n$. 
For a general distribution $P$, let $P_\lambda$ denote the corresponding exponentially tilted distribution. The following theorem establishes that, for any target Cox coefficient $q$ sufficiently close to $\beta(P)$, there exists a local tilt that calibrates the Cox coefficient to $q$ without substantially changing the distribution. 

\begin{theorem}[Local existence and stability of Cox calibration]
\label{thm:cox_calibration} 
Suppose Assumptions~\ref{ass:cox_parameter}-- \ref{ass:cox_w1_times} hold, and $\Sigma(P_0)$ is positive definite. Further assume that $(\lambda,\beta)\mapsto U_{P_\lambda}(\beta)$ is continuously differentiable in the neighborhood of $(0,\beta(P))$, with uniformly bounded and uniformly continuous first derivatives over $P\in\mathcal P_0$. 
Then there exist constants $\varepsilon>0$, $\delta>0$, and $C<\infty$ such that, for every $P\in\mathcal P_0$ satisfying $W_1(P,P_0)\le\varepsilon$ and every target $q\in\mathbb R^p$ satisfying $\|q-\beta(P)\|\le\delta$, there exists a unique local calibration parameter $\lambda(q)$ such that
$U_{P_{\lambda(q)}}(q)=0,
\beta(P_{\lambda(q)})=q,$
and
\[
\|\lambda(q)\|
+d_{\mathrm{TV}}(P_{\lambda(q)},P)
+W_1(P_{\lambda(q)},P)
\le C\|q-\beta(P)\|.
\]
\end{theorem}

In particular, suppose $W_1(\widehat P_n,P_0)=O_p(r_n), r_n\to0$. Applying the calibration to $\widehat P_n$ with target $q=\beta(P_n)$ yields a distribution $\overline P_n$ satisfying 
\[\beta(\overline P_n)=\beta(P_n),\qquad W_1(\overline P_n,P_0) = O_p(r_n).\]
Thus, by \Cref{thm:cox_pvalue_preservation_cali}, the above calibration exactly preserves the original data Cox coefficient while retaining the generator's distributional convergence rate. In practice, this calibration can be implemented with samples from $\widehat P_n$ to numerically solve the score equation, followed by a finite-pool Sample-Importance-Resampling (SIR) approximation to the resulting tilted distribution. The  calibration algorithm is summarized in \Cref{app:algorithm} Algorithm~\ref{alg:cox_calibration}.

\section{Simulation}\label{sec:sim}

We next examine our theoretical results through simulations.

\subsection{Simulation Study Verifying Convergence Rate}
\label{subsec:simulation-convergence}

We first studied how joint-distributional error changes with training sample size for SurvivalGAN and two implementations of our proposed TSGSurv using GAN (TSGSurv(GAN)) and diffusion (TSGSurv(diff)).

\noindent\textbf{Data generation process.} We generated data from the nonlinear accelerated failure-time model
\[
X\sim N(0,1),\qquad
\log T=0.5X+0.3X^2+0.5\varepsilon,\qquad
\varepsilon\sim N(0,1).
\]
The censoring time followed a shifted Weibull distribution with shape parameter being 2. The shift was the 0.05 quantile of the marginal event-time distribution and the scale was selected to produce 50\% censoring.

\noindent\textbf{Experiment 1: Distributional convergence rate verification.}
To study the distributional convergence rate, we fitted SurvivalGAN, TSGSurv(GAN), and TSGSurv(diff) on 35 replicates at each training size \(n\in\{2{,}000,4{,}000,6{,}000,8{,}000,12{,}000,18{,}000\}\). Each fitted model generated $m=200{,}000$ synthetic data, which were compared with a fixed, disjoint reference sample of the same size. We reported the median $W_1$ and RBF-kernel MMD distances~\citep{gretton2012kernel,peyre2019computational} between standardized $(X,Y,\Delta)$ vectors across the 35 fits. Because computing $W_1$ is computationally demanding at the sample sizes considered, we use sliced $W_1$ as a tractable approximation throughout this article. 


\begin{figure}[!h]
    \centering
    \includegraphics[width=1\linewidth]{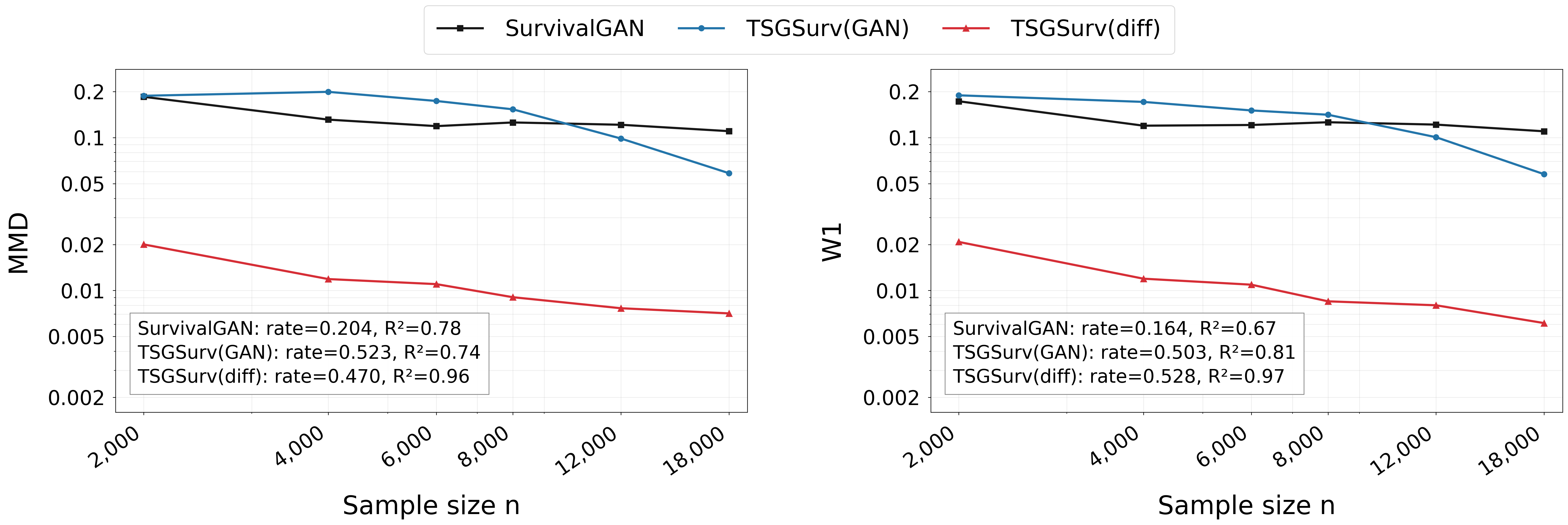}
    \caption{Empirical convergence rates of MMD and $W_1$.} 
    \label{fig:simulation-joint-convergence}
\end{figure}

As shown in \Cref{fig:simulation-joint-convergence}, both TSGSurv(GAN) and TSGSurv(diff) framework achieves an empirical convergence rate close to $n^{-1/2}$, for both metrics, consistent with \Cref{thm:rate_2_stage}. In contrast, SurvivalGAN improves much more slowly, with a clear plateau in both metrics, consistent with the irreducible approximation error established in \Cref{lem:deterministic_error2}.

\noindent\textbf{Experiment 2: Cox $\beta$ convergence rate verification.}
We next examined whether the improvement in distributional fidelity was accompanied by improved Cox estimation for TSGSurv(diff), which is easier to train in practice with better performance based on Experiment 1 compared to TSGSurv(GAN). Using the same setting, we compared $\beta(\widehat P_{n,m})$ and $\Sigma(\widehat P_{n,m})$ with their large-sample approximations for the population counterparts $\beta(P_0)$ and $\Sigma(P_0)$  across training sample sizes. 
\begin{figure}[!h]
    \centering
    \includegraphics[width=0.6\linewidth]{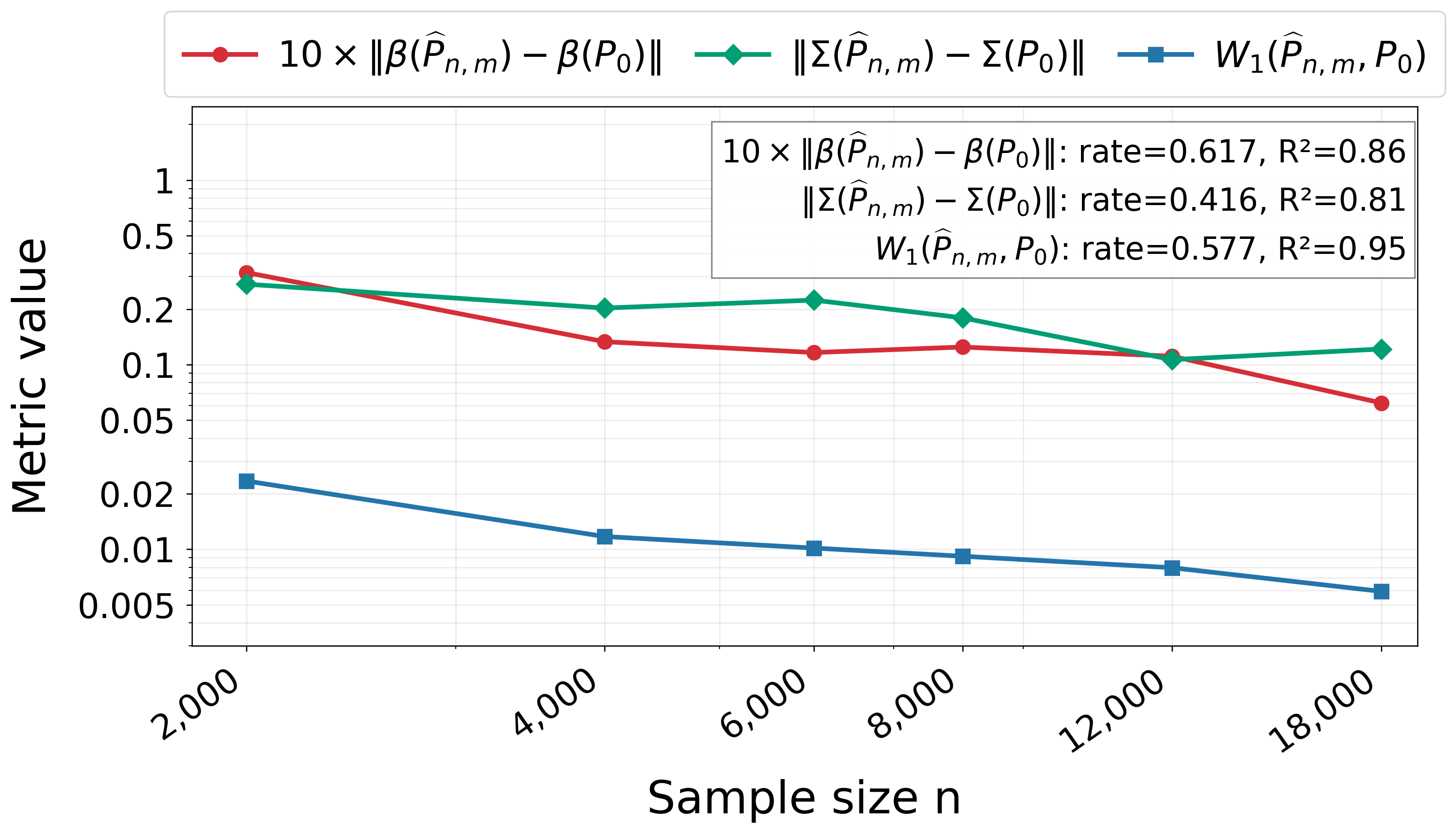}
    \caption{Empirical convergence of Cox coefficient error, $W_1$, and sandwich-covariance error for TSGSurv(diff). For presentation purposes, $\|\beta(\widehat P_{n,m})-\beta(P_0)\|$ is multiplied by 10.}
    \label{fig:simulation5.1-beta-bias-rate}
\end{figure}

As shown in \Cref{fig:simulation5.1-beta-bias-rate}, the empirical error rates for both $\beta$ and $\Sigma$ are close to our theoretical results in \Cref{thm:cox_wasserstein_stability} and \Cref{thm:cox_covariance_stability}.



\subsection{Simulation Study Assessing Cox-specific Properties}
\label{subsec:simulation-evaluation}

To study the estimation and inferential properties of the Cox model, we generated data under a 30-dimensional proportional-hazards model, in which the true Cox $\beta$ is known.

\noindent\textbf{Data generation process.}
We generated $X\sim N_{30}(0,\Sigma)$ with $\Sigma_{jk}=0.6^{|j-k|}$ and linear predictor $\eta=\beta^\top X$. The coefficients were generated as $\beta_j\overset{\mathrm{i.i.d.}}{\sim}\operatorname{Uniform}(-0.5,0.5)$, $j=1,\cdots,30$. The resulting $\beta$ was held fixed across settings S1, S3, and S4. For S2, we imposed sparsity by retaining five randomly selected nonzero coefficients and rescaled them so that all four settings had the same $\Var(\beta^\top X)$. Then, given $U\sim\operatorname{Unif}(0,1)$, the event time was $T=\left\{\frac{-\log U}{\exp(\eta)}\right\}^{1/0.7},$ corresponding to the Weibull baseline cumulative hazard $H_0(t)=t^{0.7}$. Censoring was independent, with $C\sim\operatorname{Unif}(0,c)$, and the observed data were $Y=\min(T,C)$ and $\Delta=I(T\leq C)$. 

For each setting, we generated 100 replicates of size $n=10{,}000$. Each method was fitted separately to every original sample and produced 10 synthetic samples of size $m=10n=100{,}000$. The naive and corrected statistics were always computed from the same synthetic sample. We compared TSGSurv(diff) with SurvivalGAN, SurvDiff, and Ashhad's conditional outcome generator. Cox models included only covariates with nonzero data-generating coefficients. Each metric was first summarized by the median over the 10 synthetic samples belonging to an original sample and then by the median over the 100 original samples. \Cref{tab:simulation-settings} summarizes the sparsity and event rate of each setting, with additional details in \Cref{app:protocol-details}.

\begin{table}[!h]
\centering
\caption{Data generation mechanisms.}
\label{tab:simulation-settings}
\small
\begin{tabular}{lll}
\toprule
Setting & Sparsity of $\beta$ & Event rate \\
\midrule
S1 & $\|\beta\|_0=30$ & 0.439 \\
S2 & $\|\beta\|_0=5$ & 0.440 \\
S3 & $\|\beta\|_0=30$ & 0.200 \\
S4 & $\|\beta\|_0=30$ & 0.700 \\
\bottomrule
\end{tabular}
\end{table}


\noindent\textbf{Distributional and censor-aware dependence fidelity.}
Event-rate gap is $|\widehat P_{\mathrm{syn}}(\Delta=1)-\widehat P_{\mathrm{real}}(\Delta=1)|$.
KM ISE and NA ISE integrate squared differences between the synthetic and original Kaplan--Meier survival curves and Nelson--Aalen cumulative hazards\citep{kaplan1958nonparametric,nelson1972theory,aalen1978nonparametric}.
IPCW Kendall MAE is the mean absolute difference between covariate specific,  inverse-probability-of-censoring weighted Kendall associations, and censored dCor gap is the absolute difference in the corresponding censor aware distance  correlation summary. Smaller values indicate better fidelity for every metric in \Cref{tab:simulation-fidelity}.
\noindent\textbf{Distributional and censor-aware dependence fidelity.}
Event-rate gap is $|\widehat P_{\mathrm{syn}}(\Delta=1)-\widehat P_{\mathrm{real}}(\Delta=1)|$.
KM ISE and NA ISE integrate squared differences between the synthetic and original Kaplan--Meier survival curves and Nelson--Aalen cumulative hazards \citep{kaplan1958nonparametric,nelson1972theory,aalen1978nonparametric}.
IPCW Kendall MAE is the mean absolute difference between covariate-specific, inverse-probability-of-censoring-weighted Kendall associations, and censored dCor gap is the absolute difference in the corresponding censor-aware distance correlation summary. Smaller values indicate better fidelity for every metric in \Cref{tab:simulation-fidelity}.

\begin{table}[!h]
\centering
\caption{Distributional and censor-aware dependence fidelity in simulation settings S1--S4.}
\label{tab:simulation-fidelity}
\resizebox{\textwidth}{!}{%
\begin{tabular}{llrrrrrrr}
\toprule
Setting & Method & $W_1$ & MMD & Event-rate gap & KM ISE & NA ISE & IPCW Kendall MAE & Censored dCor gap \\
\midrule
S1 & TSGSurv(diff) & \textbf{0.0257} & \textbf{0.0240} & 0.00101 & $1.97\times10^{-5}$ & $5.85\times10^{-5}$ & \textbf{0.0225} & \textbf{0.0304} \\
 & SurvivalGAN & 0.0814 & 0.0639 & 0.00106 & $2.33\times10^{-3}$ & $6.73\times10^{-3}$ & 0.0248 & 0.0582 \\
 & SurvDiff & 0.215 & 0.144 & 0.0556 & $1.79\times10^{-3}$ & $6.09\times10^{-3}$ & 0.0864 & 0.166 \\
 & Ashhad & 0.169 & 0.0807 & \textbf{0.000} & $\boldsymbol{1.28\times10^{-5}}$ & $\boldsymbol{1.83\times10^{-5}}$ & 0.0294 & 0.121 \\
\midrule
S2 & TSGSurv(diff) & \textbf{0.0256} & \textbf{0.0239} & 0.00104 & $1.71\times10^{-5}$ & $5.06\times10^{-5}$ & \textbf{0.0223} & \textbf{0.0305} \\
 & SurvivalGAN & 0.0863 & 0.0667 & 0.00126 & $2.31\times10^{-3}$ & $6.60\times10^{-3}$ & 0.0248 & 0.0612 \\
 & SurvDiff & 0.225 & 0.143 & 0.0514 & $1.81\times10^{-3}$ & $5.85\times10^{-3}$ & 0.0771 & 0.178 \\
 & Ashhad & 0.183 & 0.0838 & \textbf{0.000} & $\boldsymbol{1.30\times10^{-5}}$ & $\boldsymbol{1.86\times10^{-5}}$ & 0.0295 & 0.126 \\
\midrule
S3 & TSGSurv(diff) & \textbf{0.0171} & \textbf{0.0184} & 0.000860 & $1.17\times10^{-5}$ & $2.09\times10^{-5}$ & \textbf{0.0355} & \textbf{0.0559} \\
 & SurvivalGAN & 0.0710 & 0.0561 & 0.00130 & $1.30\times10^{-3}$ & $2.12\times10^{-3}$ & 0.0388 & 0.0707 \\
 & SurvDiff & 0.151 & 0.120 & 0.0301 & $1.32\times10^{-3}$ & $2.23\times10^{-3}$ & 0.0988 & 0.237 \\
 & Ashhad & 0.0591 & 0.0210 & \textbf{0.000} & $\boldsymbol{7.37\times10^{-6}}$ & $\boldsymbol{9.57\times10^{-6}}$ & 0.0449 & 0.142 \\
\midrule
S4 & TSGSurv(diff) & \textbf{0.0189} & \textbf{0.0183} & 0.000998 & $\boldsymbol{6.22\times10^{-6}}$ & $\boldsymbol{1.67\times10^{-5}}$ & \textbf{0.0167} & \textbf{0.0229} \\
 & SurvivalGAN & 0.137 & 0.0933 & 0.000958 & $2.18\times10^{-3}$ & $8.54\times10^{-3}$ & 0.0235 & 0.0679 \\
 & SurvDiff & 0.674 & 0.142 & 0.100 & $2.36\times10^{-3}$ & $7.36\times10^{-3}$ & 0.0923 & 0.0414 \\
 & Ashhad & 0.377 & 0.163 & \textbf{0.000} & $2.13\times10^{-4}$ & $2.98\times10^{-4}$ & 0.0532 & 0.162 \\
\bottomrule
\end{tabular}}
\end{table}

TSGSurv(diff) attains the smallest $W_1$ and MMD in all four settings and provides the best censor-aware dependence preservation. Ashhad exactly reproduces the event proportion,  
but its distributional and dependence errors are materially larger. The contrast is most pronounced in S4, where SurvDiff and Ashhad deteriorate under the higher event rate while TSGSurv(diff) remains stable.

The relatively large errors for Ashhad may partly reflect the difficulty of learning the high-dimensional conditional distribution $X\mid(Y,\Delta)$ when $Y$ is continuous. 
In the high-event-rate setting S4, the upper bound $c$ in $C\sim\operatorname{Uniform}(0,c)$ is large, so the observed times among censored records may be spread over a wider range. This reduces the sample density at particular values of $Y$ and may make continuous conditional generation more difficult. In the low-event-rate setting S3, the event group is smaller, creating class imbalance, but the smaller value of \(c\) concentrates the censored records over a narrower time range. The observed pattern therefore suggests that both event-status imbalance and local sparsity over the conditioning-time support may contribute. 

SurvDiff’s performance may reflect a different trade-off. Its training objective combines a diffusion loss with a Cox-based ranking loss. The ranking component emphasizes relative risk ordering rather than the exact joint distribution of event and censoring times, and the two objectives may not always reinforce each other. This additional model-specific objective may therefore weaken distributional learning. 

\noindent\textbf{Cox estimation and inferential fidelity.}
In \Cref{tab:simulation-cox}, the two RMSEs compare the estimated coefficients using synthetic data with the true parameter and the estimate using original data. C-index and integrated Brier score (IBS) gaps compare synthetic-trained and original-trained Cox models on the same original sample. Wald agreement is the proportion of coefficients for which the original and synthetic analyses reach the same two-sided decision at level 0.05. Because $m=10n$, the corrected synthetic standard error is $\hat{s}_{\mathrm{corrected},j}=\sqrt{11}\hat{s}_{\mathrm{syn},j}$. Breslow ISE compares fitted cumulative baseline hazards. Wald agreement is the proportion of coefficient-level decisions that agree between the original and synthetic analyses at the two-sided level $0.05$, pooled across all original samples and synthetic replicates. For each synthetic replicate, the final column averages $-\log_{10}|p_{\mathrm{corrected},j}-p_{\mathrm{real},j}|$ across modeled coefficients. 

\begin{table}[!h]
\centering
\caption{Cox effect and inferential fidelity in simulation settings S1--S4.}
\label{tab:simulation-cox}
\resizebox{\textwidth}{!}{%
\begin{tabular}{llrrrrrrrr}
\toprule
Setting & Method & \shortstack{$\beta(\widehat P_{n,m})$ vs. $\beta(P_0)$\\RMSE} & \shortstack{$\beta(\widehat P_{n,m})$ vs. $\beta(P_n)$\\RMSE} & \shortstack{C-index\\gap} & \shortstack{IBS\\gap} & \shortstack{Naive Wald\\agreement} & \shortstack{Corrected Wald\\agreement} & \shortstack{Breslow\\ISE} & \shortstack{Corrected mean\\$-\log_{10}|\Delta p|\uparrow$} \\
\midrule
S1 & TSGSurv(diff) & \textbf{0.0314} & \textbf{0.0197} & \textbf{0.000294} & \textbf{0.000258} & \textbf{94.2\%} & \textbf{97.5\%} & \textbf{$9.93\times10^{-5}$} & \textbf{38.2} \\
 & SurvivalGAN & 0.0944 & 0.0931 & 0.00594 & 0.00858 & 90.9\% & 89.8\% & 0.0115 & 32.2 \\
 & SurvDiff & 0.108 & 0.108 & 0.0110 & 0.0345 & 91.6\% & 91.1\% & 0.0364 & 37.8 \\
 & Ashhad & 0.199 & 0.198 & 0.0158 & 0.0172 & 90.0\% & 85.5\% & 0.00633 & 19.9 \\
\midrule
S2 & TSGSurv(diff) & \textbf{0.0240} & \textbf{0.0190} & \textbf{$4.61\times10^{-5}$} & \textbf{$7.89\times10^{-5}$} & \textbf{100\%} & \textbf{100\%} & \textbf{0.000128} & \textbf{290} \\
 & SurvivalGAN & 0.0474 & 0.0437 & 0.000259 & 0.00425 & \textbf{100\%} & \textbf{100\%} & 0.0115 & 286 \\
 & SurvDiff & 0.243 & 0.241 & 0.00346 & 0.0338 & 98.8\% & 96.8\% & 0.0389 & 214 \\
 & Ashhad & 0.460 & 0.458 & 0.00330 & 0.0240 & \textbf{100\%} & \textbf{100\%} & 0.00902 & 177 \\
\midrule
S3 & TSGSurv(diff) & \textbf{0.0418} & \textbf{0.0282} & \textbf{0.000325} & \textbf{0.000292} & \textbf{91.9\%} & \textbf{95.1\%} & \textbf{$4.32\times10^{-5}$} & 17.3 \\
 & SurvivalGAN & 0.138 & 0.139 & 0.0105 & 0.00755 & 88.1\% & 84.5\% & 0.000961 & 11.8 \\
 & SurvDiff & 0.0946 & 0.0940 & 0.00991 & 0.0191 & 89.9\% & 90.3\% & 0.00153 & \textbf{19.2} \\
 & Ashhad & 0.132 & 0.132 & 0.00723 & 0.00563 & 89.7\% & 88.6\% & 0.00223 & 12.1 \\
\midrule
S4 & TSGSurv(diff) & \textbf{0.0238} & \textbf{0.0166} & \textbf{0.000150} & \textbf{0.000130} & \textbf{94.6\%} & \textbf{97.2\%} & \textbf{$4.75\times10^{-5}$} & 58.3 \\
 & SurvivalGAN & 0.151 & 0.148 & 0.00108 & 0.0137 & 92.8\% & 93.0\% & 0.112 & \textbf{71.5} \\
 & SurvDiff & 0.127 & 0.126 & 0.0139 & 0.0313 & 92.4\% & 91.9\% & 0.0227 & 49.5 \\
 & Ashhad & 0.265 & 0.263 & 0.0434 & 0.0393 & 88.9\% & 81.3\% & 0.0184 & 19.7 \\
\bottomrule
\end{tabular}}
\end{table}

TSGSurv(diff) has the smallest coefficient RMSE, C-index gap, IBS gap, and Breslow ISE in every setting, and it attains the highest or tied-highest naive Wald agreement. After variance correction, it has the highest corrected Wald agreement in S1, S3, and S4 and remains tied at the ceiling value of $100\%$ in S2. For the corrected mean $-\log_{10}|\Delta p|$, TSGSurv(diff) performs best in S1 and S2, SurvDiff performs best in S3, and SurvivalGAN performs best in S4. In the sparse setting S2, all five modeled effects are strong, giving $100\%$ decision agreement for every method. The corresponding negative log absolute differences are strongly influenced by near-zero $p$-values and should therefore be interpreted together with the effect-estimation and predictive metrics.

For TSGSurv(diff), the variance correction improves Wald agreement from $94.2\%$ to $97.5\%$ in S1, from $91.9\%$ to $95.3\%$ in S3, and from $94.6\%$ to $97.2\%$ in S4. In S2, both the naive and corrected analyses already attain the ceiling value of $100\%$. These results provide empirical evidence for the practical effectiveness of the proposed variance correction and are consistent with the validity and inferential fidelity results in \Cref{thm:cox_pvalue_preservation}.

For the competing generators, the correction either leaves agreement unchanged or reduces it in several settings. This contrast does not imply that variance correction is method-specific. \Cref{thm:cox_pvalue_preservation} applies to any fitted generator satisfying the required distributional and score function assumptions. However, the competing generators may not satisfy these assumptions. For example, SurvivalGAN does not satisfy the distributional convergence assumption as established in \Cref{lem:deterministic_error2}. We therefore report their corrected agreements as descriptive comparisons.


\noindent\textbf{Coefficient attenuation in the sparse setting.}
To examine the coefficient-level patterns, we focus on S2, the only sparse setting, whose five nonzero coefficients permit a concise coefficient-by-coefficient comparison. \Cref{tab:simulation-s2-coefficients} reports the absolute differences between the true coefficients and the median Cox estimates obtained from each method's synthetic samples.
\begin{table}[!h]
\centering
\caption{Absolute errors
$\left|\beta_j(\widehat P_{n,m})-\beta_j(P_0)\right|$
for nonzero coefficients in setting S2.}
\label{tab:simulation-s2-coefficients}
\small
\begin{tabular}{lccccc}
\toprule
Method & $x_{2}$ & $x_{6}$ & $x_{15}$ & $x_{19}$ & $x_{27}$ \\
\midrule
TSGSurv(diff) & \textbf{0.001} & \textbf{0.002} & \textbf{0.017} & \textbf{0.006} & \textbf{0.001} \\
SurvivalGAN & 0.012 & 0.013 & 0.019 & 0.008 & 0.009 \\
SurvDiff & 0.213 & 0.212 & 0.281 & 0.196 & 0.248 \\
Ashhad & 0.451 & 0.451 & 0.568 & 0.352 & 0.454 \\
\bottomrule
\end{tabular}
\end{table}

Substantial coefficient attenuation coexists with complete Wald decision agreement because all five modeled effects are sufficiently strong that even markedly attenuated estimates remain statistically significant. TSGSurv(diff) has the smallest absolute error for every nonzero coefficient. SurvivalGAN also retains the five coefficients relatively closely, although its errors are consistently larger. SurvDiff exhibits moderate attenuation toward zero, whereas Ashhad produces substantial attenuation for all five covariates despite its conditional generation design.

Both our framework and the Ashhad method avoid deterministic time prediction and use valid factorizations of the observed-data distribution. But Ashhad requires a potentially high-dimensional covariate generator conditional on continuous follow-up time and event status, whereas our framework jointly generates covariates and time within each event-status group. Conditioning on continuous time can place additional demands on learning how covariate distributions vary across follow-up, particularly in sparsely observed regions.  

Coefficient discrepancies, baseline-hazard errors, and inferential comparisons capture different consequences of such distortion and therefore complement joint-distributional and censor-aware dependence metrics. In particular, the complete Wald agreement in S2 should be interpreted together with the coefficient errors, because near-zero $p$-values can conceal substantial attenuation in estimated effect sizes.
\subsection{Simulation Study Assessing Corrected and Calibrated Cox Inference}
\label{subsec:simulation-inference}

In \Cref{subsec:simulation-evaluation}, the score residual $c_n$ of each fitted generator is unknown, so the separate roles of the variance correction and the distribution calibration cannot be identified. We therefore conducted two controlled generator settings in which the generator, and hence $c_n$, is known by construction. Both experiments share the same data-generating process and differ only in the fitted generator. They verify the mechanisms described in \Cref{prop:cox_naive_invalid}, \Cref{thm:cox_pvalue_preservation,thm:cox_pvalue_preservation_cali,thm:cox_calibration}, and \Cref{alg:cox_calibration}. 

\noindent\textbf{Data-generating process.}
We generated $n=1,000$ samples by $X\sim\operatorname{Bernoulli}(0.5)$ and
$T\mid X=x\sim\operatorname{Exp}(e^{\beta_0x})$, with $\beta_0=0$.
Administrative censoring was imposed at $\tau=1$, so that
$Y=\min(T,1)$ and $\Delta=\ind(T\leq 1)$. The expected event rate is
$1-e^{-1}\approx0.632$.

We considered two cases, according to whether calibration was needed.
We first considered the oracle generator described following
\Cref{ass:cox_weak_calibration}. The oracle generator $\widehat P_n$
draws $X^*\sim\operatorname{Bernoulli}(1/2)$ and
$T^*\mid X^*=x
\sim
\operatorname{Exp}\!\left(e^{\beta(P_n)x}\right),$
with the same administrative censoring at $\tau=1$. It satisfies
$\beta(\widehat P_n)=\beta(P_n)$ and therefore $c_n=0$.
Let $m=10n$, \Cref{prop:cox_naive_invalid} gives $
Z_{\mathrm{syn}}\xrightarrow{d}N(0,1+m/n)=N(0,11),$
so that the limiting null rejection probability of the naive test is
\[
2\left[1-\Phi\left\{
\frac{z_{0.975}}{\sqrt{11}}
\right\}\right]\approx0.555.
\]
By contrast, \Cref{thm:cox_pvalue_preservation} gives a limiting null
rejection probability of $0.05$ for the corrected test.

We next considered a perturbed generator with a root-$n$ score residual. It uses
the same covariate distribution and censoring mechanism as the oracle
generator, but draws
\[
T^*\mid X^*=x
\sim
\operatorname{Exp}\!\left[
 \exp\left\{
 \left(\beta(P_n)
 +2\{\Sigma(P_n)/n\}^{1/2}\right)x
 \right\}
\right].
\]
Consequently,
$\beta(\widehat P_n)-\beta(P_n) = 2\{\Sigma(P_n)/n\}^{1/2}.$
The generator remains consistent, with
$W_1(\widehat P_n,P_0)=O_p(n^{-1/2})$. However, its score residual is
of exact order $n^{-1/2}$, rather than
$o_p(s_{n,m})$, so \Cref{ass:cox_weak_calibration} fails. In particular,
the standardized location shift is
\[
\frac{
2\{\Sigma(P_n)/n\}^{1/2}
}{
\{\Sigma(P_n)(1/n+1/m)\}^{1/2}
}
=
\frac{2}{\sqrt{1+n/m}}.
\]
It follows that
$Z_{\mathrm{corrected}}
\xrightarrow{d}
N\!\left(2/\sqrt{1+n/m},1\right).$
For $m/n=10$, the limiting mean is approximately $1.91$. Thus, the
variance correction fixes the scale but not the location of the Wald
statistic, resulting in a limiting null rejection probability of
approximately $0.479$.

We then applied \Cref{alg:cox_calibration} to the perturbed generator
with target $q=\beta(P_n)$, yielding the calibrated distribution
$\overline P_n$. Synthetic samples from its numerical approximation
were generated by SIR. Implementation
details are provided in \Cref{app:protocol-details}.

\noindent\textbf{Calibration performance.} After generating each synthetic sample, we fitted a Cox model with
$X$ as the only covariate and tested $H_0:\beta=0$ at level
$\alpha=0.05$ using sandwich standard errors. We use $1{,}000$ replicates to
construct the oracle and perturbed generators, and calibrate the
perturbed generator. We analyzed synthetic samples from the oracle
generator, the uncalibrated perturbed generator, and the calibrated
distribution. 
For each synthetic sample, we computed both the naive and corrected
Wald statistics. For samples drawn from the calibrated distribution,
these analyses are labeled ``Synthetic, calibrated" and
``Synthetic, calibrated + corrected"
, respectively. Because
$\beta_0=0$ is the tested value, the coverage probability of each
two-sided $95\%$ Wald interval is one minus its null rejection
probability. Hence, we report the rejection rates only.

\begin{table}[!h]
\centering
\caption{Null rejection rates and estimates for corrected and calibrated Cox inference.} 
\label{tab:simulation-inference}
\resizebox{\textwidth}{!}{%
\begin{tabular}{llcccccc}
\toprule
Generator & Analysis & Rejection (95\% CI) & Theory & Mean coefficient estimate & SD of coefficient estimate & RMS SE & Agreement \\
\midrule
Oracle: $c_n=0$
& Original data
& 0.048 (0.036, 0.063)
& 0.050 & $-0.003$ & 0.078 & 0.080 & -- \\

& Synthetic, naive
& 0.536 (0.505, 0.567)
& 0.555 & $-0.004$ & 0.081 & 0.025 & 0.512 \\

& \textbf{Synthetic, corrected}
& \textbf{0.039 (0.029, 0.053)}
& 0.050 & $-0.004$ & 0.081 & 0.084 & \textbf{0.969} \\
\midrule

Perturbed: $c_n\asymp n^{-1/2}$
& Original data
& 0.048 (0.036, 0.063)
& 0.050 & $-0.003$ & 0.078 & 0.080 & -- \\

& Synthetic, naive
& 0.913 (0.894, 0.929)
& 0.912 & $\phantom{-}0.157$ & 0.082 & 0.025 & 0.103 \\

& Synthetic, corrected
& 0.492 (0.461, 0.523)
& 0.479 & $\phantom{-}0.157$ & 0.082 & 0.082 & 0.504 \\

& Synthetic, calibrated
& 0.559 (0.528, 0.589)
& 0.555 & $-0.004$ & 0.081 & 0.025 & 0.489 \\

& \textbf{Synthetic, calibrated + corrected}
& \textbf{0.052 (0.040, 0.068)}
& 0.050 & $-0.004$ & 0.081 & 0.082 & \textbf{0.974} \\
\bottomrule
\end{tabular}}
\end{table}

\begin{figure}[!h]
    \centering
    \includegraphics[width=0.9\linewidth]{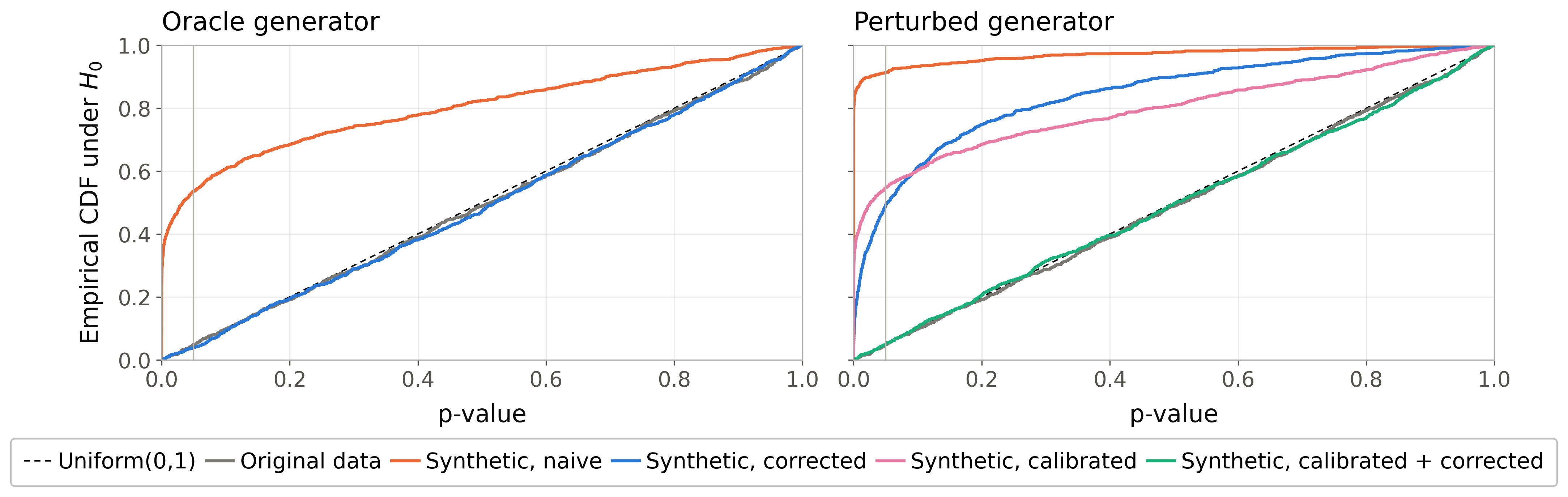}
    \caption{Empirical null distributions of the Cox $p$-values of oracle and perturbed generators. The dashed diagonal is the Uniform$(0,1)$ reference, and the vertical line marks $\alpha=0.05$.}
    \label{fig:simulation-inference-pvalue}
\end{figure}

\Cref{tab:simulation-inference} and \Cref{fig:simulation-inference-pvalue} summarize the two generator settings. Here the coefficient estimate is $\beta(P_n)$ for the original-data analysis, $\beta(\widehat P_{n,m})$ for the uncalibrated synthetic analyses, and $\beta(\overline P_{n,m})$ for the calibrated synthetic analyses. SD denotes the across-replication standard deviation of the Cox coefficient estimates, whereas RMS SE is $\{R^{-1}\sum_{r=1}^R\widehat s_r^2\}^{1/2}$, where $\widehat s_r$ is the standard error reported in replication $r$. \Cref{fig:simulation-inference-rejection} in \Cref{app:protocol-details} plots the rejection rates against their theoretical values.

For the oracle generator, $c_n=0$, so
\Cref{ass:cox_weak_calibration} holds. Although the generator exactly
preserves the original-data Cox coefficient, the naive synthetic test
rejects the true null in $53.6\%$ of replications because it ignores the
uncertainty inherited from the original sample. The variance correction
reduces the rejection rate to $3.9\%$, close to the nominal level. Thus,
when the score-calibration condition holds, variance correction is
sufficient for valid inference.

For the perturbed generator, $c_n\asymp n^{-1/2}$, so
\Cref{ass:cox_weak_calibration} fails. Variance correction alone cannot
remove this generator discrepancy, and the corrected test still rejects
in $49.2\%$ of replications. Calibration removes the discrepancy, and
combining calibration with variance correction restores the rejection
rate to $5.2\%$. Calibration alone remains invalid because it does not
account for the uncertainty inherited from the original sample.
Therefore, both calibration and variance correction are needed in this
setting.

Additional simulations varying $m/n$ support the theoretical results:
naive inference becomes increasingly distorted as the synthetic sample size grows, while variance correction alone cannot remove a root-$n$ generator discrepancy. In contrast, corrected inference under the oracle generator and calibrated and corrected inference under the perturbed generator remain close to the nominal level. Detailed results are provided in \Cref{app:protocol-details}.

\section{Real Data}\label{sec:real}

\subsection{Datasets}
\label{subsec:real-data}
\noindent\textbf{NHANES.} The National Health and Nutrition Examination Survey
(NHANES), conducted by the National Center for Health Statistics of the U.S.
Centers for Disease Control and Prevention, assesses the health and nutritional
status of the U.S. civilian, noninstitutionalized population\citep{nchs2019nhanes}.
Our harmonized analytic cohort combines NHANES III (1988--1994) with the ten
two-year continuous-NHANES cycles from 1999--2000 through 2017--2018. After
complete-case selection, it contains 57,974 participants (13,528 from NHANES
III and 44,446 from continuous NHANES), 13 continuous covariates (age, family income-to-
poverty ratio, body-mass index, weight, height, waist circumference, pulse, systolic blood pressure,
diastolic blood pressure, total cholesterol, HDL cholesterol, glycated hemoglobin, and hemoglobin), follow-up in
months, and all-cause mortality status. The cohort includes 11,250 deaths and
46,724 censored records (80.6\% censoring). 
Details of the source releases, downloaded files, and cohort-construction procedure are provided in Appendix~\ref{app:data-source}.

\noindent\textbf{SUPPORT.}
The Study to Understand Prognoses and Preferences for Outcomes and Risks of
Treatments (SUPPORT) enrolled seriously ill hospitalized adults at five U.S.
tertiary-care academic centers\citep{knaus1995support}. We analyze the 8,873
patients in the preprocessed release, including 6,036 deaths and 2,837 censored
records (32.0\% censoring), using the observed survival time, event indicator, and eight continuous covariates: age, mean
arterial blood pressure, heart rate, respiratory rate, temperature, serum sodium, white blood cell count, and serum creatinine. The exact \texttt{pycox} release and upstream data file are documented in Appendix~\ref{app:data-source}.

\subsection{Convergence Rate Verification}
\label{subsec:real-convergence}

\begin{figure}[htbp]
    \centering
    \includegraphics[width=0.6\linewidth]{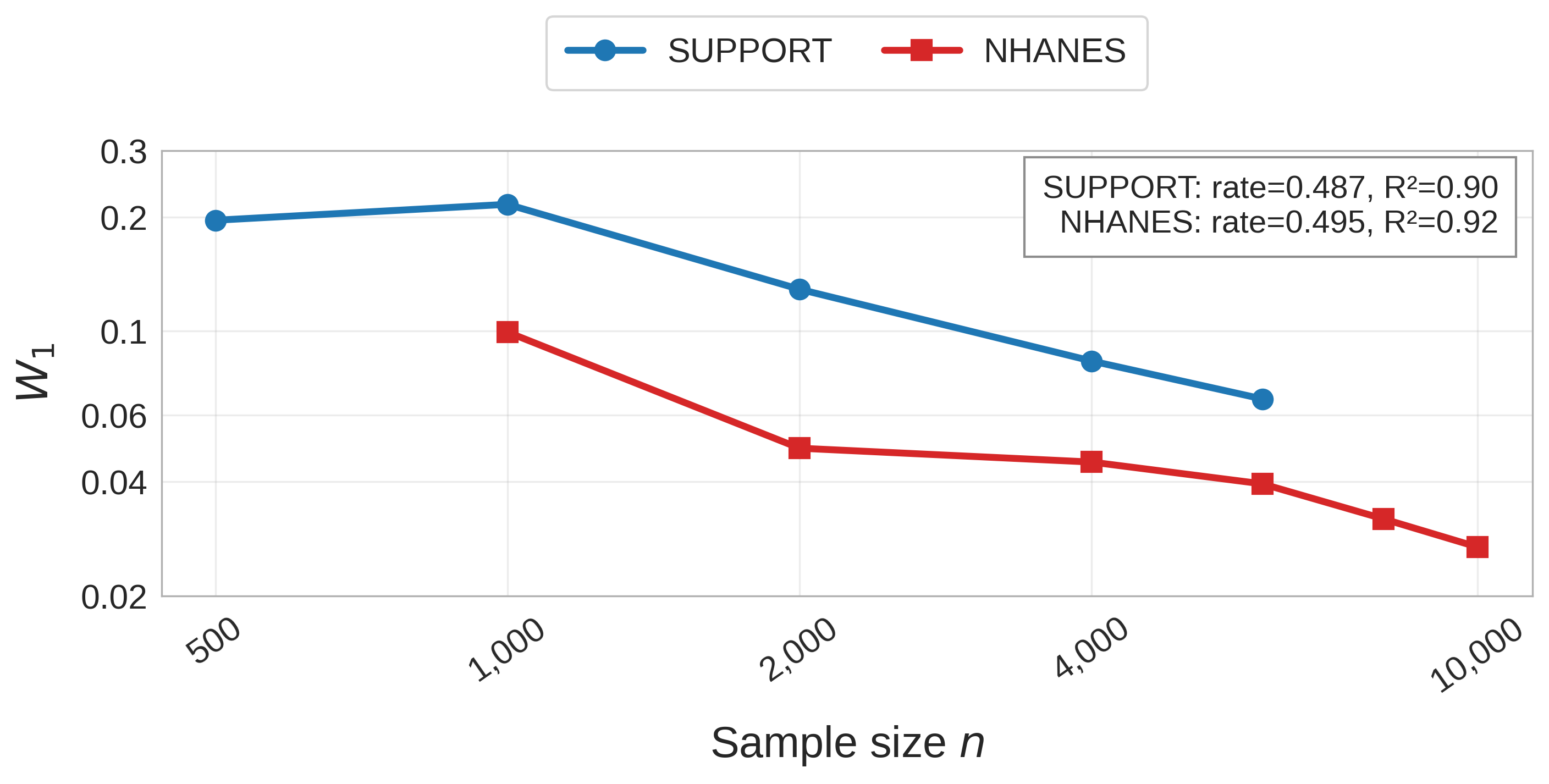}
    \caption{Mean $W_1$ distance for TSGSurv(diff) on real data. }
    \label{fig:real-joint-convergence}
\end{figure}

To study the distributional convergence rate in real data, we fitted TSGSurv(diff) 10 times at each training size. For NHANES, we considered \( n\in\{1{,}000,2{,}000,4{,}000,6{,}000,8{,}000,10{,}000\}\) and m=12{,}000. For SUPPORT, we considered \( n\in\{500,1{,}000,2{,}000,4{,}000,6{,}000\} \) and \(m=2{,}000\). For every training size and dataset, we generated 10 replicates and computed $W_1$ distance after standardizing with the corresponding real reference data.
As shown in \Cref{fig:real-joint-convergence}, the $W_1$ distance decreases with training size in both datasets, with empirical rates close to the theoretical rate $n^{-1/2}$.

\subsection{Cox-specific Properties Assessment}
\label{subsec:real-evaluation}

Having established the empirical distributional convergence of TSGSurv(diff), we next examine whether this distributional fidelity translates into reliable downstream Cox analysis. We further investigate whether the proposed calibration procedure improves the preservation of Cox regression estimates and inferential conclusions.
The evaluation used original and synthetic sample sizes $n=m=20{,}000$ for NHANES and $n=m=6{,}000$ for SUPPORT, with 100 synthetic replicates. We report the same metrics used in \Cref{subsec:simulation-evaluation}, except population-target coefficient RMSEs as $P_0$ and $\beta(P_0)$ are unknown. 


All competing generators were evaluated without calibration. We additionally evaluated the targeted calibration for the fitted TSGSurv(diff) generator, together with its uncalibrated counterpart. For NHANES, the randomly selected calibration target covariates were age (Age), family income-to-poverty ratio (Income-to-poverty ratio), glycated hemoglobin (HbA1c), and systolic blood pressure (Systolic BP). For SUPPORT, they were age (Age), heart rate (Heart rate), serum creatinine (Serum creatinine), and mean arterial blood pressure (Mean arterial BP). The exponential tilt targeted four prespecified coefficients in \Cref{subsec:real-convergence}. In the Cox-specific evaluation reported below, all Cox-based metrics were calculated based on the full Cox model, containing all 13 covariates for NHANES and all eight covariates for SUPPORT. Numerical implementation details are provided in \Cref{app:experiment-details}.

\begin{table}[htbp]
\centering
\caption{Distributional and censor-aware dependence fidelity in real data.}
\label{tab:real-fidelity}
\resizebox{\textwidth}{!}{%
\begin{tabular}{llrrrrrrr}
\toprule
Dataset & Method & $W_1$ & MMD & Event-rate gap & KM ISE & NA ISE &
IPCW Kendall MAE & Censored dCor gap \\
\midrule
NHANES
 & TSGSurv(diff) & 0.02245 & 0.02127 & 0.00183 &
   $4.37\times10^{-5}$ & $1.12\times10^{-4}$ &
   0.0345 & \textbf{0.0297} \\
 & TSGSurv(diff), calibrated & \textbf{0.02237} & \textbf{0.02110} &
   0.00208 & $5.39\times10^{-5}$ & $1.41\times10^{-4}$ &
   \textbf{0.0341} & 0.0340 \\
 & SurvivalGAN & 0.108 & 0.0811 & 0.00170 &
   $1.49\times10^{-3}$ & $2.83\times10^{-3}$ &
   0.0409 & 0.0888 \\
 & SurvDiff & 0.220 & 0.178 & 0.0385 &
   $1.64\times10^{-3}$ & $3.51\times10^{-3}$ &
   0.163 & 0.0442 \\
 & Ashhad & 0.0535 & 0.0455 & \textbf{0.000} &
   $\boldsymbol{9.29\times10^{-6}}$ &
   $\boldsymbol{2.03\times10^{-5}}$ &
   0.0402 & 0.0353 \\
\midrule
SUPPORT
 & TSGSurv(diff) & \textbf{0.0450} & \textbf{0.0227} & 0.00500 &
   $8.04\times10^{-5}$ & $7.61\times10^{-4}$ &
   \textbf{0.0198} & 0.0315 \\
 & TSGSurv(diff), calibrated & 0.0456 & 0.0231 & 0.00408 &
   $5.36\times10^{-5}$ & $5.27\times10^{-4}$ &
   0.0200 & 0.0286 \\
 & SurvivalGAN & 0.0725 & 0.0547 & 0.00850 &
   $6.19\times10^{-3}$ & $4.38\times10^{-2}$ &
   0.0378 & 0.111 \\
 & SurvDiff & 0.148 & 0.0892 & 0.0642 &
   $5.44\times10^{-3}$ & $3.94\times10^{-2}$ &
   0.0353 & 0.0636 \\
 & Ashhad & 0.0920 & 0.0600 & \textbf{0.000} &
   $\boldsymbol{2.90\times10^{-5}}$ &
   $\boldsymbol{1.46\times10^{-4}}$ &
   0.0261 & \textbf{0.0198} \\
\bottomrule
\end{tabular}}
\end{table}

As shown in \Cref{tab:real-fidelity}, on NHANES, calibration improves the $W_1$, MMD, and  IPCW Kendall MAE of TSGSurv(diff), while the uncalibrated version has the  smallest censored distance-correlation gap. Ashhad exactly preserves the event  rate by construction and has the smallest KM and NA ISEs. On SUPPORT, TSGSurv(diff) has the smallest $W_1$,  MMD, and IPCW Kendall MAE. Calibration reduces its event-rate gap and marginal survival-curve errors, but remains almost the same distributional and dependence  discrepancies. SurvDiff and SurvivalGAN generally have the largest event-rate and survival-curve discrepancies. These results show  that calibration can have dataset-dependent effects on distributional fidelity.

\begin{table}[htbp]
\centering
\caption{Cox effect and inferential fidelity in real data.}
\label{tab:real-cox-matched}
\resizebox{\textwidth}{!}{%
\begin{tabular}{llrrrrrrr}
\toprule
Dataset & Method &
\shortstack{$\beta(\widehat P_{n,m})$ vs. $\beta(P_n)$\\RMSE} &
\shortstack{C-index\\gap} &
\shortstack{IBS\\gap} &
\shortstack{Naive Wald\\agreement} &
\shortstack{Corrected Wald\\agreement} &
\shortstack{Breslow\\ISE} &
\shortstack{Corrected mean\\$-\log_{10}|\Delta p|\uparrow$} \\
\midrule
NHANES
 & TSGSurv(diff) & 0.159 & \textbf{0.000235} & 0.000201 &
   \textbf{80.8\%} & \textbf{82.4\%} &
   $1.75\times10^{-4}$ & \textbf{26.5} \\
 & TSGSurv(diff), calibrated & 0.174 & 0.000414 &
   \textbf{0.000164} & 78.3\% & 81.8\% &
   $\boldsymbol{1.73\times10^{-4}}$ & 23.0 \\
 & SurvivalGAN & \textbf{0.139} & 0.00218 & 0.00886 &
   62.2\% & 61.5\% &
   $1.14\times10^{-2}$ & 26.0 \\
 & SurvDiff & 0.432 & 0.102 & 0.0655 &
   45.9\% & 42.6\% &
   $1.04\times10^{-2}$ & 0.499 \\
 & Ashhad & 0.154 & 0.000752 & 0.000799 &
   75.8\% & 79.5\% &
   $4.62\times10^{-4}$ & 26.0 \\
\midrule
SUPPORT
 & TSGSurv(diff) & 0.0395 & 0.00368 & 0.00175 &
   \textbf{75.5\%} & 78.0\% &
   $8.30\times10^{-4}$ & 3.31 \\
 & TSGSurv(diff), calibrated & \textbf{0.0348} & 0.00433 &
   \textbf{0.00145} & 74.5\% & \textbf{79.5\%} &
   $6.68\times10^{-4}$ & 3.04 \\
 & SurvivalGAN & 0.0510 & \textbf{0.00179} & 0.00892 &
   72.3\% & 78.8\% &
   $4.75\times10^{-2}$ & \textbf{4.09} \\
 & SurvDiff & 0.0695 & 0.0130 & 0.00657 &
   64.4\% & 72.4\% &
   $2.29\times10^{-2}$ & 2.38 \\
 & Ashhad & 0.0518 & 0.00870 & 0.00188 &
   65.5\% & 68.0\% &
   $\boldsymbol{2.45\times10^{-4}}$ & 3.90 \\
\bottomrule
\end{tabular}}
\end{table}

\Cref{tab:real-cox-matched} shows that, on NHANES, SurvivalGAN achieves the smallest coefficient RMSE, whereas uncalibrated TSGSurv(diff) has the smallest C-index gap and the highest naive and corrected Wald agreement. Calibrated TSGSurv(diff) achieves the smallest IBS gap and Breslow ISE. On SUPPORT, calibrated TSGSurv(diff) achieves the smallest coefficient RMSE and IBS gap, and the highest corrected Wald agreement. Calibration also reduces the coefficient RMSE from $0.0395$ to $0.0348$ and Breslow ISE from $8.30\times10^{-4}$ to $6.68\times10^{-4}$, while SurvivalGAN and Ashhad achieve the smallest C-index gap and Breslow ISE, respectively. Overall, TSGSurv(diff) demonstrates strong Cox estimation and inferential fidelity across both datasets, with calibration preserving its competitive performance and providing additional improvements in several metrics. Notably, these evaluations use the full Cox model rather than the prespecified calibration targets, suggesting that targeted calibration can also benefit broader downstream analyses.

These results highlight the practical value of calibration in real-world applications. Across both NHANES and SUPPORT, calibration generally improves or maintains the fidelity of Cox coefficient estimates, even when the uncalibrated generator already exhibits strong distributional fidelity. Since the proposed calibration procedure requires access to the target coefficient estimates from the original data, we recommend applying calibration whenever these estimates are available, as it can further improve the preservation of target Cox coefficients.


\Cref{tab:nhanes-full-cox-coefficients,tab:support-full-cox-coefficients} provide coefficient-level comparisons of the original and synthetic Cox analyses. In each original-data cell, the first line reports the Cox coefficient estimate with its sandwich standard error in parentheses, and the second line reports the exact two sided Wald $p$-value. For each synthetic method, the first line reports the arithmetic means of the coefficient estimate and its sandwich standard error across 100 independently generated synthetic datasets. The second line reports
$100^{-1}\sum_{r=1}^{100}\mathbf 1\{p_{rj}\leq0.05\}$. Corrected columns use $\widehat{\mathrm{SE}}_{\mathrm{corrected}}=\sqrt{1+m/n}\,\widehat{\mathrm{SE}}_{\mathrm{naive}}=\sqrt{2}\,\widehat{\mathrm{SE}}_{\mathrm{naive}}$ before computing the replicate-specific Wald $p$-values.

\begin{table}[!h]
\centering
\caption{Full-model Cox coefficient and inference comparison for NHANES.}
\label{tab:nhanes-full-cox-coefficients}
\setlength{\tabcolsep}{2.2pt}
\renewcommand{\arraystretch}{1.32}
\resizebox{\textwidth}{!}{%
\begin{tabular}{l c *{5}{cc}}
\toprule
Covariate & Original data &
\multicolumn{2}{c}{TSGSurv(diff)} &
\multicolumn{2}{c}{TSGSurv(diff), calibrated} &
\multicolumn{2}{c}{SurvivalGAN} &
\multicolumn{2}{c}{SurvDiff} &
\multicolumn{2}{c}{Ashhad} \\
\cmidrule(lr){3-4}\cmidrule(lr){5-6}\cmidrule(lr){7-8}
\cmidrule(lr){9-10}\cmidrule(lr){11-12}
& & Naive & Corrected & Naive & Corrected & Naive & Corrected &
Naive & Corrected & Naive & Corrected \\
\midrule

\covariatecell{Age} &
\shortstack{$1.49$\\$(0.0334)$\\$p<10^{-300}$} &
\shortstack{$1.61$\\$(0.0315)$\\$100\%$} &
\shortstack{$1.61$\\$(0.0446)$\\$100\%$} &
\shortstack{$\textbf{1.51}$\\$(\textbf{0.0326})$\\$\textbf{100\%}$} &
\shortstack{$\textbf{1.51}$\\$(\textbf{0.0462})$\\$\textbf{100\%}$} &
\shortstack{$1.35$\\$(0.0280)$\\$100\%$} &
\shortstack{$1.35$\\$(0.0396)$\\$100\%$} &
\shortstack{$0.171$\\$(0.0151)$\\$55\%$} &
\shortstack{$0.171$\\$(0.0213)$\\$46\%$} &
\shortstack{$1.74$\\$(0.0342)$\\$100\%$} &
\shortstack{$1.74$\\$(0.0484)$\\$100\%$} \\
\lightrowrule

\covariatecell{Income-to-poverty ratio} &
\shortstack{$-0.225$\\$(0.0183)$\\$p=9.85\!\times\!10^{-35}$} &
\shortstack{$-0.263$\\$(0.0192)$\\$100\%$} &
\shortstack{$-0.263$\\$(0.0272)$\\$100\%$} &
\shortstack{$\textbf{-0.234}$\\$(\textbf{0.0192})$\\$\textbf{100\%}$} &
\shortstack{$\textbf{-0.234}$\\$(\textbf{0.0272})$\\$\textbf{100\%}$} &
\shortstack{$-0.374$\\$(0.0199)$\\$100\%$} &
\shortstack{$-0.374$\\$(0.0281)$\\$100\%$} &
\shortstack{$-0.0250$\\$(0.0147)$\\$37\%$} &
\shortstack{$-0.0250$\\$(0.0208)$\\$27\%$} &
\shortstack{$-0.258$\\$(0.0200)$\\$100\%$} &
\shortstack{$-0.258$\\$(0.0283)$\\$100\%$} \\
\lightrowrule

\covariatecell{BMI} &
\shortstack{$-0.285$\\$(0.174)$\\$p=0.101$} &
\shortstack{$-0.495$\\$(0.183)$\\$70\%$} &
\shortstack{$-0.495$\\$(0.258)$\\$60\%$} &
\shortstack{$-0.546$\\$(0.186)$\\$76\%$} &
\shortstack{$-0.546$\\$(0.262)$\\$56\%$} &
\shortstack{$-0.420$\\$(0.0400)$\\$100\%$} &
\shortstack{$-0.420$\\$(0.0565)$\\$100\%$} &
\shortstack{$-0.0484$\\$(0.0205)$\\$38\%$} &
\shortstack{$-0.0484$\\$(0.0290)$\\$34\%$} &
\shortstack{$\textbf{-0.376}$\\$(\textbf{0.187})$\\$\textbf{58\%}$} &
\shortstack{$\textbf{-0.376}$\\$(\textbf{0.265})$\\$\textbf{41\%}$} \\
\lightrowrule

\covariatecell{Weight} &
\shortstack{$-0.0467$\\$(0.193)$\\$p=0.809$} &
\shortstack{$0.129$\\$(0.203)$\\$35\%$} &
\shortstack{$0.129$\\$(0.287)$\\$21\%$} &
\shortstack{$0.0740$\\$(0.208)$\\$41\%$} &
\shortstack{$0.0740$\\$(0.294)$\\$27\%$} &
\shortstack{$-0.237$\\$(0.0410)$\\$98\%$} &
\shortstack{$-0.237$\\$(0.0580)$\\$94\%$} &
\shortstack{$\textbf{0.0116}$\\$(\textbf{0.0213})$\\$\textbf{41\%}$} &
\shortstack{$\textbf{0.0116}$\\$(\textbf{0.0302})$\\$\textbf{31\%}$} &
\shortstack{$-0.129$\\$(0.210)$\\$36\%$} &
\shortstack{$-0.129$\\$(0.297)$\\$23\%$} \\
\lightrowrule

\covariatecell{Height} &
\shortstack{$0.159$\\$(0.0914)$\\$p=0.0821$} &
\shortstack{$0.0849$\\$(0.0959)$\\$38\%$} &
\shortstack{$0.0849$\\$(0.136)$\\$23\%$} &
\shortstack{$0.0804$\\$(0.0979)$\\$43\%$} &
\shortstack{$0.0804$\\$(0.138)$\\$28\%$} &
\shortstack{$0.197$\\$(0.0252)$\\$100\%$} &
\shortstack{$0.197$\\$(0.0356)$\\$100\%$} &
\shortstack{$0.0140$\\$(0.0165)$\\$33\%$} &
\shortstack{$0.0140$\\$(0.0234)$\\$29\%$} &
\shortstack{$\textbf{0.189}$\\$(\textbf{0.0992})$\\$\textbf{53\%}$} &
\shortstack{$\textbf{0.189}$\\$(\textbf{0.140})$\\$\textbf{32\%}$} \\
\lightrowrule

\covariatecell{Waist circumference} &
\shortstack{$0.302$\\$(0.0528)$\\$p=1.05\!\times\!10^{-8}$} &
\shortstack{$\textbf{0.349}$\\$(\textbf{0.0611})$\\$\textbf{95\%}$} &
\shortstack{$\textbf{0.349}$\\$(\textbf{0.0865})$\\$\textbf{93\%}$} &
\shortstack{$0.449$\\$(0.0620)$\\$99\%$} &
\shortstack{$0.449$\\$(0.0877)$\\$98\%$} &
\shortstack{$0.569$\\$(0.0358)$\\$100\%$} &
\shortstack{$0.569$\\$(0.0507)$\\$100\%$} &
\shortstack{$0.0335$\\$(0.0167)$\\$36\%$} &
\shortstack{$0.0335$\\$(0.0236)$\\$27\%$} &
\shortstack{$0.440$\\$(0.0618)$\\$96\%$} &
\shortstack{$0.440$\\$(0.0873)$\\$96\%$} \\
\lightrowrule

\covariatecell{Pulse} &
\shortstack{$0.127$\\$(0.0157)$\\$p=4.34\!\times\!10^{-16}$} &
\shortstack{$0.137$\\$(0.0173)$\\$100\%$} &
\shortstack{$0.137$\\$(0.0244)$\\$100\%$} &
\shortstack{$0.132$\\$(0.0173)$\\$100\%$} &
\shortstack{$0.132$\\$(0.0244)$\\$99\%$} &
\shortstack{$0.131$\\$(0.0197)$\\$98\%$} &
\shortstack{$0.131$\\$(0.0278)$\\$94\%$} &
\shortstack{$0.0146$\\$(0.0147)$\\$28\%$} &
\shortstack{$0.0146$\\$(0.0208)$\\$16\%$} &
\shortstack{$\textbf{0.125}$\\$(\textbf{0.0193})$\\$\textbf{98\%}$} &
\shortstack{$\textbf{0.125}$\\$(\textbf{0.0273})$\\$\textbf{97\%}$} \\
\lightrowrule

\covariatecell{Systolic BP} &
\shortstack{$0.155$\\$(0.0172)$\\$p=1.88\!\times\!10^{-19}$} &
\shortstack{$0.136$\\$(0.0185)$\\$100\%$} &
\shortstack{$0.136$\\$(0.0262)$\\$100\%$} &
\shortstack{$\textbf{0.156}$\\$(\textbf{0.0188})$\\$\textbf{99\%}$} &
\shortstack{$\textbf{0.156}$\\$(\textbf{0.0266})$\\$\textbf{99\%}$} &
\shortstack{$0.302$\\$(0.0198)$\\$100\%$} &
\shortstack{$0.302$\\$(0.0280)$\\$100\%$} &
\shortstack{$0.0443$\\$(0.0150)$\\$41\%$} &
\shortstack{$0.0443$\\$(0.0213)$\\$38\%$} &
\shortstack{$0.111$\\$(0.0212)$\\$91\%$} &
\shortstack{$0.111$\\$(0.0300)$\\$86\%$} \\
\lightrowrule

\covariatecell{Diastolic BP} &
\shortstack{$-0.0280$\\$(0.0157)$\\$p=0.0757$} &
\shortstack{$-0.0170$\\$(0.0169)$\\$30\%$} &
\shortstack{$-0.0170$\\$(0.0239)$\\$15\%$} &
\shortstack{$\textbf{-0.0292}$\\$(\textbf{0.0172})$\\$\textbf{50\%}$} &
\shortstack{$\textbf{-0.0292}$\\$(\textbf{0.0243})$\\$\textbf{25\%}$} &
\shortstack{$-0.130$\\$(0.0219)$\\$96\%$} &
\shortstack{$-0.130$\\$(0.0310)$\\$93\%$} &
\shortstack{$-0.0244$\\$(0.0153)$\\$27\%$} &
\shortstack{$-0.0244$\\$(0.0216)$\\$21\%$} &
\shortstack{$0.0102$\\$(0.0195)$\\$38\%$} &
\shortstack{$0.0102$\\$(0.0276)$\\$26\%$} \\
\lightrowrule

\covariatecell{Total cholesterol} &
\shortstack{$-0.0537$\\$(0.0198)$\\$p=0.00655$} &
\shortstack{$-0.0658$\\$(0.0196)$\\$85\%$} &
\shortstack{$-0.0658$\\$(0.0277)$\\$67\%$} &
\shortstack{$\textbf{-0.0585}$\\$(\textbf{0.0201})$\\$\textbf{78\%}$} &
\shortstack{$\textbf{-0.0585}$\\$(\textbf{0.0284})$\\$\textbf{54\%}$} &
\shortstack{$-0.111$\\$(0.0221)$\\$89\%$} &
\shortstack{$-0.111$\\$(0.0313)$\\$76\%$} &
\shortstack{$-0.0128$\\$(0.0149)$\\$28\%$} &
\shortstack{$-0.0128$\\$(0.0211)$\\$14\%$} &
\shortstack{$-0.0789$\\$(0.0209)$\\$89\%$} &
\shortstack{$-0.0789$\\$(0.0296)$\\$78\%$} \\
\lightrowrule

\covariatecell{HDL cholesterol} &
\shortstack{$-0.0229$\\$(0.0194)$\\$p=0.237$} &
\shortstack{$-0.0353$\\$(0.0209)$\\$39\%$} &
\shortstack{$-0.0353$\\$(0.0295)$\\$24\%$} &
\shortstack{$\textbf{-0.0313}$\\$(\textbf{0.0211})$\\$\textbf{34\%}$} &
\shortstack{$\textbf{-0.0313}$\\$(\textbf{0.0298})$\\$\textbf{15\%}$} &
\shortstack{$-0.0602$\\$(0.0222)$\\$63\%$} &
\shortstack{$-0.0602$\\$(0.0314)$\\$51\%$} &
\shortstack{$-0.00560$\\$(0.0150)$\\$21\%$} &
\shortstack{$-0.00560$\\$(0.0212)$\\$16\%$} &
\shortstack{$-0.0492$\\$(0.0226)$\\$53\%$} &
\shortstack{$-0.0492$\\$(0.0320)$\\$31\%$} \\
\lightrowrule

\covariatecell{HbA1c} &
\shortstack{$0.128$\\$(0.0149)$\\$p=6.32\!\times\!10^{-18}$} &
\shortstack{$\textbf{0.138}$\\$(\textbf{0.0173})$\\$\textbf{98\%}$} &
\shortstack{$\textbf{0.138}$\\$(\textbf{0.0245})$\\$\textbf{95\%}$} &
\shortstack{$0.139$\\$(0.0170)$\\$99\%$} &
\shortstack{$0.139$\\$(0.0241)$\\$98\%$} &
\shortstack{$0.167$\\$(0.0167)$\\$100\%$} &
\shortstack{$0.167$\\$(0.0236)$\\$100\%$} &
\shortstack{$0.0144$\\$(0.0156)$\\$24\%$} &
\shortstack{$0.0144$\\$(0.0221)$\\$13\%$} &
\shortstack{$0.110$\\$(0.0202)$\\$98\%$} &
\shortstack{$0.110$\\$(0.0286)$\\$95\%$} \\
\lightrowrule

\covariatecell{Hemoglobin} &
\shortstack{$-0.0549$\\$(0.0201)$\\$p=0.00614$} &
\shortstack{$-0.0656$\\$(0.0218)$\\$84\%$} &
\shortstack{$-0.0656$\\$(0.0308)$\\$59\%$} &
\shortstack{$-0.0754$\\$(0.0222)$\\$87\%$} &
\shortstack{$-0.0754$\\$(0.0313)$\\$67\%$} &
\shortstack{$-0.0869$\\$(0.0234)$\\$78\%$} &
\shortstack{$-0.0869$\\$(0.0331)$\\$67\%$} &
\shortstack{$-0.00173$\\$(0.0150)$\\$8\%$} &
\shortstack{$-0.00173$\\$(0.0212)$\\$4\%$} &
\shortstack{$\textbf{-0.0495}$\\$(\textbf{0.0239})$\\$\textbf{51\%}$} &
\shortstack{$\textbf{-0.0495}$\\$(\textbf{0.0339})$\\$\textbf{34\%}$} \\
\bottomrule
\end{tabular}%
}
\end{table}

\begin{table}[!h]
\centering
\caption{Full-model Cox coefficient and inference comparison for SUPPORT.}
\label{tab:support-full-cox-coefficients}
\setlength{\tabcolsep}{2.2pt}
\renewcommand{\arraystretch}{1.32}
\resizebox{\textwidth}{!}{%
\begin{tabular}{l c *{5}{cc}}
\toprule
Covariate & Original data &
\multicolumn{2}{c}{TSGSurv(diff)} &
\multicolumn{2}{c}{TSGSurv(diff), calibrated} &
\multicolumn{2}{c}{SurvivalGAN} &
\multicolumn{2}{c}{SurvDiff} &
\multicolumn{2}{c}{Ashhad} \\
\cmidrule(lr){3-4}\cmidrule(lr){5-6}\cmidrule(lr){7-8}
\cmidrule(lr){9-10}\cmidrule(lr){11-12}
& & Naive & Corrected & Naive & Corrected & Naive & Corrected &
Naive & Corrected & Naive & Corrected \\
\midrule

\covariatecell{Age} &
\shortstack{$0.211$\\$(0.0169)$\\$p=1.01\!\times\!10^{-35}$} &
\shortstack{$\textbf{0.213}$\\$(\textbf{0.0179})$\\$\textbf{100\%}$} &
\shortstack{$\textbf{0.213}$\\$(\textbf{0.0253})$\\$\textbf{100\%}$} &
\shortstack{$0.209$\\$(0.0178)$\\$100\%$} &
\shortstack{$0.209$\\$(0.0252)$\\$100\%$} &
\shortstack{$0.242$\\$(0.0180)$\\$100\%$} &
\shortstack{$0.242$\\$(0.0254)$\\$100\%$} &
\shortstack{$0.160$\\$(0.0172)$\\$100\%$} &
\shortstack{$0.160$\\$(0.0244)$\\$100\%$} &
\shortstack{$0.273$\\$(0.0191)$\\$100\%$} &
\shortstack{$0.273$\\$(0.0271)$\\$100\%$} \\
\lightrowrule

\covariatecell{Mean arterial BP} &
\shortstack{$-0.0609$\\$(0.0174)$\\$p=0.000464$} &
\shortstack{$-0.0904$\\$(0.0182)$\\$99\%$} &
\shortstack{$-0.0904$\\$(0.0258)$\\$95\%$} &
\shortstack{$-0.0628$\\$(0.0182)$\\$94\%$} &
\shortstack{$-0.0628$\\$(0.0257)$\\$78\%$} &
\shortstack{$-0.0786$\\$(0.0189)$\\$85\%$} &
\shortstack{$-0.0786$\\$(0.0267)$\\$76\%$} &
\shortstack{$-0.00349$\\$(0.0177)$\\$19\%$} &
\shortstack{$-0.00349$\\$(0.0250)$\\$9\%$} &
\shortstack{$\textbf{-0.0611}$\\$(\textbf{0.0185})$\\$\textbf{77\%}$} &
\shortstack{$\textbf{-0.0611}$\\$(\textbf{0.0262})$\\$\textbf{59\%}$} \\
\lightrowrule

\covariatecell{Heart rate} &
\shortstack{$0.0788$\\$(0.0185)$\\$p=2.06\!\times\!10^{-5}$} &
\shortstack{$0.0729$\\$(0.0196)$\\$88\%$} &
\shortstack{$0.0729$\\$(0.0277)$\\$77\%$} &
\shortstack{$0.0750$\\$(0.0196)$\\$97\%$} &
\shortstack{$0.0750$\\$(0.0277)$\\$86\%$} &
\shortstack{$0.155$\\$(0.0197)$\\$100\%$} &
\shortstack{$0.155$\\$(0.0279)$\\$99\%$} &
\shortstack{$\textbf{0.0799}$\\$(\textbf{0.0185})$\\$\textbf{90\%}$} &
\shortstack{$\textbf{0.0799}$\\$(\textbf{0.0261})$\\$\textbf{79\%}$} &
\shortstack{$0.0575$\\$(0.0195)$\\$68\%$} &
\shortstack{$0.0575$\\$(0.0276)$\\$48\%$} \\
\lightrowrule

\covariatecell{Respiratory rate} &
\shortstack{$0.0193$\\$(0.0175)$\\$p=0.272$} &
\shortstack{$0.00602$\\$(0.0186)$\\$11\%$} &
\shortstack{$0.00602$\\$(0.0263)$\\$1\%$} &
\shortstack{$0.00868$\\$(0.0186)$\\$14\%$} &
\shortstack{$0.00868$\\$(0.0263)$\\$6\%$} &
\shortstack{$\textbf{0.0291}$\\$(\textbf{0.0210})$\\$\textbf{47\%}$} &
\shortstack{$\textbf{0.0291}$\\$(\textbf{0.0296})$\\$\textbf{32\%}$} &
\shortstack{$-0.0336$\\$(0.0188)$\\$51\%$} &
\shortstack{$-0.0336$\\$(0.0265)$\\$29\%$} &
\shortstack{$-0.0401$\\$(0.0201)$\\$53\%$} &
\shortstack{$-0.0401$\\$(0.0284)$\\$32\%$} \\
\lightrowrule

\covariatecell{Body temperature} &
\shortstack{$0.00257$\\$(0.0180)$\\$p=0.887$} &
\shortstack{$0.0392$\\$(0.0189)$\\$52\%$} &
\shortstack{$0.0392$\\$(0.0267)$\\$37\%$} &
\shortstack{$0.0370$\\$(0.0188)$\\$52\%$} &
\shortstack{$0.0370$\\$(0.0266)$\\$24\%$} &
\shortstack{$0.0226$\\$(0.0200)$\\$42\%$} &
\shortstack{$0.0226$\\$(0.0283)$\\$26\%$} &
\shortstack{$\textbf{-0.0155}$\\$(\textbf{0.0181})$\\$\textbf{28\%}$} &
\shortstack{$\textbf{-0.0155}$\\$(\textbf{0.0256})$\\$\textbf{7\%}$} &
\shortstack{$0.0382$\\$(0.0201)$\\$48\%$} &
\shortstack{$0.0382$\\$(0.0284)$\\$32\%$} \\
\lightrowrule

\covariatecell{Serum sodium} &
\shortstack{$-0.00998$\\$(0.0173)$\\$p=0.563$} &
\shortstack{$0.0457$\\$(0.0201)$\\$55\%$} &
\shortstack{$0.0457$\\$(0.0285)$\\$40\%$} &
\shortstack{$0.0392$\\$(0.0203)$\\$57\%$} &
\shortstack{$0.0392$\\$(0.0287)$\\$36\%$} &
\shortstack{$-0.0203$\\$(0.0198)$\\$46\%$} &
\shortstack{$-0.0203$\\$(0.0279)$\\$27\%$} &
\shortstack{$-0.0388$\\$(0.0181)$\\$55\%$} &
\shortstack{$-0.0388$\\$(0.0256)$\\$33\%$} &
\shortstack{$\textbf{-0.00148}$\\$(\textbf{0.0217})$\\$\textbf{49\%}$} &
\shortstack{$\textbf{-0.00148}$\\$(\textbf{0.0306})$\\$\textbf{28\%}$} \\
\lightrowrule

\covariatecell{White blood cell count} &
\shortstack{$0.0162$\\$(0.0166)$\\$p=0.329$} &
\shortstack{$-0.0168$\\$(0.0137)$\\$55\%$} &
\shortstack{$-0.0168$\\$(0.0193)$\\$46\%$} &
\shortstack{$-0.0138$\\$(0.0137)$\\$63\%$} &
\shortstack{$-0.0138$\\$(0.0194)$\\$39\%$} &
\shortstack{$0.0491$\\$(0.0197)$\\$64\%$} &
\shortstack{$0.0491$\\$(0.0279)$\\$42\%$} &
\shortstack{$-0.0541$\\$(0.0225)$\\$60\%$} &
\shortstack{$-0.0541$\\$(0.0318)$\\$40\%$} &
\shortstack{$\textbf{0.0211}$\\$(\textbf{0.0214})$\\$\textbf{35\%}$} &
\shortstack{$\textbf{0.0211}$\\$(\textbf{0.0303})$\\$\textbf{17\%}$} \\
\lightrowrule

\covariatecell{Serum creatinine} &
\shortstack{$0.0556$\\$(0.0159)$\\$p=0.000475$} &
\shortstack{$0.0655$\\$(0.0174)$\\$90\%$} &
\shortstack{$0.0655$\\$(0.0246)$\\$76\%$} &
\shortstack{$\textbf{0.0601}$\\$(\textbf{0.0171})$\\$\textbf{91\%}$} &
\shortstack{$\textbf{0.0601}$\\$(\textbf{0.0242})$\\$\textbf{77\%}$} &
\shortstack{$0.0798$\\$(0.0172)$\\$92\%$} &
\shortstack{$0.0798$\\$(0.0244)$\\$82\%$} &
\shortstack{$0.191$\\$(0.0206)$\\$100\%$} &
\shortstack{$0.191$\\$(0.0291)$\\$100\%$} &
\shortstack{$0.0492$\\$(0.0188)$\\$64\%$} &
\shortstack{$0.0492$\\$(0.0265)$\\$46\%$} \\
\bottomrule
\end{tabular}%
}
\end{table}
The coefficient-level results reveal several patterns that are obscured by the aggregate metrics. On NHANES, TSGSurv(diff) closely reproduces the major Cox associations observed in the original data, with calibration further improving or largely preserving the coefficient estimates for nearly all prespecified target covariates. The tables also reveal method-specific distortions: SurvDiff substantially attenuates most coefficient estimates, while SurvivalGAN tends to overestimate several effects and produces unusually small standard errors for some nonsignificant anthropometric variables, leading to frequent rejections inconsistent with the original-data conclusions. Ashhad generally preserves the directions of major associations but exhibits larger discrepancies in several coefficient magnitudes. Overall, TSGSurv(diff) provides more balanced coefficient and inferential fidelity, with calibration offering additional improvements for selected covariates.

On SUPPORT, TSGSurv(diff) again closely reproduces the principal Cox effects, with calibration improving or maintaining the accuracy of all four prespecified target coefficients. SurvivalGAN substantially overestimates the heart rate effect, while SurvDiff exhibits considerable coefficient distortions, especially for mean arterial blood pressure and serum creatinine. Ashhad preserves several coefficient estimates reasonably well but shows noticeable discrepancies for age and heart rate. Across both datasets, variance correction generally reduces significance frequencies by accounting for additional sampling uncertainty, although it does not eliminate coefficient estimation bias. Notably, the coefficients benefiting most consistently from calibration largely coincide with the prespecified calibration targets, even though the evaluations are conducted using the full Cox model. These results suggest that targeted calibration can improve the preservation of scientifically relevant effects while maintaining competitive performance in broader downstream analyses.

\begin{figure}[!h]
\centering
\includegraphics[width=\textwidth]{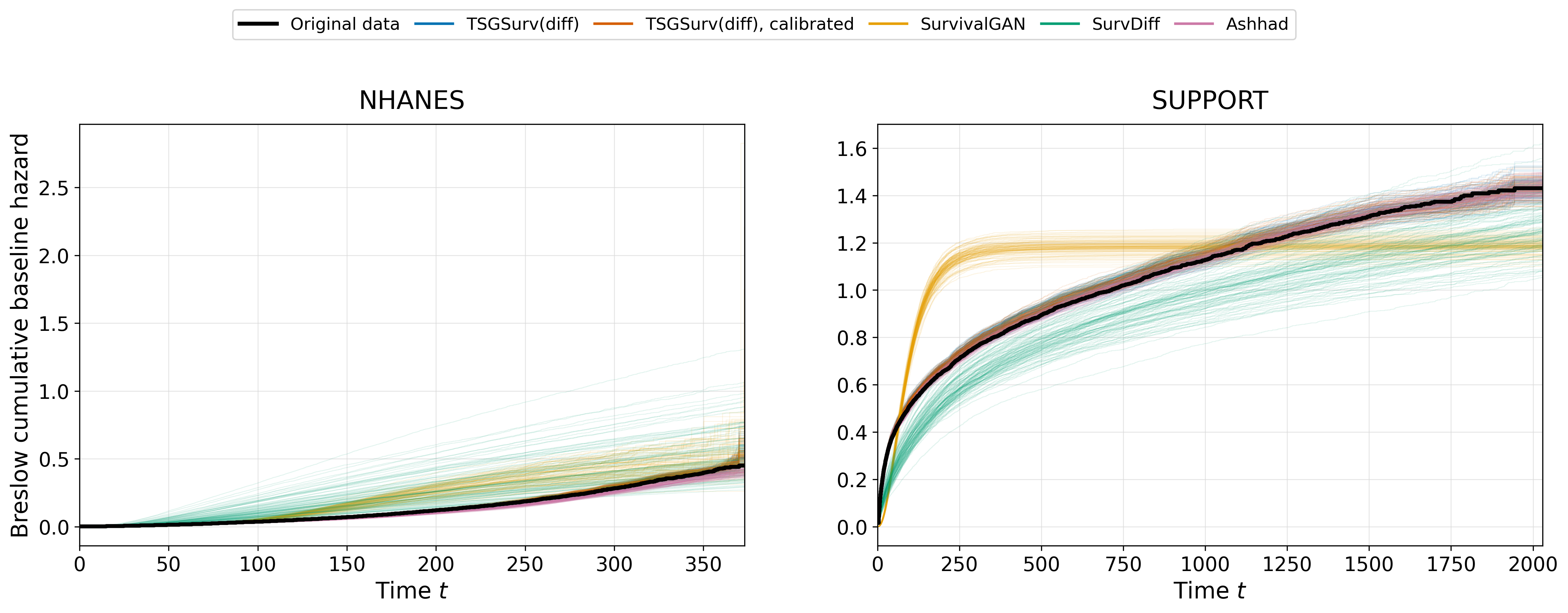}
\caption{Breslow cumulative baseline hazards fitted to the real  data and the synthetic datasets. }
\label{fig:real-baseline-hazards}
\end{figure}

Figure~\ref{fig:real-baseline-hazards} compares the Breslow cumulative baseline hazard estimated from the original data with those estimated from the 100 synthetic replicates generated by each method. Both TSGSurv(diff) and Ashhad closely tracks the original-data curve throughout the observed follow-up range on both NHANES and SUPPORT. In contrast, SurvDiff systematically overestimates the cumulative baseline hazard on NHANES and underestimates it on SUPPORT. SurvivalGAN exhibits a different form of discrepancy: its NHANES curves deviate increasingly near the end of follow-up, while its SUPPORT curves rise rapidly at early times and then plateau. These curves complement the Breslow ISE by showing when and how each method deviates from the original-data estimate.

\section{Discussion}\label{sec:dis}

Our study develops a general framework for faithful generation of censored survival data and establishes theoretical connections between distributional fidelity and downstream statistical inference. We first identify an irreducible approximation error in deterministic time generation and propose a two-stage generative framework with distributional convergence guarantees. We then establish the stability of Cox regression coefficients and their asymptotic covariance under distributional perturbations. Importantly, we show that coefficients and covariance preservation alone do not guarantee valid statistical inference, and develop variance correction and distribution calibration procedures to address this limitation. Extensive simulations and real-data applications demonstrate the effectiveness of the proposed framework in preserving distributional fidelity and downstream Cox regression results. Despite these contributions, several limitations remain and motivate future research.


\noindent\textbf{Cox model misspecification.} 
Our theoretical results do not necessarily require the proportional hazards assumption to hold. Under model misspecification, the population Cox coefficient may still be well defined as a pseudo-true parameter, and our stability results remain applicable under the stated regularity conditions. Similarly, the sandwich covariance characterizes the asymptotic variability of the Cox estimator around this pseudo-true parameter. 
However, preservation of the pseudo-true coefficient does not necessarily imply preservation of the underlying time-varying covariate effects or other features of the true hazard function. Further research could investigate how distributional fidelity translates into preservation of these features under model misspecification.

\noindent\textbf{Extension to other survival functionals.} Our theoretical analysis primarily focuses on finite-dimensional Cox regression coefficients, their associated covariance matrices, and associated hypothesis tests. Other important survival quantities, including baseline cumulative hazards, survival probabilities, and individualized survival predictions, are not covered by our theoretical guarantees. These targets may require additional regularity conditions to establish stability under distributional perturbations. Extending our results to these quantities would provide a more comprehensive theoretical foundation for downstream survival analysis using synthetic data.


\noindent\textbf{Dependence on original-data information.} Our distribution calibration procedure requires the original data Cox estimate as a calibration target, which may not be available when researchers have access only to synthetic data. Although this requirement does not necessarily require direct access to individual-level records, sharing the original data estimate may be restricted in privacy-sensitive applications. An important future direction is to develop calibration procedures using privacy-preserving estimates or other limited summaries of the original data, while quantifying how the uncertainty induced by these summaries affects inferential validity and fidelity. 


\bibliographystyle{plainnat}
\bibliography{ref}

\appendix
\section{Code and Data Availability}
\label{app:data-source}

Code for all simulations and real data experiments can be found in the GitHub repository:
\url{https://github.com/Maeve816/Generative_survival}.
The real-data analyses use only publicly released, de-identified records. The construction details below identify the exact releases and transformations used to produce the analytic files.
\subsection{NHANES source files}

We downloaded the public examination and laboratory files for NHANES III (1988--1994) and for each of the ten continuous-NHANES cycles from 1999--2000 through 2017--2018 from the \href{https://wwwn.cdc.gov/nchs/nhanes/continuousnhanes/}{NCHS continuous-NHANES portal} and the \href{https://wwwn.cdc.gov/nchs/nhanes/nhanes3/datafiles.aspx}{NCHS NHANES III data-files portal}. Mortality outcomes came from the NCHS 2019 Public-Use Linked Mortality Files, downloaded from the \href{https://ftp.cdc.gov/pub/health_statistics/nchs/datalinkage/linked_mortality/}{NCHS linked-mortality FTP release}. Thus, the mortality follow-up vintage is held fixed at the 2019 public release for NHANES III and all continuous-NHANES cycles.  

For continuous NHANES, the raw inputs are the SAS transport (XPT) component files for demographics (\texttt{DEMO}), body measures (\texttt{BMX}), blood pressure (\texttt{BPX}), diabetes (\texttt{DIQ}), smoking (\texttt{SMQ}), and blood-pressure questionnaire data (\texttt{BPQ}), together with the  cycle-specific laboratory files containing total and HDL cholesterol, glycohemoglobin, blood chemistry, and complete blood count measurements.  Files were left-joined one-to-one on the participant identifier \texttt{SEQN}.
Systolic and diastolic blood pressure were computed as the row means of up tofour available examination readings.  

NHANES III was processed separately because its public examination and laboratory records are fixed-width rather than cycle-specific XPT files. We used the NCHS NHANES III release 1A examination and laboratory files (\texttt{1a\_exam.dat} and \texttt{1a\_lab.dat}), recoded the documented repeated 8/9 missing-value codes to missing, and linked these records by \texttt{SEQN} to \path{NHANES_III_MORT_2019_PUBLIC.dat}. The same eligibility, vital status, and positive-follow-up restrictions were then applied.

The two sources were harmonized to the same 13 continuous covariates: age, family income-to-poverty ratio, body-mass index, weight, height, waist circumference, pulse, systolic blood pressure, diastolic blood pressure, total cholesterol, HDL cholesterol, glycated hemoglobin, and hemoglobin. Units were harmonized before appending the surveys. We performed complete-case selection on these covariates, follow-up in months $Y$,  and all-cause mortality status $\Delta$, without imputation, yielding 13,528 NHANES III records and 44,446 continuous-NHANES records. The generative-model benchmark treats the resulting 57,974 records as an unweighted empirical distribution.

\subsection{SUPPORT source files}

We used \texttt{pycox.datasets.support.read\_df()}. This loader obtains \path{support/support_train_test.h5} from the \href{https://github.com/jaredleekatzman/DeepSurv/tree/master/experiments/data/support}{DeepSurv SUPPORT data release}, concatenates the original training and test partitions, maps the outcome \texttt{t} and \texttt{e} to \texttt{duration} and \texttt{event}, and caches the result as a Feather data frame.

The distributed frame has 8,873 records, 14 covariates (age, sex, race, number of comorbidities, diabetes, dementia, cancer status, mean arterial blood pressure, heart rate, respiratory rate, body temperature, serum sodium, white blood cell count, and serum creatinine), duration, and event, with no missing values. We retained the eight continuous variables (age, mean arterial blood pressure, heart rate, respiratory rate, body temperature, serum sodium, white blood cell count, and serum creatinine). We renamed duration as $Y$ and retained event as the binary  $\Delta$.

\section{Proofs}
\label{app:proof}

In this appendix we prove the results from
\Cref{sec:method2} and \Cref{sec:theory}:

\subsection{ Proof of \Cref{lem:deterministic_error2}~(Irreducible error under deterministic time generation)}
\label{proof:deterministic_error2}

\begin{proof}
Let $(\widetilde X,\widetilde\Delta)\sim\widehat P_{X,\Delta,n}$. By the definitions in \Cref{subsec:deterministic_error2}, $\widehat P_n^{\mathrm{reg}}=\mathcal L\{\widetilde X, f_n(\widetilde X,\widetilde\Delta),\widetilde\Delta\}$ and $P^\dagger=\mathcal L\{X,f_*(X,\Delta),\Delta\}$, where $(X,\Delta)\sim P_{X,\Delta}$. So by triangle inequality, we have 
\[W_1\bigl(\widehat P_n^{\mathrm{reg}},
P^\dagger\bigr) \le
W_1\bigl(\widehat P_n^{\mathrm{reg}},
\mathcal L\{\widetilde X,f_*(\widetilde X,\widetilde\Delta),
\widetilde\Delta\}\bigr) +
W_1\bigl(
\mathcal L\{\widetilde X,f_*(\widetilde X,\widetilde\Delta),
\widetilde\Delta\}, P^\dagger\bigr)
\]

To bound the first regression-error term, draw a single pair $(\widetilde X,\widetilde\Delta)$ from $\widehat P_{X,\Delta,n}$ and use it to form both $\{\widetilde X,f_n(\widetilde X,\widetilde\Delta),\widetilde\Delta\}$ and $\{\widetilde X,f_*(\widetilde X,\widetilde\Delta), \widetilde\Delta\}$. Their $X$ and $\Delta$ components agree, and the distance between them is
$|f_n(\widetilde X,\widetilde\Delta)-f_*(\widetilde X,\widetilde\Delta)|$. 
The expected transportation cost of this coupling, conditional on the fitted generator and regression, is
$\int |f_n(x,\delta)-f_*(x,\delta)|
\,d\widehat P_{X,\Delta,n}(x,\delta).$
Since $W_1$ is the infimum of the expected cost over all couplings, the regression-error rate gives
$$W_1\bigl(\widehat P_n^{\mathrm{reg}},
\mathcal L\{\widetilde X,f_*(\widetilde X,\widetilde\Delta),
\widetilde\Delta\}\bigr) \le \int |f_n(x,\delta)-f_*(x,\delta)|
\,d\widehat P_{X,\Delta,n}(x,\delta)
=O_p(n^{-\gamma_r}).$$
For the second term, take any coupling of $(\widetilde X,\widetilde\Delta)\sim\widehat P_{X,\Delta,n}$ and $(X,\Delta)\sim P_{X,\Delta}$. Forming the corresponding full vectors using $f_*$ gives a coupling of $\mathcal L\{\widetilde X,f_*(\widetilde X,\widetilde\Delta), \widetilde\Delta\}$ and $P^\dagger$. By the assumed Lipschitz property, the distance between these vectors is at most $L$ times the distance between the coupled inputs, where $L$ is a Lipschitz constant for $(x,\delta)\mapsto\{x,f_*(x,\delta),\delta\}$. Hence,
$$W_1\bigl(\mathcal L\{\widetilde X,f_*(\widetilde X,\widetilde\Delta),
\widetilde\Delta\},P^\dagger\bigr)
\le L W_1(\widehat P_{X,\Delta,n},P_{X,\Delta})
=O_p(n^{-\gamma_g}).
$$
Consequently, the triangle inequality gives
\[
W_1(\widehat P_n^{\mathrm{reg}},P^\dagger)
=O_p(n^{-\gamma_g}+n^{-\gamma_r}).
\]
Recall $ b \coloneqq W_1(P^\dagger,P_0).$ Then the reverse triangle inequality now implies
\begin{align*}
\big|W_1(\widehat P_n^{\mathrm{reg}},P_0)-b\big|
&=\big|W_1(\widehat P_n^{\mathrm{reg}},P_0)
-W_1(P^\dagger,P_0)\big|\\
&\le W_1(\widehat P_n^{\mathrm{reg}},P^\dagger)
=O_p(n^{-\gamma_g}+n^{-\gamma_r}),
\end{align*}
which establishes
\[
W_1(\widehat P_n^{\mathrm{reg}},P_0)
=b+O_p(n^{-\gamma_g}+n^{-\gamma_r}).
\]
Finally, under $P^\dagger$, the conditional distribution  $log(Y)|(X,\Delta)$ is a point mass and therefore has
conditional variance zero almost surely. If $\operatorname{Var}_{P_0}(log(Y)\mid X,\Delta)>0$ on a set of positive $P_{X,\Delta}$-probability, then $P^\dagger\ne P_0$. Since $W_1$ is a metric on the distributions under consideration, $b \coloneqq W_1(P^\dagger,P_0)>0$. The preceding rate then gives $W_1(\widehat P_n^{\mathrm{reg}},P_0)\xrightarrow{p}b>0$.
\end{proof}

\subsection{Proof of \Cref{thm:rate_2_stage}~
(Convergence rate of the two-stage generative framework)}
\label{proof:rate_2_stage}
\begin{proof}
For $\delta\in\{0,1\}$, recall $P_{X,Y\mid\Delta=\delta}$ and
$\widehat G_{\delta,n_\delta}$ are the true conditional distribution and its fitted generative distribution. So we have
$P_0=\pi P_{X,Y\mid\Delta=1}
 +(1-\pi)P_{X,Y\mid\Delta=0},
\widehat P_n=\widehat\pi_n\widehat G_{1,n_1}
 +(1-\widehat\pi_n)\widehat G_{0,n_0}.$
Define the intermediate distribution $
P_n^*=\widehat\pi_n P_{X,Y\mid\Delta=1}
 +(1-\widehat\pi_n)P_{X,Y\mid\Delta=0}.$ In these mixtures, each stratum-specific law is understood as a law of $(X,Y,\Delta)$ with its third coordinate fixed at the corresponding value of $\Delta$. By the triangle inequality,
\[
W_1(\widehat P_n,P_0)
\leq W_1(\widehat P_n,P_n^*)+W_1(P_n^*,P_0).
\]

For the first term, both mixtures have weight $\widehat\pi_n$. The mixture inequality
\[W_1\!\left(\lambda P+(1-\lambda)Q,\,
         \lambda R+(1-\lambda)S\right)
\leq \lambda W_1(P,R)+(1-\lambda)W_1(Q,S)\]
therefore gives
$W_1(\widehat P_n,P_n^*)
\leq {}
\widehat\pi_n\,W_1\!\left(
\widehat G_{1,n_1},P_{X,Y\mid\Delta=1}\right)
+(1-\widehat\pi_n)\,W_1\!\left(
\widehat G_{0,n_0},P_{X,Y\mid\Delta=0}\right).$
Since $\pi\in(0,1)$, it also follows that $n_1/n\to\pi$ and $n_0/n\to1-\pi$ in probability. Thus both group sizes are of order $n$, and the common conditional-generator rate gives
\[
W_1(\widehat P_n,P_n^*)
=O_p\!\left(n^{-\gamma}\right).
\]

For the second term, the component distributions are fixed and only the weights differ. The identity
\[W_1\!\left(\lambda P+(1-\lambda)Q,\,
         \lambda'P+(1-\lambda')Q\right)
=|\lambda-\lambda'|\,W_1(P,Q)\]
gives $W_1(P_n^*,P_0) =|\widehat\pi_n-\pi|\, W_1\!\left(P_{X,Y\mid\Delta=1},          P_{X,Y\mid\Delta=0}\right),$
where the distance $W_1(P_{X,Y\mid\Delta=1},P_{X,Y\mid\Delta=0})$ is finite and does not depend on $n$. Since $\widehat\pi_n=n^{-1}\sum_{i=1}^n\Delta_i$, the central limit theorem for the original i.i.d. records implies $|\widehat\pi_n-\pi|=O_p(n^{-1/2}).$ Combining the bounds yields
\[
W_1(\widehat P_n,P_0)
=O_p\!\left(n^{-\gamma}+n^{-1/2}\right)
=O_p\!\left(n^{-\min\{\gamma,1/2\}}\right).
\]
\end{proof}

\subsection{Proof of \Cref{thm:cox_population_stability}~(Local stability of the Cox parameter under $\mathrm{TV}$)}
\label{proof:cox_population_stability}
\begin{proof}\quad
By Assumptions~\ref{ass:cox_parameter} and \ref{ass:cox_identification}, choose $r>0$ such that the closed ball
$\mathcal B_r=\{\beta:\|\beta-\beta(P_0)\|\leq r\} \subset \mathcal N_{\beta(P_0)}\cap\operatorname{int}(B)$. All bounds below are uniform over
$\beta\in\mathcal B_r$ and $t\in[0,\tau]$. The theorem restricts
$P$ to $\mathcal P_0$ such that Assumptions~\ref{ass:cox_covariates} and
\ref{ass:cox_risksets} can be applied to both $P$ and $P_0$.
 With the convention $d_{\mathrm{TV}}(P,P_0)=\sup_A|P(A)-P_0(A)|$ for any measurable set $A$, every bounded measurable function $h$ satisfies
$\|\mathbb E_Ph-\mathbb E_{P_0}h\|\leq2\|h\|_\infty
d_{\mathrm{TV}}(P,P_0)$, where $\|h\|_\infty$ is the supremum of its
pointwise norm. 

\noindent\textbf{Step 1: Risk-set moments.}\quad We first control the risk-set moments, which determine both the score function and its information matrix.

By \Cref{ass:cox_parameter} and \Cref{ass:cox_covariates}, \(\|X\|\leq K\) and \(\sup_{\beta\in B}\|\beta\|<\infty\). Since \(0\leq R(t)\leq1\),
$$\sup_{\beta\in B,\,t\in[0,\tau]}
\left\|R(t)e^{\beta^\top X}X^{\otimes k}\right\|
\leq e^{K\sup_{\beta\in B}\|\beta\|}K^k<\infty,\qquad
k=0,1,2.$$
Applying the total-variation bound to $S_P^{(k)}(\beta,t) = \mathbb{E}_P \left[ R(t)\exp(\beta^\top X)X^{\otimes k} \right]$ gives
\begin{equation}
\sup_{\beta\in\mathcal B_r,\,t\in[0,\tau]}
\|S_P^{(k)}(\beta,t)-S_{P_0}^{(k)}(\beta,t)\|
\leq C_S\,d_{\mathrm{TV}}(P,P_0),\qquad k=0,1,2
\label{eq:cox_tv_riskset_bound}
\end{equation} for some $0 < C_S  < \infty$.
By the definition of $\overline X_P$,
$$\overline X_P(\beta,t)-\overline X_{P_0}(\beta,t)
={}\frac{S_P^{(1)}(\beta,t)-S_{P_0}^{(1)}(\beta,t)}{S_P^{(0)}(\beta,t)}
+S_{P_0}^{(1)}(\beta,t)
\frac{S_{P_0}^{(0)}(\beta,t)-S_P^{(0)}(\beta,t)}
{S_P^{(0)}(\beta,t)S_{P_0}^{(0)}(\beta,t)}.$$

\Cref{ass:cox_risksets} bounds the first denominator below by \(a\) and the product in the second denominator below by \(a^2\), and Assumptions~\ref{ass:cox_parameter}--\ref{ass:cox_covariates} bound $S_{P_0}^{(1)}$. The two numerator differences satisfy \eqref{eq:cox_tv_riskset_bound}, so
\begin{equation}
\sup_{\beta\in\mathcal B_r,\,t\in[0,\tau]}
\|\overline X_P(\beta,t)-\overline X_{P_0}(\beta,t)\|
\leq C_{\overline X} d_{\mathrm{TV}}(P,P_0)
\label{eq:cox_tv_xbar_bound}
\end{equation}  for some $0 < C_{\overline X} < \infty$.
The same ratio calculation with $S^{(2)}$ in place of $S^{(1)}$ shows that
\begin{equation}
\sup_{\beta\in\mathcal B_r,\,t\in[0,\tau]}
\left\|\frac{S_P^{(2)}(\beta,t)}{S_P^{(0)}(\beta,t)}
-\frac{S_{P_0}^{(2)}(\beta,t)}{S_{P_0}^{(0)}(\beta,t)}\right\|
\leq C_{S2} d_{\mathrm{TV}}(P,P_0)
\label{eq:cox_tv_second_ratio_bound}
\end{equation}
for some $0 < C_{S2} < \infty$.

\noindent\textbf{Step 2: Score function.}\quad 
The risk-set bounds allow us to compare the scores under $P$ and $P_0$ at the same parameter value.
Since $0\leq Y\leq\tau$, the definition of $N(t)$ in \Cref{subsec:cox_stability} gives $U_P(\beta)=\mathbb E_P[\Delta\{X-\overline X_P(\beta,Y)\}]$.
Adding and subtracting the expectation under $P$ with $\overline X_{P_0}$ yields
$$U_P(\beta)-U_{P_0}(\beta)
={}\mathbb E_P\!\left[\Delta\{\overline X_{P_0}(\beta,Y)
-\overline X_P(\beta,Y)\}\right]
+(\mathbb E_P-\mathbb E_{P_0})\!\left[
\Delta\{X-\overline X_{P_0}(\beta,Y)\}\right].$$
The norm of the first term is at most the right-hand side of \eqref{eq:cox_tv_xbar_bound}. Moreover, $\overline X_{P_0}(\beta,t)$ is a weighted mean of covariates with $\|X\|\leq K$ under \Cref{ass:cox_covariates}. Hence $\|\overline X_{P_0}(\beta,t)\|\leq K$ and $\|\Delta\{X-\overline X_{P_0}(\beta,Y)\}\|_\infty\leq2K$. 

The total-variation bound therefore controls the second term by $4K d_{\mathrm{TV}}(P,P_0)$. Consequently, for some $0 < C_U<\infty$,
\begin{equation}
\sup_{\beta\in\mathcal B_r}\|U_P(\beta)-U_{P_0}(\beta)\|
\leq C_U d_{\mathrm{TV}}(P,P_0).
\label{eq:cox_tv_score_bound}
\end{equation}

\noindent\textbf{Step 3: Information matrices.}\quad 
A bound on the score difference alone does not control the distance between the roots of the two scores. We therefore compare the information matrices to obtain a positive lower bound on $J_P=-\nabla_\beta U_P$.
Differentiating the score, we have
\[
J_P(\beta)=\mathbb E_P\!\left[\Delta\left\{
\frac{S_P^{(2)}(\beta,Y)}{S_P^{(0)}(\beta,Y)}
-\overline X_P(\beta,Y)\overline X_P(\beta,Y)^\top
\right\}\right].
\]
Since \(S_P^{(2)}/S_P^{(0)}\) is a nonnegative weighted average of \(XX^\top\), \Cref{ass:cox_covariates} and \Cref{ass:cox_risksets} give
$\left\|S_P^{(2)}/S_P^{(0)}\right\|\leq K^2, \left\|S_{P_0}^{(2)}/S_{P_0}^{(0)}\right\|\leq K^2.$ Also, $\|\overline X_P\| \le K$ and $\|\overline X_{P_0}\| \le K$, so 
$\|\overline X_P\overline X_P^\top\| \le K^2, \|\overline X_{P_0}\overline X_{P_0}^\top\| \le K^2,
\|\overline X_P\overline X_P^\top
-\overline X_{P_0}\overline X_{P_0}^\top\|
\leq2K\|\overline X_P-\overline X_{P_0}\|
\leq 2KC_{\overline X} d_{\mathrm{TV}}(P,P_0),$ with the arguments $(\beta,t)$ understood. Thus,
the fixed $P_0$ expression has norm $$\|\Delta\left\{ \frac{S_P^{(2)}(\beta,Y)}{S_P^{(0)}(\beta,Y)} -\overline X_P(\beta,Y)\overline X_P(\beta,Y)^\top \right\}\|$$ at most $2K^2$, so its expectation under $P$ differs from that under $P_0$ by at most $4K^2d_{\mathrm{TV}}(P,P_0)$. Combining this expectation bound with \eqref{eq:cox_tv_second_ratio_bound} and \eqref{eq:cox_tv_xbar_bound} gives, for some $C_J<\infty$,
\begin{equation}
\sup_{\beta\in\mathcal B_r}\|J_P(\beta)-J_{P_0}(\beta)\|
\leq C_J d_{\mathrm{TV}}(P,P_0).
\label{eq:cox_tv_information_bound}
\end{equation}
\Cref{ass:cox_information} states that
$\inf_{\beta\in\mathcal N_{\beta(P_0)}}
\lambda_{\min}\{J_{P_0}(\beta)\}\geq\kappa.$
Both information matrices are symmetric. For a symmetric matrix \(A\), the smallest eigenvalue satisfies
$\lambda_{\min}(A)=\min_{\|u\|=1}u^\top Au.$
Now fix \(\beta\in\mathcal B_r\subset\mathcal N_{\beta(P_0)}\) and any unit vector \(u\). The assumption gives \(u^\top J_{P_0}(\beta)u\geq\kappa\). Although \(J_P(\beta)-J_{P_0}(\beta)\) need not be positive semidefinite, \eqref{eq:cox_tv_information_bound} gives
\[
\left|u^\top\{J_P(\beta)-J_{P_0}(\beta)\}u\right|
\leq\|J_P(\beta)-J_{P_0}(\beta)\|
\leq C_Jd_{\mathrm{TV}}(P,P_0).
\]
Consequently,
$u^\top J_P(\beta)u
=u^\top J_{P_0}(\beta)u
  +u^\top\{J_P(\beta)-J_{P_0}(\beta)\}u
\geq\kappa-C_Jd_{\mathrm{TV}}(P,P_0).$
This holds for every unit vector \(u\). Taking the minimum over \(u\) therefore yields
\[
\lambda_{\min}\{J_P(\beta)\}
\geq\kappa-C_Jd_{\mathrm{TV}}(P,P_0).
\]
Choose \(\varepsilon>0\) such that \(C_J\varepsilon\leq\kappa/2\). Then \(d_{\mathrm{TV}}(P,P_0)\leq\varepsilon\) implies
\begin{equation}
\inf_{\beta\in\mathcal B_r}\lambda_{\min}\{J_P(\beta)\}
\geq\frac{\kappa}{2}>0.
\label{eq:cox_information_lower_bound}
\end{equation}

\noindent\textbf{Step 4: Existence and uniqueness of the root.}\quad
The theorem also requires $\beta(P)$ to be well defined. Using the lower bound on $J_P$, we show that $U_P$ has a unique root in $\mathcal B_r$.
By \Cref{ass:cox_identification}, $U_{P_0}\{\beta(P_0)\}=0$. Evaluating \eqref{eq:cox_tv_score_bound} at $\beta(P_0)$ therefore gives
\begin{equation}
\|U_P\{\beta(P_0)\}\|\leq C_Ud_{\mathrm{TV}}(P,P_0).
\label{eq:cox_score_at_beta0}
\end{equation}
For a boundary point $\beta(P_0)+v$, where $\|v\|=r$, the fundamental theorem of calculus and $J_P=-\nabla_\beta U_P$ give
\[
U_P\{\beta(P_0)+v\}
=U_P\{\beta(P_0)\}
-\int_0^1J_P\{\beta(P_0)+sv\}v\,ds.
\]
Let \(u=v/r\), so that \(\|u\|=1\). Since \(\beta(P_0)+sv\in\mathcal B_r\) for every \(s\in[0,1]\),
\[
v^\top J_P\{\beta(P_0)+sv\}v
=r^2u^\top J_P\{\beta(P_0)+sv\}u \geq r^2\lambda_{\min}\{J_P(\beta(P_0)+sv)\}
\geq\frac{\kappa r^2}{2}.
\]
Multiplying the integral identity by \(v^\top\), applying Cauchy–Schwarz to its first term $v^\top U_P\{\beta(P_0)\} \leq \left|v^\top U_P\{\beta(P_0)\}\right| \leq \|v\|\,\|U_P\{\beta(P_0)\}\|$ , and using \eqref{eq:cox_score_at_beta0} and \eqref{eq:cox_information_lower_bound} gives
\begin{equation}
v^\top U_P\{\beta(P_0)+v\}
\leq rC_Ud_{\mathrm{TV}}(P,P_0)-\frac{\kappa r^2}{2}.
\label{eq:cox_boundary_score_bound}
\end{equation}
Reduce \(\varepsilon\) if necessary so that \(C_U\varepsilon\leq\kappa r/4\). Then, whenever \(d_{\mathrm{TV}}(P,P_0)\leq\varepsilon\), the right-hand side is at most \(-\kappa r^2/4<0\).


At a boundary point $\beta(P_0)+v$ with $\|v\|=r$, the path $\beta(P_0)+(1-h)v$ moves toward the center as $h$ increases from zero. For $0<h<1$, its distance from $\beta(P_0)$ is $(1-h)r<r$. Since $\nabla_\beta\ell_P=U_P$, \eqref{eq:cox_boundary_score_bound} gives
\[
-v^\top U_P\{\beta(P_0)+v\}
=\left.\frac{d}{dh}\ell_P\{\beta(P_0)+(1-h)v\}\right|_{h=0}>0.
\]
Thus moving toward the center increases the population Cox partial log-likelihood. Its maximum over the closed ball therefore lies in the interior, where $U_P(\beta)=0$. The lower bound in \eqref{eq:cox_information_lower_bound} makes this root unique in $\mathcal B_r$. Because the population Cox partial log-likelihood is concave, a second stationary point in $\mathcal N_{\beta(P_0)}$ would contradict strict concavity near this root. We denote the root by $\beta(P)$.

\noindent\textbf{Step 5: Parameter bound.}\quad 
With the existence and uniqueness of $\beta(P)$ established, we use the information bound once more to control its distance from $\beta(P_0)$. Finally, put $v=\beta(P)-\beta(P_0)$. We have $0 =U_P\{\beta(P_0)\} -\int_0^1J_P\{\beta(P_0)+sv\}v\,ds.$ Multiplying by \(v^\top\), we obtain
$v^\top U_P\{\beta(P_0)\}
=\int_0^1v^\top J_P\{\beta(P_0)+sv\}v\,ds.$

The line segment \(\{\beta(P_0)+sv:0\leq s\leq1\}\) lies in \(\mathcal B_r\). Hence \eqref{eq:cox_information_lower_bound}, together with \eqref{eq:cox_score_at_beta0}, implies
\[
\begin{aligned}
\frac{\kappa}{2}\|v\|^2
&\leq\int_0^1v^\top J_P\{\beta(P_0)+sv\}v\,ds\\
&=v^\top U_P\{\beta(P_0)\}\\
&\leq\|v\|\,\|U_P\{\beta(P_0)\}\|\\
&\leq\|v\|C_Ud_{\mathrm{TV}}(P,P_0).
\end{aligned}
\]
If \(v\neq0\), dividing by \(\|v\|\) yields
\[
\|\beta(P)-\beta(P_0)\|
=\|v\|
\leq\frac{2C_U}{\kappa}d_{\mathrm{TV}}(P,P_0).
\]
The same inequality holds when \(v=0\).
\end{proof}

\subsection{Proof of \Cref{thm:cox_wasserstein_stability}~(Local stability of the Cox parameter under $W_1$)}
\label{proof:cox_wasserstein_stability}
\begin{proof}\quad
By Assumptions~\ref{ass:cox_parameter} and \ref{ass:cox_identification}, choose $r>0$ such that
\[
\mathcal B_r=\{\beta:\|\beta-\beta(P_0)\|\leq r\}
\subset\mathcal N_{\beta(P_0)}\cap\operatorname{int}(B).
\]
We use $W_1$ with ground metric $\|X-X'\|+|Y-Y'|+|\Delta-\Delta'|$. All bounds below are uniform over $\beta\in\mathcal B_r$. Pointwise boundedness and time-Lipschitz statements hold over $t\in[0,\tau]$; perturbation bounds between risk-set quantities under $P$ and $P_0$ are integrated over $t$.

\noindent\textbf{Step 1: Integrated risk-set moments.}\quad 
Recall the risk-set moments determine both the score function and its information matrix. Unlike total variation, $W_1$ does not directly control the discontinuous risk-set indicator $R(t)=\mathbf 1(Y\ge t)$ at a fixed time. Since the score function averages risk-set quantities over event times, we instead establish integrated risk-set bounds, which can be controlled through a coupling argument under \Cref{ass:cox_w1_times}.

Fix $\beta\in\mathcal B_r$. Let $\pi\in\Pi(P_0,P)$ be an arbitrary coupling of $(X,Y,\Delta)\sim P_0$ and $(X',Y',\Delta')\sim P$, and write $\mathbb E_\pi$ for expectation under this joint distribution. Since the first marginal of $\pi$ is $P_0$ and its second marginal is $P$, $S_{P_0}^{(k)}(\beta,t)=\mathbb E_\pi[e^{\beta^\top X}X^{\otimes k}\mathbf 1(Y\geq t)]$ and $S_P^{(k)}(\beta,t)=\mathbb E_\pi[e^{\beta^\top X'}{X'}^{\otimes k}\mathbf 1(Y'\geq t)]$. Assumptions~\ref{ass:cox_parameter} and \ref{ass:cox_covariates} imply that, for each $k=0,1,2$, $X\mapsto e^{\beta^\top X}X^{\otimes k}$ is uniformly bounded by some fixed $0<B_k<\infty$ on $\{\|X\|\leq K\}$, over $\beta\in\mathcal B_r$. The mean-value theorem and Cauchy--Schwarz, together with boundedness of the covariates and $\beta$, give a fixed Lipschitz constant $0<L_k<\infty$ for the same map, uniformly over $\beta\in\mathcal B_r$. For $k=0,1,2$, the resulting comparison is
\begin{align*}
&~~~~\int_0^\tau\|S_P^{(k)}(\beta,t)-S_{P_0}^{(k)}(\beta,t)\|\,dt\\
&=\int_0^\tau\left\|\mathbb E_\pi\!\left[
e^{\beta^\top X'}{X'}^{\otimes k}\mathbf 1(Y'\geq t)
-e^{\beta^\top X}X^{\otimes k}\mathbf 1(Y\geq t)
\right]\right\|\,dt \\
&\leq\int_0^\tau\mathbb E_\pi\!\left\|
e^{\beta^\top X'}{X'}^{\otimes k}\mathbf 1(Y'\geq t)
-e^{\beta^\top X}X^{\otimes k}\mathbf 1(Y\geq t)
\right\|\,dt
\tag*{\scriptsize (Jensen's inequality)}\\
&=\mathbb E_\pi\int_0^\tau\left\|
e^{\beta^\top X'}{X'}^{\otimes k}\mathbf 1(Y'\geq t)
-e^{\beta^\top X}X^{\otimes k}\mathbf 1(Y\geq t)
\right\|\,dt
\tag*{\scriptsize (Tonelli's theorem)}\\
&=\mathbb E_\pi\int_0^\tau\left\| 
e^{\beta^\top X'}{X'}^{\otimes k}
\{\mathbf 1(Y'\geq t)-\mathbf 1(Y\geq t)\}
+\{e^{\beta^\top X'}{X'}^{\otimes k}
-e^{\beta^\top X}X^{\otimes k}\}\mathbf 1(Y\geq t)
\right\|\,dt\\
&\leq B_k\mathbb E_\pi\int_0^\tau
|\mathbf 1(Y'\geq t)-\mathbf 1(Y\geq t)|\,dt
+L_k\tau\,\mathbb E_\pi\|X'-X\|\\
&=B_k\mathbb E_\pi|Y'-Y|
+L_k\tau\,\mathbb E_\pi\|X'-X\|\\
&\leq\max\{B_k,L_k\tau\}\,\mathbb E_\pi
[\|X'-X\|+|Y'-Y|+|\Delta'-\Delta|]\\
\end{align*}

Because the coupling $\pi$ was arbitrary, taking the infimum over all couplings yields, for some fixed $0<C_S<\infty$ valid for $k=0,1,2$,
\begin{equation}
\sup_{\beta\in\mathcal B_r}\int_0^\tau
\|S_P^{(k)}(\beta,t)-S_{P_0}^{(k)}(\beta,t)\|\,dt
\leq C_S W_1(P,P_0),\qquad k=0,1,2.
\label{eq:cox_w1_riskset_bound}
\end{equation} 
\Cref{ass:cox_risksets} gives $S_P^{(0)}(\beta,t)\geq a$ and $S_{P_0}^{(0)}(\beta,t)\geq a$; their product is at least $a^2$. At each $(\beta,t)$,
\[
\overline X_P-\overline X_{P_0}
=\frac{S_P^{(1)}-S_{P_0}^{(1)}}{S_P^{(0)}}
+S_{P_0}^{(1)}\frac{S_{P_0}^{(0)}-S_P^{(0)}}
{S_P^{(0)}S_{P_0}^{(0)}}.
\]
Arguments $(\beta,t)$ are suppressed only in this identity. Assumptions~\ref{ass:cox_parameter} and \ref{ass:cox_covariates} uniformly bound $S_{P_0}^{(1)}$. Integrating and applying \eqref{eq:cox_w1_riskset_bound} yields, for some fixed $0<C_{\overline X}<\infty$,
\begin{equation}
\sup_{\beta\in\mathcal B_r}\int_0^\tau
\|\overline X_P(\beta,t)-\overline X_{P_0}(\beta,t)\|\,dt
\leq C_{\overline X}W_1(P,P_0).
\label{eq:cox_w1_xbar_bound}
\end{equation}
The same ratio calculation with $S^{(2)}$ in place of $S^{(1)}$ gives, for some fixed $0<C_{S2}<\infty$,
\begin{equation}
\sup_{\beta\in\mathcal B_r}\int_0^\tau
\left\|\frac{S_P^{(2)}(\beta,t)}{S_P^{(0)}(\beta,t)}
-\frac{S_{P_0}^{(2)}(\beta,t)}{S_{P_0}^{(0)}(\beta,t)}\right\|\,dt
\leq C_{S2}W_1(P,P_0).
\label{eq:cox_w1_second_ratio_bound}
\end{equation}
We also need time regularity under $P_0$ when changing the distribution in an expectation. For $0\leq t<t'\leq\tau$, \Cref{ass:cox_w1_times} gives $P_0(t\leq Y<t')\leq M(t'-t)$. Consequently,
\[
\|S_{P_0}^{(k)}(\beta,t)-S_{P_0}^{(k)}(\beta,t')\|
\leq B_kM(t'-t),\qquad k=0,1,2.
\]
The possible administrative censoring atom at $\tau$ belongs to both risk sets and does not enter $[t,t')$. The denominator lower bound and the ratio calculations above make $\overline X_{P_0}(\beta,t)$ and $S_{P_0}^{(2)}(\beta,t)/S_{P_0}^{(0)}(\beta,t)$ uniformly Lipschitz in $t$.

\noindent\textbf{Step 2: Score function.}\quad
The risk-set bounds allow us to compare the scores under $P$ and $P_0$ at the same parameter value. Unlike the total-variation argument, this step uses the event-time density to pass from an integrated risk-set bound to a score bound. Since $0\leq Y,Y'\leq\tau$, the definition of $N(t)$ in \Cref{subsec:cox_stability} gives $U_{P_0}(\beta)=\mathbb E_{P_0}[\Delta\{X-\overline X_{P_0}(\beta,Y)\}]$ and $U_P(\beta)=\mathbb E_P[\Delta'\{X'-\overline X_P(\beta,Y')\}]$. Adding and subtracting the expectation under $P$ with $\overline X_{P_0}$ gives
\[
\begin{aligned}
&~~~~U_P(\beta)-U_{P_0}(\beta)\\
&=\mathbb E_P\!\left[\Delta'\{\overline X_{P_0}(\beta,Y')
-\overline X_P(\beta,Y')\}\right]
+\mathbb E_P\!\left[\Delta'\{X'-\overline X_{P_0}(\beta,Y')\}\right]
-\mathbb E_{P_0}\!\left[\Delta\{X-\overline X_{P_0}(\beta,Y)\}\right]\\
&=\mathbb E_P\!\left[\Delta'\{\overline X_{P_0}(\beta,Y')
-\overline X_P(\beta,Y')\}\right]+
\mathbb E_\pi\!\left[
\Delta'\{X'-\overline X_{P_0}(\beta,Y')\}
-\Delta\{X-\overline X_{P_0}(\beta,Y)\}
\right].
\end{aligned}
\]

For any measurable set $A\subset[0,\tau)$, \Cref{ass:cox_w1_times} gives the event-time measure bound
\begin{equation}
P(Y'\in A,\Delta'=1)
\leq P(Y'\in A)
\leq M|A|,
\label{eq:cox_w1_event_time_measure_bound}
\end{equation}
where $|A|$ denotes the Lebesgue measure of $A$. Thus, the event-time subdistribution $P(Y'\in dt,\Delta'=1)$ has a density bounded by $M$ on $[0,\tau)$. Since the same assumption excludes an event atom at $\tau$, \eqref{eq:cox_w1_xbar_bound} gives
$$\left\|
\mathbb E_P\left[
\Delta'\{
\overline X_{P_0}(\beta,Y')
-\overline X_P(\beta,Y')
\}
\right]
\right\|
\leq
M\int_0^\tau
\left\|
\overline X_{P_0}(\beta,t)
-\overline X_P(\beta,t)
\right\|\,dt
\leq
MC_{\overline X}W_1(P,P_0).$$

For the second term, Step 1 shows that $\overline X_{P_0}(\beta,t)$ is uniformly Lipschitz. \Cref{ass:cox_covariates} gives $\|X\|,\|X'\|\leq K$, while $\overline X_{P_0}$, as a weighted covariate mean, has norm at most $K$. Thus the single-record map $(X,Y,\Delta)\mapsto\Delta\{X-\overline X_{P_0}(\beta,Y)\}$ is bounded and Lipschitz in the stated ground metric, uniformly over $\beta\in\mathcal B_r$. The Wasserstein bound for expectations controls this difference by $C_{U,0}W_1(P,P_0)$ for some fixed $0<C_{U,0}<\infty$. Combining the two errors gives, for some fixed $0<C_U<\infty$,
\begin{equation}
\sup_{\beta\in\mathcal B_r}\|U_P(\beta)-U_{P_0}(\beta)\|
\leq C_UW_1(P,P_0).
\label{eq:cox_w1_score_bound}
\end{equation}

\noindent\textbf{Step 3: Information matrices.}\quad
A bound on the score difference alone does not control the distance between the roots of the two scores. We therefore compare the information matrices to obtain a positive lower bound on $J_P=-\nabla_\beta U_P$. Recall
\[
J_P(\beta)=\mathbb E_P\!\left[\Delta'\left\{
\frac{S_P^{(2)}(\beta,Y')}{S_P^{(0)}(\beta,Y')}
-\overline X_P(\beta,Y')\overline X_P(\beta,Y')^\top
\right\}\right].
\]
Adding and subtracting
$\mathbb E_P\!\left[\Delta'\left\{
\frac{S_{P_0}^{(2)}(\beta,Y')}{S_{P_0}^{(0)}(\beta,Y')}
-\overline X_{P_0}(\beta,Y')\overline X_{P_0}(\beta,Y')^\top
\right\}\right]$
gives\[
\begin{aligned}
J_P(\beta)-J_{P_0}(\beta)
={}&
\mathbb E_P\!\left[\Delta'\left\{
\frac{S_P^{(2)}(\beta,Y')}{S_P^{(0)}(\beta,Y')}
-\frac{S_{P_0}^{(2)}(\beta,Y')}{S_{P_0}^{(0)}(\beta,Y')}
\right\}\right]\\
&-\mathbb E_P\!\left[\Delta'\left\{
\overline X_P(\beta,Y')\overline X_P(\beta,Y')^\top
-\overline X_{P_0}(\beta,Y')\overline X_{P_0}(\beta,Y')^\top
\right\}\right]\\
&+\mathbb E_P\!\left[\Delta'\left\{
\frac{S_{P_0}^{(2)}(\beta,Y')}{S_{P_0}^{(0)}(\beta,Y')}
-\overline X_{P_0}(\beta,Y')\overline X_{P_0}(\beta,Y')^\top
\right\}\right]\\
&-\mathbb E_{P_0}\!\left[\Delta\left\{
\frac{S_{P_0}^{(2)}(\beta,Y)}{S_{P_0}^{(0)}(\beta,Y)}
-\overline X_{P_0}(\beta,Y)\overline X_{P_0}(\beta,Y)^\top
\right\}\right].
\end{aligned}
\]
The first two terms account for changes in the risk-set quantities. The last two terms compare the same \(P_0\)-based integrand evaluated under the joint observed-data distributions \(P\) and \(P_0\). Under $P$, all risk-set quantities are evaluated at $(\beta,Y')$, while evaluated at $(\beta,Y)$  under $P_0$.
For the first term, \eqref{eq:cox_w1_event_time_measure_bound} and \eqref{eq:cox_w1_second_ratio_bound} give $MC_{S2}W_1(P,P_0)$. For the second, $\|\overline X_P\|,\|\overline X_{P_0}\|\leq K$ imply, at each $(\beta,t)$,
\[
\|\overline X_P\overline X_P^\top
-\overline X_{P_0}\overline X_{P_0}^\top\|
\leq2K\|\overline X_P-\overline X_{P_0}\|.
\]
The event-time measure bound \eqref{eq:cox_w1_event_time_measure_bound} and \eqref{eq:cox_w1_xbar_bound} therefore control the second term by $2KMC_{\overline X}W_1(P,P_0)$. 

The difference of the last two expectations can be written as
\[
\mathbb E_\pi\Bigg[
\Delta'\left\{
\frac{S_{P_0}^{(2)}(\beta,Y')}{S_{P_0}^{(0)}(\beta,Y')}
-\overline X_{P_0}(\beta,Y')\overline X_{P_0}(\beta,Y')^\top
\right\}
-\Delta\left\{
\frac{S_{P_0}^{(2)}(\beta,Y)}{S_{P_0}^{(0)}(\beta,Y)}
-\overline X_{P_0}(\beta,Y)\overline X_{P_0}(\beta,Y)^\top
\right\}
\Bigg].
\]
By the time-Lipschitz bounds established in Step 1,  the fixed $P_0$ integrand is bounded and uniformly Lipschitz in time, and the discrete event-indicator coordinate controls changes in event status. Its expectation difference is then bounded by $C_{J,0}W_1(P,P_0)$ for some fixed $0<C_{J,0}<\infty$. Thus, for some fixed $0<C_J<\infty$,
\begin{equation}
\sup_{\beta\in\mathcal B_r}\|J_P(\beta)-J_{P_0}(\beta)\|
\leq C_J W_1(P,P_0).
\label{eq:cox_w1_information_bound}
\end{equation}
\Cref{ass:cox_information} gives $\lambda_{\min}\{J_{P_0}(\beta)\}\geq\kappa$ on $\mathcal B_r$. Applying \eqref{eq:cox_w1_information_bound} directly,
\[
\lambda_{\min}\{J_P(\beta)\}
\geq\lambda_{\min}\{J_{P_0}(\beta)\}
-\|J_P(\beta)-J_{P_0}(\beta)\|
\geq\kappa-C_JW_1(P,P_0).
\]
Choose $\varepsilon>0$ so that $C_J\varepsilon\leq\kappa/2$. Then
\begin{equation}
\inf_{\beta\in\mathcal B_r}\lambda_{\min}\{J_P(\beta)\}
\geq\kappa/2
\quad\text{whenever }W_1(P,P_0)\leq\varepsilon.
\label{eq:cox_w1_information_lower_bound}
\end{equation}

\noindent\textbf{Step 4: Existence and uniqueness of the root.}\quad
The theorem also requires $\beta(P)$ to be well defined. Using the lower bound on $J_P$, we show that $U_P$ has a unique root in $\mathcal B_r$. \Cref{ass:cox_identification} gives $U_{P_0}\{\beta(P_0)\}=0$. Evaluating \eqref{eq:cox_w1_score_bound} at $\beta(P_0)$ yields
\begin{equation}
\|U_P\{\beta(P_0)\}\|\leq C_UW_1(P,P_0).
\label{eq:cox_w1_score_at_beta0}
\end{equation}
Take a boundary point $\beta(P_0)+v$, where $\|v\|=r$. The fundamental theorem of calculus and $J_P=-\nabla_\beta U_P$ give
\[
v^\top U_P\{\beta(P_0)+v\}
=v^\top U_P\{\beta(P_0)\}
-\int_0^1v^\top J_P\{\beta(P_0)+sv\}v\,ds.
\]
Cauchy--Schwarz and \eqref{eq:cox_w1_score_at_beta0} bound the first term by $rC_UW_1(P,P_0)$. The path lies in $\mathcal B_r$, so \eqref{eq:cox_w1_information_lower_bound} bounds the integral below by $\kappa r^2/2$. Hence
\begin{equation}
v^\top U_P\{\beta(P_0)+v\}
\leq rC_UW_1(P,P_0)-\kappa r^2/2.
\label{eq:cox_w1_boundary_score_bound}
\end{equation}
Reduce $\varepsilon$ if necessary so that $C_U\varepsilon\leq\kappa r/4$. The right-hand side is then at most $-\kappa r^2/4<0$. The population Cox partial log-likelihood $\ell_P$ is continuous on the compact ball $\mathcal B_r$, and its gradient is $U_P$. At a boundary point, its directional derivative toward the center is $-v^\top U_P\{\beta(P_0)+v\}>0$ by \eqref{eq:cox_w1_boundary_score_bound}. Its maximum over the ball therefore lies in the interior, where $U_P=0$.

The lower bound in \eqref{eq:cox_w1_information_lower_bound} makes the population Cox partial log-likelihood strictly concave on $\mathcal B_r$, so this root is unique there. \Cref{subsec:cox_stability} states that $\ell_P$ is concave on $\mathbb R^p$. If another stationary point existed, both points would be global maximizers. The segment joining them would then consist of maximizers, including a nontrivial portion inside $\mathcal B_r$, contradicting strict concavity there. Thus no other stationary point exists. We denote this unique root by $\beta(P)$.

\noindent\textbf{Step 5: Parameter bound.}\quad
With the existence and uniqueness of $\beta(P)$ established, we use the information bound once more to control its distance from $\beta(P_0)$. Put $v=\beta(P)-\beta(P_0)$. Since $U_P\{\beta(P)\}=0$, integration along the segment joining the two parameters gives
\[
v^\top U_P\{\beta(P_0)\}
=\int_0^1v^\top J_P\{\beta(P_0)+sv\}v\,ds.
\]
The segment lies in $\mathcal B_r$. Applying \eqref{eq:cox_w1_information_lower_bound} to the integral and \eqref{eq:cox_w1_score_at_beta0} to the score gives
\[
\begin{aligned}
\frac{\kappa}{2}\|v\|^2 \leq v^\top U_P\{\beta(P_0)\} \leq\|v\|\,\|U_P\{\beta(P_0)\}\| \leq C_U\|v\|W_1(P,P_0).
\end{aligned}
\]
Dividing by $\|v\|$ when $v\neq0$, with the $v=0$ case immediate, yields
\[
\|\beta(P)-\beta(P_0)\|
\leq\frac{2C_U}{\kappa}W_1(P,P_0).
\]
Taking $0<C=2C_U/\kappa<\infty$ completes the proof.
\end{proof}

\subsection{Proof of \Cref{cor:cox_synthetic}~(Estimation rate and preservation of coefficient directions)}
\label{proof:cox_synthetic}
\begin{proof}\quad
\noindent\textbf{Step 1: Population error.}\quad
The corollary assumes $\widehat P_n\in\mathcal P_0$. In the total-variation case, $d_{\mathrm{TV}}(\widehat P_n,P_0)=O_p(r_n)$ and $r_n\to0$ in \Cref{thm:cox_population_stability}. In the $W_1$ case, $W_1(\widehat P_n,P_0)=O_p(r_n)$ in \Cref{thm:cox_wasserstein_stability}. Applying the corresponding theorem yields
\begin{equation}
\|\beta(\widehat P_n)-\beta(P_0)\|=O_p(r_n).
\label{eq:cox_cor_population_rate}
\end{equation}

\noindent\textbf{Step 2: Empirical sampling error.}\quad
Conditional on the original sample, the synthetic records are i.i.d. from $\widehat P_n$.  $\beta(\widehat P_n)$ lies in the fixed local parameter set, and Assumptions~\ref{ass:cox_covariates} and \ref{ass:cox_risksets} give common covariate and risk-set denominator bounds. The information lower bounds in \eqref{eq:cox_information_lower_bound} and \eqref{eq:cox_w1_information_lower_bound} provide a common nonsingular neighborhood for the fitted laws under the respective distance conditions. Together with the uniform empirical regularity condition in the corollary, the classical empirical Cox estimator result applies conditionally with uniform constants \citep{andersen1982cox,andersen1993statistical}, giving
\begin{equation}
\|\beta(\widehat P_{n,m})-\beta(\widehat P_n)\|
=O_p(m^{-1/2}).
\label{eq:cox_cor_synthetic_sampling_rate}
\end{equation}
Applying the same result to the i.i.d. original records from $P_0$ gives $\|\beta(P_n)-\beta(P_0)\|=O_p(n^{-1/2})$. Combining \eqref{eq:cox_cor_population_rate} and \eqref{eq:cox_cor_synthetic_sampling_rate} with the triangle inequality gives
\begin{equation}
\|\beta(\widehat P_{n,m})-\beta(P_0)\|
\leq\|\beta(\widehat P_{n,m})-\beta(\widehat P_n)\|
+\|\beta(\widehat P_n)-\beta(P_0)\|=O_p(r_n+m^{-1/2}).
\label{eq:cox_cor_total_rate}
\end{equation}
Using this bound and the original-sample rate in another application of the triangle inequality gives
\[
\|\beta(\widehat P_{n,m})-\beta(P_n)\|
\leq\|\beta(\widehat P_{n,m})-\beta(P_0)\|
+\|\beta(P_n)-\beta(P_0)\|=O_p(r_n+m^{-1/2}+n^{-1/2}).
\]

\noindent\textbf{Step 3: Coefficient directions.}\quad
Recall that $\mathcal J\coloneqq \{j:\beta_j(P_0)\neq 0\}$. Then define \(\eta \coloneqq \min_{j\in\mathcal J}|\beta_j(P_0)|\). For any $j\in\mathcal J$, if the signs of $\beta_j(\widehat P_{n,m})$ and $\beta_j(P_0)$ differ, then
\[
|\beta_j(\widehat P_{n,m})-\beta_j(P_0)|
\geq|\beta_j(P_0)|\geq\eta >0.
\]
Hence
\[\Pr\!\left\{
\operatorname{sign}\{\beta_j(\widehat P_{n,m})\}
\neq\operatorname{sign}\{\beta_j(P_0)\}
\text{ for some }j\in\mathcal J
\right\}\leq
\Pr\!\left\{
\|\beta(\widehat P_{n,m})-\beta(P_0)\|\geq\eta
\right\}
\longrightarrow0.
\]
The convergence follows from \eqref{eq:cox_cor_total_rate}.
\end{proof}

\subsection{Proof of \Cref{thm:cox_covariance_stability}~(Local stability of the Cox sandwich covariance)}
\label{proof:cox_covariance_stability}
\begin{proof}\quad
Choose $\mathcal B_r$ as in the proof of \Cref{thm:cox_wasserstein_stability} and take $W_1(P,P_0)$ sufficiently small. That theorem gives, for a fixed $0<C_\beta<\infty$,
\[
\|\beta(P)-\beta(P_0)\|\leq C_\beta W_1(P,P_0),
\]
and \eqref{eq:cox_w1_information_lower_bound} gives $\inf_{\beta\in\mathcal B_r}\lambda_{\min}\{J_P(\beta)\} \geq\kappa/2$. All bounds involving $\beta$ below are uniform over $\mathcal B_r$. We first compare the influence-function expressions, then their second moments and the information matrices.

\noindent\textbf{Step 1: Changes in the risk-set integrands.}\quad
Fix $\beta\in\mathcal B_r$. Recall from \eqref{eq:cox_xi} that, for $D'=(X',Y',\Delta')\sim P$ and with $\beta(P)$ temporarily replaced by the common coefficient $\beta$, 
\[ \xi(D') = 
\underbrace{\Delta'\{X'-\overline X_P(\beta,Y')\}}_{\text{direct term}}
-\underbrace{e^{\beta^\top X'}\int_0^\tau
\mathbf 1(Y'\geq t)
\frac{X'-\overline X_P(\beta,t)}{S_P^{(0)}(\beta,t)}
\,\mathbb E_P\{dN(t)\}}_{\text{cumulative term}}.
\]
We first compare the $P$ and $P_0$ expressions at this common coefficient. Step~3 accounts for the subsequent change between $\beta(P)$ and $\beta(P_0)$.

For the direct term, 
\[
\begin{aligned}
&~~~~~~~\mathbb E_P\!\left[\Delta'
\|\overline X_P(\beta,Y')-\overline X_{P_0}(\beta,Y')\|\right]\\
&\quad\leq M\int_0^\tau
\|\overline X_P(\beta,t)-\overline X_{P_0}(\beta,t)\|\,dt~&&\text{\scriptsize  \Cref{eq:cox_w1_event_time_measure_bound} and \Cref{ass:cox_w1_times}}&\\
&\quad\leq MC_{\overline X}W_1(P,P_0).~&&\text{\scriptsize \Cref{eq:cox_w1_xbar_bound}}&.
\end{aligned}
\]

For the cumulative term, 
\[
\begin{aligned}
&~~~~e^{\beta^\top X'}\int_0^\tau \mathbf 1(Y'\geq t)
\frac{X'-\overline X_P(\beta,t)}{S_P^{(0)}(\beta,t)}
\,\mathbb E_P\{dN(t)\}-
e^{\beta^\top X'}\int_0^\tau \mathbf 1(Y'\geq t)
\frac{X'-\overline X_{P_0}(\beta,t)}{S_{P_0}^{(0)}(\beta,t)}
\,\mathbb E_{P_0}\{dN(t)\}\\
&=
e^{\beta^\top X'}\int_0^\tau \mathbf 1(Y'\geq t)
\left\{
\frac{X'-\overline X_P(\beta,t)}{S_P^{(0)}(\beta,t)}
-
\frac{X'-\overline X_{P_0}(\beta,t)}{S_{P_0}^{(0)}(\beta,t)}
\right\}\mathbb E_P\{dN(t)\}\\
&~~~~\quad+
e^{\beta^\top X'}\int_0^\tau \mathbf 1(Y'\geq t)
\frac{X'-\overline X_{P_0}(\beta,t)}{S_{P_0}^{(0)}(\beta,t)}
(\mathbb E_P-\mathbb E_{P_0})\{dN(t)\}.
\end{aligned}
\] In step 1, we only consider the first term on the right side. Step 2 account for the second term.

Keep $\mathbb E_P\{dN(t)\}$ fixed and change its risk-set integrand. At each $(\beta,t)$, adding and subtracting $(X'-\overline X_{P_0}(\beta,t))/S_P^{(0)}(\beta,t)$ gives
\[
\begin{aligned}
&~~~~~~~e^{\beta^\top X'}\mathbf 1(Y'\geq t)
\left\{
\frac{X'-\overline X_P(\beta,t)}{S_P^{(0)}(\beta,t)}
-\frac{X'-\overline X_{P_0}(\beta,t)}{S_{P_0}^{(0)}(\beta,t)}
\right\}\\
&\quad=e^{\beta^\top X'}\mathbf 1(Y'\geq t)
\frac{\overline X_{P_0}(\beta,t)-\overline X_P(\beta,t)}
{S_P^{(0)}(\beta,t)}+e^{\beta^\top X'}\mathbf 1(Y'\geq t)
\{X'-\overline X_{P_0}(\beta,t)\}
\frac{S_{P_0}^{(0)}(\beta,t)-S_P^{(0)}(\beta,t)}
{S_P^{(0)}(\beta,t)S_{P_0}^{(0)}(\beta,t)}.
\end{aligned}
\]
\Cref{ass:cox_risksets} bounds the first denominator below by $a$ and the product in the second denominator below by $a^2$. \Cref{ass:cox_covariates} gives $\|X'\|\leq K$, and $\|\overline X_{P_0}(\beta,t)\| \leq K$. Taking norms of the fraction comparison therefore gives
$$\left\|
\frac{X'-\overline X_P(\beta,t)}{S_P^{(0)}(\beta,t)}
-\frac{X'-\overline X_{P_0}(\beta,t)}{S_{P_0}^{(0)}(\beta,t)}
\right\|\leq
\frac{\|\overline X_P(\beta,t)-\overline X_{P_0}(\beta,t)\|}{a}+\frac{2K|S_P^{(0)}(\beta,t)-S_{P_0}^{(0)}(\beta,t)|}{a^2}.$$ Equation~\eqref{eq:cox_w1_event_time_measure_bound} bounds the density of $\mathbb E_P\{dN(t)\}$ by $M$ on $[0,\tau)$, while \Cref{ass:cox_w1_times} excludes an event atom at $\tau$. Integrating the last inequality against this measure, applying \eqref{eq:cox_w1_xbar_bound} to its first term and \eqref{eq:cox_w1_riskset_bound} with $k=0$ to its second, gives
$$M\int_0^\tau\left\{
\frac{\|\overline X_P(\beta,t)-\overline X_{P_0}(\beta,t)\|}{a}
+\frac{2K|S_P^{(0)}(\beta,t)-S_{P_0}^{(0)}(\beta,t)|}{a^2}
\right\}dt\leq
\left(\frac{MC_{\overline X}}{a}
+\frac{2KMC_S}{a^2}\right)W_1(P,P_0).$$
This bound holds for every $X'$ with $\|X'\|\leq K$ and every upper limit $Y'\in[0,\tau]$.

Finally, $\mathbf 1(Y'\geq t)\leq1$, and the fixed constant $B_0$ from Step~1 of the proof of \Cref{thm:cox_wasserstein_stability} bounds $e^{\beta^\top X'}$ uniformly over $\beta\in\mathcal B_r$ and $\|X'\|\leq K$. Restoring both factors therefore gives
\begin{equation}
\begin{aligned}
&~~~~~~~\mathbb E_P\left\|
e^{\beta^\top X'}\int_0^\tau\mathbf 1(Y'\geq t)
\left\{
\frac{X'-\overline X_P(\beta,t)}{S_P^{(0)}(\beta,t)}
-\frac{X'-\overline X_{P_0}(\beta,t)}
{S_{P_0}^{(0)}(\beta,t)}
\right\}\mathbb E_P\{dN(t)\}
\right\|\\
&\quad\leq B_0M\int_0^\tau\left\{
\frac{\|\overline X_P(\beta,t)-\overline X_{P_0}(\beta,t)\|}{a}
+\frac{2K|S_P^{(0)}(\beta,t)-S_{P_0}^{(0)}(\beta,t)|}{a^2}
\right\}dt\\
&\quad\leq B_0\left(\frac{MC_{\overline X}}{a}
+\frac{2KMC_S}{a^2}\right)W_1(P,P_0).
\end{aligned}
\label{eq:cox_w1_cov_risk_integral_bound}
\end{equation}
This controls the risk-set change within the cumulative term. Step~2 controls the change from $\mathbb E_P\{dN(t)\}$ to $\mathbb E_{P_0}\{dN(t)\}$.

\noindent\textbf{Step 2: Change in the event-time measure.}\quad
We now keep the $P_0$ risk-set integrand fixed and change $\mathbb E_{P_0}\{dN(t)\}$ to $\mathbb E_P\{dN(t)\}$. Both expressions are evaluated at the same record $D'=(X',Y',\Delta')$. Thus the $P_0$-based integrand still contains $X'$.

The same calculation applies to
\[
h(t)=\frac{1}{S_{P_0}^{(0)}(\beta,t)}
\quad\text{and}\quad
h(t)=\frac{\overline X_{P_0}(\beta,t)}
{S_{P_0}^{(0)}(\beta,t)}.
\]
These are the two parts obtained by expanding
$\frac{X'-\overline X_{P_0}(\beta,t)}
{S_{P_0}^{(0)}(\beta,t)}
=X'\frac{1}{S_{P_0}^{(0)}(\beta,t)}
-\frac{\overline X_{P_0}(\beta,t)}
{S_{P_0}^{(0)}(\beta,t)}.$
The time regularity established in Step~1 of the proof of \Cref{thm:cox_wasserstein_stability} and \Cref{ass:cox_risksets} make each $h$ uniformly bounded by a fixed $B_h$ and uniformly Lipschitz in time with a fixed $L_h$.

Let $\pi$ couple $D=(X,Y,\Delta)\sim P_0$ and $D'=(X',Y',\Delta')\sim P$. For a fixed $y\in[0,\tau]$, the marginal identities give
$\int_0^y h(t)(\mathbb E_P-\mathbb E_{P_0})\{dN(t)\}
=\mathbb E_\pi\!\left[
\Delta'h(Y')\mathbf 1(Y'\leq y)
-\Delta h(Y)\mathbf 1(Y\leq y)
\right].$ The integral includes an event at its upper limit $y$.

To evaluate the cumulative difference at a random upper limit without reusing the coupled variable $Y'$, let $Y^\circ$ be independent of $(D,D')\sim\pi$ and have the marginal distribution of $Y'$ under $P$. Then
\[
\begin{aligned}
&~~~~~~~\mathbb E_{Y^\circ}\left\|
\int_0^{Y^\circ}h(t)(\mathbb E_P-\mathbb E_{P_0})\{dN(t)\}
\right\|\\
&\quad=\int_0^\tau
\left\|\mathbb E_\pi\!\left[
\Delta'h(Y')\mathbf 1(Y'\leq y)
-\Delta h(Y)\mathbf 1(Y\leq y)
\right]\right\|P(Y^\circ\in dy)\\
&\quad\leq\mathbb E_{\pi,Y^\circ}\left\|
\Delta'h(Y')\mathbf 1(Y'\leq Y^\circ)
-\Delta h(Y)\mathbf 1(Y\leq Y^\circ)
\right\|.
\end{aligned}
\]
For fixed coupled values $Y,Y'$, the two indicators differ only when $Y^\circ$ lies between $Y$ and $Y'$. Any atom at $Y^\circ=\tau$ contributes nothing because both indicators equal one there. \Cref{ass:cox_w1_times} therefore gives
\[
\mathbb E_{Y^\circ}
\left|\mathbf 1(Y'\leq Y^\circ)-\mathbf 1(Y\leq Y^\circ)\right|
\leq M|Y'-Y|.
\]
Adding and subtracting $\Delta h(Y)\mathbf 1(Y'\leq Y^\circ)$ inside the coupled integrand gives
\[
\begin{aligned}
&~~~~~~~\Delta'h(Y')\mathbf 1(Y'\leq Y^\circ)
-\Delta h(Y)\mathbf 1(Y\leq Y^\circ)\\
&\quad=\{\Delta'h(Y')-\Delta h(Y)\}\mathbf 1(Y'\leq Y^\circ)
+\Delta h(Y)\{\mathbf 1(Y'\leq Y^\circ)-\mathbf 1(Y\leq Y^\circ)\}.
\end{aligned}
\]
Since $\|h\|\leq B_h$, $h$ is $L_h$-Lipschitz in time,  $\|\Delta'h(Y')-\Delta h(Y)\| \leq B_h|\Delta'-\Delta|+L_h|Y'-Y|$ for  $\Delta,\Delta'\in\{0,1\}$. Applying these two bounds gives
$$\mathbb E_{Y^\circ}\left\|
\int_0^{Y^\circ}h(t)(\mathbb E_P-\mathbb E_{P_0})\{dN(t)\}
\right\|
\leq B_h\mathbb E_\pi|\Delta'-\Delta|
+(L_h+MB_h)\mathbb E_\pi|Y'-Y|.$$

Because $\pi$ was arbitrary, taking the infimum over couplings gives
\begin{equation}
\mathbb E_{Y^\circ}\left\|
\int_0^{Y^\circ}h(t)(\mathbb E_P-\mathbb E_{P_0})\{dN(t)\}
\right\|
\leq C_hW_1(P,P_0)
\label{eq:cox_w1_cov_event_integral_bound}
\end{equation}
for a fixed $0 < C_h<\infty$ valid for either choice of $h$. Because $Y^\circ$ and $Y'$ have the same marginal distribution, the left-hand side is the same random-upper-limit expectation required below.

Finally, apply \eqref{eq:cox_w1_cov_event_integral_bound} to both terms in the decomposition of $(X'-\overline X_{P_0})/S_{P_0}^{(0)}$. Using $\|X'\|\leq K$ and $e^{\beta^\top X'}\leq B_0$ gives
$$\mathbb E_P\left\|
e^{\beta^\top X'}\int_0^{Y'}
\frac{X'-\overline X_{P_0}(\beta,t)}
{S_{P_0}^{(0)}(\beta,t)}
(\mathbb E_P-\mathbb E_{P_0})\{dN(t)\}
\right\|
\leq B_0(K+1)C_hW_1(P,P_0).$$
This controls the change in the event-time measure within the cumulative term. Together with \eqref{eq:cox_w1_cov_risk_integral_bound}, it controls the cumulative-term difference at the common coefficient $\beta$.

\noindent\textbf{Step 3: From a common coefficient to $\xi_P$.}\quad
Steps~1 and 2 compare the $P$ and $P_0$ expressions at the common coefficient $\beta(P_0)$. We now change the coefficient in the $P$ expression from $\beta(P_0)$ to $\beta(P)$. The $P_0$ expression remains at $\beta(P_0)$, and both expressions continue to be evaluated at the same record $D'\sim P$.

The mean-value theorem, Assumptions~\ref{ass:cox_parameter} and \ref{ass:cox_covariates}, and the bound $e^{\beta^\top X'}\leq B_0$ from the proof of \Cref{thm:cox_wasserstein_stability} give $|e^{\beta(P)^\top X'}-e^{\beta(P_0)^\top X'}| \leq B_0K\|\beta(P)-\beta(P_0)\|$. The same argument applied inside the expectation defining $S_P^{(0)}$ gives $|S_P^{(0)}\{\beta(P),t\}-S_P^{(0)}\{\beta(P_0),t\}| \leq B_0K\|\beta(P)-\beta(P_0)\|$. Both bounds are uniform over $t\in[0,\tau]$.

The derivative of $\overline X_P(\beta,t)$ with respect to $\beta$ is $S_P^{(2)}(\beta,t)/S_P^{(0)}(\beta,t) -\overline X_P(\beta,t)\overline X_P(\beta,t)^\top$. This is a weighted risk-set covariance matrix. Since $\|X\|\leq K$ by \Cref{ass:cox_covariates}, its operator norm is at most $K^2$. The line segment between $\beta(P_0)$ and $\beta(P)$ lies in $\mathcal B_r$, so $\|\overline X_P\{\beta(P),t\}-\overline X_P\{\beta(P_0),t\}\| \leq K^2\|\beta(P)-\beta(P_0)\|$, uniformly over $t$.

For the direct term, the record $D'$ is unchanged. Its coefficient change therefore satisfies
\[
\bigl\|\Delta'[\overline X_P\{\beta(P),Y'\} -\overline X_P\{\beta(P_0),Y'\}] \bigr\|
\leq K^2\|\beta(P)-\beta(P_0)\|.
\]

For the cumulative term, first compare the fractions inside the integral. Adding and subtracting the fraction with numerator $X'-\overline X_P\{\beta(P_0),t\}$ and denominator $S_P^{(0)}\{\beta(P),t\}$ gives
\[
\begin{aligned}
&~~~~~~~\frac{X'-\overline X_P\{\beta(P),t\}}
{S_P^{(0)}\{\beta(P),t\}}
-\frac{X'-\overline X_P\{\beta(P_0),t\}}
{S_P^{(0)}\{\beta(P_0),t\}}\\
&\quad=
\frac{\overline X_P\{\beta(P_0),t\}
-\overline X_P\{\beta(P),t\}}
{S_P^{(0)}\{\beta(P),t\}}+
\{X'-\overline X_P\{\beta(P_0),t\}\}
\frac{S_P^{(0)}\{\beta(P_0),t\}
-S_P^{(0)}\{\beta(P),t\}}
{S_P^{(0)}\{\beta(P),t\}S_P^{(0)}\{\beta(P_0),t\}}.
\end{aligned}
\]
\Cref{ass:cox_risksets} bounds each $S_P^{(0)}$ below by $a$, so the product in the second denominator is at least $a^2$. Using the coefficient bounds above and $\|X'-\overline X_P\|\leq2K$ gives
$$\left\|
\frac{X'-\overline X_P\{\beta(P),t\}}
{S_P^{(0)}\{\beta(P),t\}}
-\frac{X'-\overline X_P\{\beta(P_0),t\}}
{S_P^{(0)}\{\beta(P_0),t\}}
\right\|\leq
\left(\frac{K^2}{a}+\frac{2B_0K^2}{a^2}\right)
\|\beta(P)-\beta(P_0)\|.$$
The cumulative integrand also contains an exponential factor. Adding and subtracting the product with $e^{\beta(P_0)^\top X'}$ and the fraction at $\beta(P)$ gives
\[
\begin{aligned}
&~~~~~~\left\|
e^{\beta(P)^\top X'}
\frac{X'-\overline X_P\{\beta(P),t\}}
{S_P^{(0)}\{\beta(P),t\}}
-e^{\beta(P_0)^\top X'}
\frac{X'-\overline X_P\{\beta(P_0),t\}}
{S_P^{(0)}\{\beta(P_0),t\}}
\right\|\\
&\quad\leq
\left|e^{\beta(P)^\top X'}-e^{\beta(P_0)^\top X'}\right|
\frac{\|X'-\overline X_P\{\beta(P),t\}\|}
{S_P^{(0)}\{\beta(P),t\}} \\
&\qquad+
e^{\beta(P_0)^\top X'}
\left\|
\frac{X'-\overline X_P\{\beta(P),t\}}
{S_P^{(0)}\{\beta(P),t\}}-\frac{X'-\overline X_P\{\beta(P_0),t\}}
{S_P^{(0)}\{\beta(P_0),t\}}
\right\|\\
&\quad\leq
B_0\left(\frac{3K^2}{a}+\frac{2B_0K^2}{a^2}\right)
\|\beta(P)-\beta(P_0)\|.
\end{aligned}
\]
Since $\mathbf 1(Y'\geq t)\leq1$ and $\mathbb E_P\{dN(t)\}$ has total mass $\mathbb E_P[\Delta']\leq1$, the last line also bounds the coefficient change in the entire cumulative term after integration.

The direct and cumulative coefficient changes together are therefore bounded by
\[
\left[ K^2+B_0\left(\frac{3K^2}{a} +\frac{2B_0K^2}{a^2}\right)
\right]\|\beta(P)-\beta(P_0)\|.
\]
At the common coefficient $\beta(P_0)$, Step~1 controls the direct term and the risk-set change within the cumulative term. Step~2 controls the event-time measure change within that cumulative term. Applying those three bounds and then the parameter bound of \Cref{thm:cox_wasserstein_stability} gives
\begin{equation}
\begin{aligned}
\mathbb E_P\|\xi_P(D')-\xi_{P_0}(D')\|
&\leq
\left[
MC_{\overline X}
+B_0\left(\frac{MC_{\overline X}}{a}
+\frac{2KMC_S}{a^2}\right)
+B_0(K+1)C_h
\right]W_1(P,P_0)\\
&\quad+
\left[
K^2+B_0\left(\frac{3K^2}{a}
+\frac{2B_0K^2}{a^2}\right)
\right]\|\beta(P)-\beta(P_0)\|\\
&\leq C_\xi W_1(P,P_0),
\end{aligned}
\label{eq:cox_w1_cov_xi_bound}
\end{equation}
for a fixed $0<C_\xi<\infty$. Both functions in \eqref{eq:cox_w1_cov_xi_bound} are evaluated at the same $P$-distributed record $D'$.

\noindent\textbf{Step 4: Centering and second moment.}\quad
Taking expectations of \eqref{eq:cox_xi}, $\mathbb E_P[\xi_P(D')]=0$. Hence $\mathbb E_P[\psi_P(D')]=0$ by \eqref{eq:cox_if}.

Equation~\eqref{eq:cox_xi}, \Cref{ass:cox_covariates}, and the denominator bound in \Cref{ass:cox_risksets} give, for every $P\in\mathcal P_0$ with $W_1(P,P_0)\leq\varepsilon$ and every record $D'=(X',Y',\Delta')$ with $\|X'\|\leq K$ and $Y'\in[0,\tau]$,
\[
\|\xi_P(D')\|
\leq 2K+\frac{2KB_0}{a}
\leq B_\xi,
\qquad 0<B_\xi<\infty.
\]
Here $B_0$ uniformly bounds $e^{\beta(P)^\top X'}$ on $\mathcal B_r$, and the event-time measure $\mathbb E_P\{dN(t)\}$ has total mass at most one. The same bound applies to $\xi_{P_0}(D')$ and $\xi_{P_0}(D)$ for records satisfying the corresponding covariate and time bounds.

Keeping the $P_0$ expression fixed, the map $D\mapsto\xi_{P_0}(D)$ is Lipschitz under the ground metric specified in the proof of \Cref{thm:cox_wasserstein_stability}, as Step~1 of that proof gives a uniform time Lipschitz bound for $\overline X_{P_0}(\beta(P_0),t)$. \Cref{ass:cox_w1_times}, applied to $P_0$, gives $P_0(Y\in A,\Delta=1)\leq M|A|$ for measurable $A\subset[0,\tau)$ and excludes an event atom at $\tau$, as in \eqref{eq:cox_w1_event_time_measure_bound}. In particular, for $y,y'\in[0,\tau]$ and $\|x\|\leq K$, the change in the cumulative integral caused by its upper limit satisfies
$$\left\|
\int_0^{y'}
\frac{x-\overline X_{P_0}(\beta(P_0),t)}
{S_{P_0}^{(0)}(\beta(P_0),t)}
\,\mathbb E_{P_0}\{dN(t)\}
-\int_0^y
\frac{x-\overline X_{P_0}(\beta(P_0),t)}
{S_{P_0}^{(0)}(\beta(P_0),t)}
\,\mathbb E_{P_0}\{dN(t)\}
\right\|\leq\frac{2KM}{a}|y'-y|.$$ 
The mean-value theorem bounds the change in $e^{\beta(P_0)^\top X}$ by $B_0\|\beta(P_0)\|\|X'-X\|$. Changing $X$ inside the cumulative integral contributes at most $a^{-1}\|X'-X\|$, since the event-time measure has total mass at most one. The direct term is also Lipschitz in the record by the time Lipschitz bound for $\overline X_{P_0}$ and the bound $\|X\|\leq K$. Therefore $D\mapsto\xi_{P_0}(D)\xi_{P_0}(D)^\top$ is bounded and Lipschitz under the same ground metric.

Adding and subtracting $\mathbb E_P[\xi_{P_0}(D')\xi_{P_0}(D')^\top]$ and applying \eqref{eq:cox_w1_cov_xi_bound} give, for a fixed
$0<C_V<\infty$,
\begin{equation}
\begin{aligned}
&~~~~~~\left\|
\mathbb E_P[\xi_P(D')\xi_P(D')^\top]
-\mathbb E_{P_0}[\xi_{P_0}(D)\xi_{P_0}(D)^\top]
\right\|\\
&\quad\leq 2B_\xi\mathbb E_P
\|\xi_P(D')-\xi_{P_0}(D')\|+
\left\|\mathbb E_P[\xi_{P_0}(D')\xi_{P_0}(D')^\top]
-\mathbb E_{P_0}[\xi_{P_0}(D)\xi_{P_0}(D)^\top]
\right\|\\
&\quad\leq C_VW_1(P,P_0).
\end{aligned}
\label{eq:cox_w1_cov_second_moment_bound}
\end{equation}
The first error is bounded by evaluating \eqref{eq:cox_w1_cov_xi_bound} at $D'$. The second is bounded by the coupling characterization of $W_1$ for the ground metric specified in the proof of \Cref{thm:cox_wasserstein_stability}, applied to the fixed bounded Lipschitz map $D\mapsto\xi_{P_0}(D)\xi_{P_0}(D)^\top$.

\noindent\textbf{Step 5: Information matrices and sandwich covariance.}\quad
The information bound \eqref{eq:cox_w1_information_bound} and the uniform Lipschitz property of $J_{P_0}(\beta)$ in
$\beta$ give
\[
\begin{aligned}
\|J_P\{\beta(P)\}-J_{P_0}\{\beta(P_0)\}\|&\leq
\|J_P\{\beta(P)\}-J_{P_0}\{\beta(P)\}\|
+\|J_{P_0}\{\beta(P)\}-J_{P_0}\{\beta(P_0)\}\|\\
&\leq C_{J,\beta}W_1(P,P_0)
\end{aligned}
\]
for a fixed $C_{J,\beta}<\infty$. The first term is controlled by evaluating \eqref{eq:cox_w1_information_bound} at $\beta(P)$. The second uses bounded covariates, the denominator lower bound, and the parameter bound in \Cref{thm:cox_wasserstein_stability}. \Cref{ass:cox_information} and
\eqref{eq:cox_w1_information_lower_bound} give $\|J_{P_0}\{\beta(P_0)\}^{-1}\|\leq1/\kappa$ and $\|J_P\{\beta(P)\}^{-1}\|\leq2/\kappa$. The inverse-matrix identity therefore yields
\begin{equation}
\left\|
J_P\{\beta(P)\}^{-1}
-J_{P_0}\{\beta(P_0)\}^{-1}
\right\|
\leq C_{J^{-1}}W_1(P,P_0),
\qquad C_{J^{-1}}=\frac{2C_{J,\beta}}{\kappa^2}.
\label{eq:cox_w1_cov_inverse_bound}
\end{equation}

For the final decomposition only, write
\[
V(P)=\mathbb E_P[\xi_P(D')\xi_P(D')^\top],
\qquad
V(P_0)=\mathbb E_{P_0}[\xi_{P_0}(D)\xi_{P_0}(D)^\top],
\] and abbreviate $J_P\{\beta(P)\}^{-1}$ and $J_{P_0}\{\beta(P_0)\}^{-1}$ by $J_P^{-1}$ and $J_{P_0}^{-1}$. A superscript $-T$ denotes the transpose of the corresponding inverse. Equation~\eqref{eq:cox_if} gives \(\Sigma(P)=J_P^{-1}V(P)J_P^{-T}\), and direct addition and subtraction yields
\begin{equation}
\Sigma(P)-\Sigma(P_0)
={}(J_P^{-1}-J_{P_0}^{-1})V(P)J_P^{-T}
+J_{P_0}^{-1}\{V(P)-V(P_0)\}J_P^{-T}+J_{P_0}^{-1}V(P_0)(J_P^{-T}-J_{P_0}^{-T}).
\label{eq:cox_w1_cov_sandwich_decomp}
\end{equation}
Equation~\eqref{eq:cox_w1_cov_second_moment_bound} gives $\|V(P)-V(P_0)\|\leq C_VW_1(P,P_0)$, while \eqref{eq:cox_w1_cov_inverse_bound} also controls the difference of the transposed inverses. Moreover, $\|V(P)\|,\|V(P_0)\|\leq B_\xi^2$. Taking norms term by term in \eqref{eq:cox_w1_cov_sandwich_decomp} therefore gives
$$\|\Sigma(P)-\Sigma(P_0)\|
\leq\frac{3B_\xi^2}{\kappa}
\|J_P^{-1}-J_{P_0}^{-1}\|
+\frac{2}{\kappa^2}\|V(P)-V(P_0)\|\leq\left(
\frac{3B_\xi^2C_{J^{-1}}}{\kappa}
+\frac{2C_V}{\kappa^2}
\right)W_1(P,P_0).$$
Absorbing the fixed quantity in parentheses into $C_\Sigma$ proves the theorem.
\end{proof}

\subsection{Proof of \Cref{prop:cox_naive_invalid}~(A coefficient-preserved generator with invalid naive inference)}
\label{proof:cox_naive_invalid}

\begin{proof}\quad
Let $\beta_0=\beta(P_0)=b^0$ under the null. We first separate the original and synthetic sampling errors and then compare their combined variance with the naive synthetic standard error.

\noindent\textbf{Step 1: Joint limit of the original and synthetic sampling errors.}\quad
Under Assumptions~\ref{ass:cox_parameter}--\ref{ass:cox_w1_times}, the standard asymptotic linear representation for the original-data Cox estimator gives
\[
A_n
=
\frac{\sqrt n\{\beta(P_n)-\beta_0\}}
{\sqrt{\Sigma(P_0)}}
\xrightarrow{d}N(0,1).
\]
Conditional on the original sample and the fitted distribution $\widehat P_n$, the synthetic records are i.i.d.\ from $\widehat P_n$. The corresponding conditional Cox expansion gives
\[
B_{n,m}
=
\frac{\sqrt m\{\beta(\widehat P_{n,m})-\beta(\widehat P_n)\}} {\sqrt{\Sigma(P_0)}}
\xrightarrow{d}N(0,1)
\]
conditionally in probability. \Cref{thm:cox_covariance_stability} and $W_1(\widehat P_n,P_0)=O_p(n^{-1/2})=o_p(1)$ give $\Sigma(\widehat P_n)\xrightarrow{p}\Sigma(P_0)$, while consistency of the synthetic sandwich covariance estimator gives $\Sigma(\widehat P_{n,m})\xrightarrow{p}\Sigma(P_0)$.

Because the conditional limiting distribution of $B_{n,m}$ is the fixed distribution $N(0,1)$, a conditional characteristic-function argument gives
\[
(A_n,B_{n,m})\xrightarrow{d}(Z_1,Z_2),
\]
where $Z_1$ and $Z_2$ are independent standard normal random variables. Thus, the synthetic sampling error is asymptotically independent of the original-sample error inherited through the fitted generator.

\noindent\textbf{Step 2: Limiting distribution of the naive synthetic statistic.}\quad
The exact coefficient-preservation condition
$\beta(\widehat P_n)=\beta(P_n)$ gives
$
\beta(\widehat P_{n,m})-\beta_0
=
\{\beta(\widehat P_{n,m})-\beta(\widehat P_n)\}
+\{\beta(P_n)-\beta_0\}.$
Therefore,
$\frac{\sqrt m\{\beta(\widehat P_{n,m})-\beta_0\}}
{\sqrt{\Sigma(P_0)}}
=
B_{n,m}+\sqrt{\frac mn}\,A_n.$
Since $m/n\to r\in(0,\infty)$, the joint convergence in Step~1 gives
$B_{n,m}+\sqrt{\frac mn}\,A_n
\xrightarrow{d}
Z_2+\sqrt r\,Z_1
\sim N(0,1+r).$

The naive statistic can be written as
\[
Z_{\mathrm{syn}}
=
\frac{B_{n,m}+\sqrt{m/n}\,A_n}
{\{\Sigma(\widehat P_{n,m})/\Sigma(P_0)\}^{1/2}}.
\]
Since $\Sigma(\widehat P_{n,m})/\Sigma(P_0)\xrightarrow{p}1$, Slutsky's theorem yields $Z_{\mathrm{syn}}\xrightarrow{d}N(0,1+r).$
The additional variance $r$ is the original-sample uncertainty inherited through $\beta(\widehat P_n)=\beta(P_n)$, which is omitted from the naive synthetic standard error.

\noindent\textbf{Step 3: Limiting rejection probability.}\quad
The two-sided naive test rejects when $|Z_{\mathrm{syn}}|>z_{1-\alpha/2}$. Therefore,
\[
\Pr\{|Z_{\mathrm{syn}}|>z_{1-\alpha/2}\}
\longrightarrow
2\left[
1-\Phi\left\{ \frac{z_{1-\alpha/2}}{\sqrt{1+r}} \right\}
\right].
\]
Since $r>0$,
$z_{1-\alpha/2}/\sqrt{1+r}<z_{1-\alpha/2}$. Hence the limiting rejection probability is strictly greater than $2\{1-\Phi(z_{1-\alpha/2})\}=\alpha$. This proves the proposition.
\end{proof}

\subsection{Proof of \Cref{prop:cox_oracle}~(Valid \(p\)-values need not reproduce the original \(p\)-value)}

\begin{proof}\quad
We first establish the joint limit of the original and oracle synthetic \(p\)-values and then calculate the probability that their testing decisions disagree.

\noindent\textbf{Step 1: Joint limit of the original and oracle synthetic \(p\)-values.}\quad
Under the null hypothesis and the standard Cox regularity conditions, the original-data Wald statistic satisfies
$Z_{\mathrm{real}}
\xrightarrow{d}
Z_1,
Z_1\sim N(0,1).$
Because the oracle synthetic records form an independent sample from \(P_0\), the synthetic-data Wald statistic similarly satisfies
$Z_{\mathrm{syn}}
\xrightarrow{d}
Z_2,
Z_2\sim N(0,1).$
The original and oracle synthetic samples are independent. Therefore, the two convergences hold jointly:
\[
(Z_{\mathrm{real}},Z_{\mathrm{syn}})
\xrightarrow{d}
(Z_1,Z_2),
\]
where \(Z_1\) and \(Z_2\) are independent standard normal random variables.

Recall that \(p(z)=2\{1-\Phi(|z|)\}\). Then
$\Pr\{p(Z_1)\leq u\}
=
\Pr\left\{ |Z_1|\geq\Phi^{-1}(1-u/2) \right\}
=
u$ for every \(u\in[0,1]\).
Hence \(p(Z_1)\sim\mathrm{Uniform}(0,1)\), and the same conclusion holds for \(p(Z_2)\). Since \(p(\cdot)\) is continuous, the continuous mapping theorem gives
\[
(p_{\mathrm{real}},p_{\mathrm{syn}})
\xrightarrow{d}
(U_1,U_2),
\]
where \(U_1\) and \(U_2\) are independent \(\mathrm{Uniform}(0,1)\) random variables.

\noindent\textbf{Step 2: Limiting probability of disagreement.}\quad 
The original and synthetic tests disagree when exactly one of their \(p\)-values is no greater than \(\alpha\). Since \(\Pr(U_1=\alpha)=\Pr(U_2=\alpha)=0\), the joint convergence in Step~1 gives
\[
\begin{split}
&\Pr\bigl(
\{p_{\mathrm{real}}\leq\alpha,\,
p_{\mathrm{syn}}>\alpha\}
\cup
\{p_{\mathrm{real}}>\alpha,\,
p_{\mathrm{syn}}\leq\alpha\}
\bigr)\longrightarrow
\Pr(U_1\leq\alpha,U_2>\alpha)
+
\Pr(U_1>\alpha,U_2\leq\alpha).
\end{split}
\]
The independence of \(U_1\) and \(U_2\) gives
\[
\Pr(U_1\leq\alpha,U_2>\alpha)
+
\Pr(U_1>\alpha,U_2\leq\alpha)
=
\alpha(1-\alpha)+(1-\alpha)\alpha
=
2\alpha(1-\alpha).
\]
Because \(\alpha\in(0,1)\), this probability is strictly positive. Thus, although both \(p\)-values are asymptotically valid, their testing decisions need not agree.
\end{proof}

\subsection{Proof of \Cref{thm:cox_pvalue_preservation}~(Consistency and rates for corrected Cox inference)}
\label{proof:cox_pvalue_preservation}

\begin{proof}\quad
\noindent\textbf{Step 1: Empirical Cox expansions.}\quad
We first record the expansions for the original-sample uncertainty $\beta(P_n)-\beta(P_0)$ and the synthetic sampling error $\beta(\widehat P_{n,m})-\beta(\widehat P_n)$. 

First recall that $\widetilde D_1,\ldots,\widetilde D_m \mid (\mathcal D_n,\widehat P_n)
\overset{\mathrm{i.i.d.}}{\sim}\widehat P_n$.
The bounded weighted threshold classes give
\[
\max_{j=0,1,2}\sup_{t\in[0,\tau]}
\left\|
S_{\widehat P_{n,m}}^{(j)}\{\beta(\widehat P_n),t\}
-
S_{\widehat P_n}^{(j)}\{\beta(\widehat P_n),t\}
\right\|=O_p(m^{-1/2}),
\]
with constants independent of the fitted law. Suppressing the
arguments $(\beta(\widehat P_n),t)$, we have
\[
\overline X_{\widehat P_{n,m}}-\overline X_{\widehat P_n}=\frac{S_{\widehat P_{n,m}}^{(1)}
-\overline X_{\widehat P_n} S_{\widehat P_{n,m}}^{(0)}}{S_{\widehat P_n}^{(0)}}+
\underbrace{\frac{\bigl(S_{\widehat P_{n,m}}^{(1)}
-\overline X_{\widehat P_n} S_{\widehat P_{n,m}}^{(0)}\bigr)\bigl(S_{\widehat P_n}^{(0)}-S_{\widehat P_{n,m}}^{(0)}\bigr)
}{S_{\widehat P_n}^{(0)} S_{\widehat P_{n,m}}^{(0)}}
}_{O_p(m^{-1})\ \text{uniformly in }t}.
\]
Here both numerator factors in the remainder are uniformly
$O_p(m^{-1/2})$, and \Cref{ass:cox_risksets} bounds both denominators
away from zero with probability tending to one. The linear term
integrated against the event-measure difference satisfies
\[
\int_0^\tau
\frac{S_{\widehat P_{n,m}}^{(1)}
-\overline X_{\widehat P_n}S_{\widehat P_{n,m}}^{(0)}}
{S_{\widehat P_n}^{(0)}}
\bigl(\mathbb E_{\widehat P_{n,m}}
      -\mathbb E_{\widehat P_n}\bigr)\{dN(t)\}
=O_p(m^{-1}),
\]
since it is a bounded second-order $V$-statistic with a kernel
centered in both arguments. Substituting the ratio expansion into
the score gives
\[
U_{\widehat P_{n,m}}\{\beta(\widehat P_n)\}
=
\frac1m\sum_{\ell=1}^m
\xi_{\widehat P_n}(\widetilde D_\ell)+O_p(m^{-1}).
\]
The same empirical process bounds give
$J_{\widehat P_{n,m}}(\beta)-J_{\widehat P_n}(\beta)=o_p(1)$
uniformly on the local parameter ball. By
\eqref{eq:cox_w1_information_lower_bound}, the empirical information
is therefore bounded away from singularity there with probability
tending to one. Taylor expansion of the empirical score gives
\[
\beta(\widehat P_{n,m})-\beta(\widehat P_n)
=
J_{\widehat P_n}\{\beta(\widehat P_n)\}^{-1}
U_{\widehat P_{n,m}}\{\beta(\widehat P_n)\}
+o_p(m^{-1/2}).
\]
Combining these identities and applying the same argument to the
original sample gives

\begin{equation}
\beta(P_n)-\beta(P_0)
=
n^{-1}\sum_{i=1}^n\psi_{P_0}(D_i)
+o_p(n^{-1/2}),
\beta(\widehat P_{n,m})-\beta(\widehat P_n)
=
m^{-1}\sum_{\ell=1}^m
\psi_{\widehat P_n}(\widetilde D_\ell)
+o_p(m^{-1/2}).
\label{eq:cox_inference_empirical_expansions}
\end{equation}
Under the same conditions, the same argument as above shows that the corresponding plug-in sandwich covariance estimators satisfy
\begin{equation}
\Sigma(P_n)-\Sigma(P_0)
=O_p(n^{-1/2}),\quad
\Sigma(\widehat P_{n,m})-\Sigma(\widehat P_n)
=O_p(m^{-1/2}),
\label{eq:cox_inference_covariance_rates}
\end{equation}
where the second relation is conditional on the original sample and the fitted generator.

\noindent\textbf{Step 2: From score calibration to coefficient calibration.}\quad
The score conditions in \Cref{ass:cox_weak_calibration} control the score of $\widehat P_n$ at the original-data estimate $\beta(P_n)$, whereas the theorem requires control of the difference between the two Cox parameters. Applying \Cref{thm:cox_wasserstein_stability} with $P=\widehat P_n$ and using $W_1(\widehat P_n,P_0)=o_p(1)$ gives $\|\beta(\widehat P_n)-\beta(P_0)\|=o_p(1)$. The first expansion in \eqref{eq:cox_inference_empirical_expansions} also gives $\|\beta(P_n)-\beta(P_0)\|=O_p(n^{-1/2})=o_p(1)$. Therefore,
uniformly over $t\in[0,1]$,
$$\|(1-t)\beta(P_n)+t\beta(\widehat P_n)-\beta(P_0)\|
\leq
(1-t)\|\beta(P_n)-\beta(P_0)\|
+t\|\beta(\widehat P_n)-\beta(P_0)\|
=o_p(1).$$
Since the radius of $\mathcal B_r$ is fixed, \eqref{eq:cox_w1_information_lower_bound} applies along this line segment. The fundamental theorem of calculus and $J_{\widehat P_n}=-\nabla_\beta U_{\widehat P_n}$ give
$$U_{\widehat P_n}\{\beta(P_n)\}
=
\left[ \int_0^1 J_{\widehat P_n}\{(1-t)\beta(P_n)+t\beta(\widehat P_n)\} \,dt \right]
\times\{\beta(\widehat P_n)-\beta(P_n)\},$$
because $U_{\widehat P_n}\{\beta(\widehat P_n)\}=0$.

To obtain a bound for the coefficient difference, multiply both sides by $\{\beta(\widehat P_n)-\beta(P_n)\}^\top$. The information lower bound \eqref{eq:cox_w1_information_lower_bound} gives
\[
\begin{aligned}
&~~~~\{\beta(\widehat P_n)-\beta(P_n)\}^\top
U_{\widehat P_n}\{\beta(P_n)\}\\
&=
\int_0^1
\{\beta(\widehat P_n)-\beta(P_n)\}^\top
J_{\widehat P_n}\{(1-t)\beta(P_n)+t\beta(\widehat P_n)\}
\{\beta(\widehat P_n)-\beta(P_n)\}\,dt\\
&\geq
\frac{\kappa}{2}
\|\beta(\widehat P_n)-\beta(P_n)\|^2.
\end{aligned}
\]
On the other hand, the Cauchy--Schwarz inequality gives
\[
\{\beta(\widehat P_n)-\beta(P_n)\}^\top U_{\widehat P_n}\{\beta(P_n)\}
\leq
\|\beta(\widehat P_n)-\beta(P_n)\| \left\|U_{\widehat P_n}\{\beta(P_n)\}\right\|.
\]
Combining these inequalities and dividing by $\|\beta(\widehat P_n)-\beta(P_n)\|$ when $\beta(\widehat P_n)\neq\beta(P_n)$ yields
\begin{equation}
\|\beta(\widehat P_n)-\beta(P_n)\|
\leq
\frac{2}{\kappa}
\left\|U_{\widehat P_n}\{\beta(P_n)\}\right\|
=
\frac{2}{\kappa}c_n.
\label{eq:cox_inference_residual_to_parameter}
\end{equation}
The inequality is immediate when $\beta(\widehat P_n)=\beta(P_n)$. Thus, the score residual and the corresponding coefficient discrepancy have the same stochastic order.

\noindent\textbf{Step 3: Corrected asymptotic normality.}\quad
We now combine the two empirical expansions~\eqref{eq:cox_inference_empirical_expansions} with the coefficient bound established in Step~2. Under \Cref{ass:cox_weak_calibration}, \eqref{eq:cox_inference_residual_to_parameter} gives $\|\beta(\widehat P_n)-\beta(P_n)\|=o_p(s_{n,m})$. Combining this bound with the two expansions in \eqref{eq:cox_inference_empirical_expansions} gives
\begin{equation}
\begin{aligned}
\beta(\widehat P_{n,m})-\beta(P_0)
={}&
n^{-1}\sum_{i=1}^n\psi_{P_0}(D_i)
+m^{-1}\sum_{\ell=1}^m
\psi_{\widehat P_n}(\widetilde D_\ell)
+o_p(s_{n,m}).
\end{aligned}
\label{eq:cox_inference_two_sample_expansion}
\end{equation}

The first leading sum in
\eqref{eq:cox_inference_two_sample_expansion} has covariance $\Sigma(P_0)/n$. By \Cref{thm:cox_covariance_stability} and $W_1(\widehat P_n,P_0)=o_p(1)$,
$\|\Sigma(\widehat P_n)-\Sigma(P_0)\|=o_p(1).$ Conditional on the original sample and fitted generator, the second leading sum has mean zero and covariance $\Sigma(\widehat P_n)/m$. Its conditional characteristic function therefore converges in probability to the corresponding nonrandom normal characteristic function. Hence the two leading sums are asymptotically independent, and their covariances add to $\left(n^{-1}+m^{-1}\right)\Sigma(P_0).$

Equations~\eqref{eq:cox_inference_covariance_rates} and \Cref{thm:cox_covariance_stability} also give $\Sigma(\widehat P_{n,m})\xrightarrow{p}\Sigma(P_0)$. Since $\Sigma_{jj}(P_0)>0$, the corrected variance estimator in \eqref{eq:cox_corrected_z} satisfies $ \frac{\widehat s_{\mathrm{corrected},j}^2} {(n^{-1}+m^{-1})\Sigma_{jj}(P_0)} \xrightarrow{p}1$ by continuous mapping theorem. Slutsky's theorem therefore gives
\[
\frac{\beta_j(\widehat P_{n,m})-\beta_j(P_0)}
{\widehat s_{\mathrm{corrected},j}}
\xrightarrow{d}N(0,1).
\]
for every relative growth rate of $m$ and $n$. Under $H_{0,j}$, the continuous mapping theorem then gives $p_{\mathrm{corrected},j}\xrightarrow{d}\operatorname{Unif}(0,1)$. Since the distribution function of $\operatorname{Unif}(0,1)$ is continuous, this convergence is equivalent to
\[
\sup_{u\in[0,1]}
\left| \Pr(p_{\mathrm{corrected},j}\leq u)-u \right|
\longrightarrow0.
\]

\noindent\textbf{Step 4: Agreement with the original analysis.}\quad
We next impose $m/n\to\infty$ and show that the corrected synthetic analysis reproduces the original analysis. In this case $s_{n,m}=n^{-1/2}(1+n/m)^{1/2}=n^{-1/2}\{1+o(1)\}$, so \eqref{eq:cox_inference_residual_to_parameter} and \Cref{ass:cox_weak_calibration} give $\|\beta(\widehat P_n)-\beta(P_n)\|=o_p(n^{-1/2})$. The conditional synthetic sampling error in \eqref{eq:cox_inference_empirical_expansions} is $O_p(m^{-1/2})=o_p(n^{-1/2})$. Consequently,
\begin{equation}
\beta(\widehat P_{n,m})-\beta(P_n)
=
o_p(n^{-1/2}).
\label{eq:cox_inference_coefficient_agreement}
\end{equation}

Let $\widehat s_{\mathrm{real},j} =\{\Sigma_{jj}(P_n)/n\}^{1/2}$. Covariance consistency, $m/n\to\infty$, and $\Sigma_{jj}(P_0)>0$ give $\frac{\widehat s_{\mathrm{real},j}} {\widehat s_{\mathrm{corrected},j}} \xrightarrow{p}1$ by continuous mapping theorem. Moreover, $\widehat s_{\mathrm{corrected},j} =n^{-1/2}\{\Sigma_{jj}(P_0)\}^{1/2}\{1+o_p(1)\}$.
The difference between the corrected and original Wald statistics can be written exactly as

\begin{equation}
\begin{aligned}
Z_{\mathrm{corrected},j}-Z_{\mathrm{real},j}
&=
\frac{\beta_j(\widehat P_{n,m})-b_j^0}
{\widehat s_{\mathrm{corrected},j}}
-
\frac{\beta_j(P_n)-b_j^0}
{\widehat s_{\mathrm{real},j}}.\\
&= \frac{\beta_j(\widehat P_{n,m})-b_j^0}
{\widehat s_{\mathrm{corrected},j}}
-
\frac{\beta_j(P_n)-b_j^0}
{\widehat s_{\mathrm{corrected},j}}+
\frac{\beta_j(P_n)-b_j^0}
{\widehat s_{\mathrm{corrected},j}}
-
\frac{\beta_j(P_n)-b_j^0}
{\widehat s_{\mathrm{real},j}}.\\
&=
\frac{\beta_j(\widehat P_{n,m})-\beta_j(P_n)}
{\widehat s_{\mathrm{corrected},j}}+
Z_{\mathrm{real},j}
\left(
\frac{\widehat s_{\mathrm{real},j}}
{\widehat s_{\mathrm{corrected},j}}-1
\right).
\end{aligned}
\label{eq:cox_inference_wald_decomposition}
\end{equation}
The first term in \eqref{eq:cox_inference_wald_decomposition} is $o_p(1)$ by \eqref{eq:cox_inference_coefficient_agreement} and the preceding expansion of $\widehat s_{\mathrm{corrected},j}$.
The second is also $o_p(1)$ because $Z_{\mathrm{real},j}=O_p(1)$ under the null and the standard-error ratio converges to one. Hence
\[
Z_{\mathrm{corrected},j}-Z_{\mathrm{real},j}=o_p(1).
\]

The map $p(z)=2\{1-\Phi(|z|)\}$ satisfies $|p(z)-p(z')| \leq \sqrt{\frac{2}{\pi}}\,|z-z'|,$ and therefore $p_{\mathrm{corrected},j}-p_{\mathrm{real},j}=o_p(1)$. For any fixed $\eta>0$, 
$$\Pr\left\{\mathbf 1(p_{\mathrm{corrected},j}\leq\alpha) \neq \mathbf 1(p_{\mathrm{real},j}\leq\alpha) \right\} \le   \Pr\{ |p_{\mathrm{corrected},j}-p_{\mathrm{real},j}|>\eta \} + \Pr\{ |p_{\mathrm{real},j}-\alpha|\leq\eta \}.$$ 
Because if neither event on the right-hand side occurs, $|p_{\mathrm{corrected},j}-p_{\mathrm{real},j}|\leq\eta$ and $|p_{\mathrm{real},j}-\alpha|>\eta$. If $p_{\mathrm{real},j}<\alpha-\eta$, then $p_{\mathrm{corrected},j}<\alpha$. If $p_{\mathrm{real},j}>\alpha+\eta$, then $p_{\mathrm{corrected},j}>\alpha$. Thus, the two decisions agree whenever neither event occurs.

The first probability converges to zero because $p_{\mathrm{corrected},j}-p_{\mathrm{real},j}=o_p(1)$. Under the null, $p_{\mathrm{real},j}\xrightarrow{d}\operatorname{Unif}(0,1)$, so
\[
\limsup_{n\to\infty}
\Pr\{ |p_{\mathrm{real},j}-\alpha|\leq\eta \} \leq 2\eta.
\]
It follows that
\[
\limsup_{n\to\infty}
\Pr\left\{ \mathbf 1(p_{\mathrm{corrected},j}\leq\alpha) \neq \mathbf 1(p_{\mathrm{real},j}\leq\alpha) \right\} 
\leq 2\eta.
\]
Letting $\eta\downarrow0$ proves the decision-agreement conclusion. The first probability converges to zero for every $\eta>0$. Under the null, $p_{\mathrm{real},j}\xrightarrow{d}\mathrm{Uniform}(0,1)$, so the limit superior of the second probability is at most $2\eta$. Taking the limit superior and then letting $\eta\downarrow0$ proves the decision-agreement conclusion.

\noindent\textbf{Step 5: Quantitative rates.}\quad
It remains to track the preceding errors under \Cref{ass:cox_weak_calibration}. First, \eqref{eq:cox_inference_residual_to_parameter} gives $\|\beta(\widehat P_n)-\beta(P_n)\|=O_p(\epsilon_n)$. Combining this result with the conditional synthetic expansion in \eqref{eq:cox_inference_empirical_expansions} yields
\begin{equation}
\|\beta(\widehat P_{n,m})-\beta(P_n)\|
=
O_p(\epsilon_n+m^{-1/2}).
\label{eq:cox_inference_coefficient_rate}
\end{equation}

Next, \eqref{eq:cox_inference_covariance_rates}, \Cref{thm:cox_covariance_stability}, and the generative error bound $W_1(\widehat P_n,P_0)=O_p(r_n)$ give
\begin{equation}
\|\Sigma(\widehat P_{n,m})-\Sigma(P_n)\|
=
O_p(r_n+n^{-1/2}+m^{-1/2}).
\label{eq:cox_inference_covariance_difference_rate}
\end{equation}
Both tested covariance components converge to the positive limit $\Sigma_{jj}(P_0)$. Using \eqref{eq:cox_inference_covariance_difference_rate} and the definitions of the two standard errors gives
\begin{align}
\left|
\Sigma_{jj}(\widehat P_{n,m})-\Sigma_{jj}(P_n)
\right|
&\leq
\left\|
\Sigma(\widehat P_{n,m})-\Sigma(P_n)
\right\|\nonumber=
O_p\left(r_n+n^{-1/2}+m^{-1/2}\right),\nonumber\\
\frac{\widehat s_{\mathrm{real},j}}
{\widehat s_{\mathrm{corrected},j}}-1
&=
\left\{
\frac{\Sigma_{jj}(P_n)}
{(1+n/m)\Sigma_{jj}(\widehat P_{n,m})}
\right\}^{1/2}-1\nonumber\\
&=
\frac{
\sqrt{\Sigma_{jj}(P_n)}
-\sqrt{(1+n/m)\Sigma_{jj}(\widehat P_{n,m})}
}{
\sqrt{(1+n/m)\Sigma_{jj}(\widehat P_{n,m})}
}\nonumber\\
&=
\frac{
\Sigma_{jj}(P_n)
-(1+n/m)\Sigma_{jj}(\widehat P_{n,m})
}{
\sqrt{(1+n/m)\Sigma_{jj}(\widehat P_{n,m})}
\left\{
\sqrt{\Sigma_{jj}(P_n)}
+\sqrt{(1+n/m)\Sigma_{jj}(\widehat P_{n,m})}
\right\}
}\nonumber\\
&=
\frac{
\Sigma_{jj}(P_n)-\Sigma_{jj}(\widehat P_{n,m})
-(n/m)\Sigma_{jj}(\widehat P_{n,m})
}{
2\Sigma_{jj}(P_0)+o_p(1)
}\nonumber\\
&=
O_p\left(
r_n+n^{-1/2}+m^{-1/2}+n/m
\right).
\label{eq:cox_inference_se_ratio_rate}
\end{align}
The penultimate equality follows because $\Sigma_{jj}(P_n)\xrightarrow{p}\Sigma_{jj}(P_0)$, $\Sigma_{jj}(\widehat P_{n,m})\xrightarrow{p}\Sigma_{jj}(P_0)$, and $n/m\to0$. Hence the denominator equals $2\Sigma_{jj}(P_0)+o_p(1)$, while $(n/m)\Sigma_{jj}(\widehat P_{n,m})=O_p(n/m)$.

Under $m/n\to\infty$, $\widehat s_{\mathrm{corrected},j} =n^{-1/2}\{\Sigma_{jj}(P_0)\}^{1/2}\{1+o_p(1)\}$ and $Z_{\mathrm{real},j}=O_p(1)$ under the null. Substituting \eqref{eq:cox_inference_coefficient_rate} and \eqref{eq:cox_inference_se_ratio_rate} into \eqref{eq:cox_inference_wald_decomposition}, and using $m^{-1/2}=O\{\sqrt{n/m}\}$ and $n/m=O\{\sqrt{n/m}\}$, gives
\[
Z_{\mathrm{corrected},j}-Z_{\mathrm{real},j}
=
O_p\left( \sqrt n\,\epsilon_n+r_n+n^{-1/2} +\sqrt{n/m} \right).
\]
The Lipschitz bound for the two-sided normal $p$-value gives the same rate:
\[
p_{\mathrm{corrected},j}-p_{\mathrm{real},j}
=
O_p\left( \sqrt n\,\epsilon_n+r_n+n^{-1/2} +\sqrt{n/m} \right).
\]
Finally, if $c_n=O_p(r_n/\sqrt n)$, $n^{-1/2}=O(r_n)$, and $\sqrt{n/m}=O(r_n)$, then $\sqrt n\,\epsilon_n=O_p(r_n)$ and every remaining term in the last two displays is $O_p(r_n)$. This proves the final assertion.
\end{proof}

\subsection{Proof of \Cref{thm:cox_pvalue_preservation_cali}~ (Consistency and rates for corrected Cox inference under distribution calibration)}
\label{proof:cox_pvalue_preservation_cali}

\begin{proof}\quad
\noindent\textbf{Step 1: Verification of the conditions in
\Cref{thm:cox_pvalue_preservation}.}\quad
Apply \Cref{thm:cox_pvalue_preservation} with the fitted generative distribution $\widehat P_n$ replaced by the calibrated distribution $\overline P_n$. Since $W_1(\overline P_n,P_0)=O_p(r_n)$ and $r_n\to0$, we have $W_1(\overline P_n,P_0)=o_p(1)$. The assumed score residual satisfies
\[ \left\| U_{\overline P_n}\{\beta(P_n)\} \right\|
=
O_p\left(\frac{r_n}{\sqrt n}\right).
\]
Thus, in part (ii) of \Cref{thm:cox_pvalue_preservation}, we may take $\epsilon_n=r_n/\sqrt n$. Since $r_n\to0$, $\epsilon_n=o(n^{-1/2})$, as required, and $\sqrt n\,\epsilon_n=r_n$. Hence, the quantitative calibration condition in part (ii) of \Cref{thm:cox_pvalue_preservation} is satisfied with $\epsilon_n=r_n/\sqrt n$.

\noindent\textbf{Step 2: Corrected statistic and $p$-value rates.}\quad 
Under the null and $m/n\to\infty$, substituting $\epsilon_n=r_n/\sqrt n$ and $W_1(\overline P_n,P_0)=O_p(r_n)$ into part (ii) of \Cref{thm:cox_pvalue_preservation} gives
\begin{align*}
&Z_{\mathrm{corrected},j}-Z_{\mathrm{real},j}
=
O_p\left(
\sqrt n\,\frac{r_n}{\sqrt n}
+r_n+n^{-1/2}
+\sqrt{n/m}
\right)=
O_p\left(
r_n+n^{-1/2}
+\sqrt{n/m}
\right),\\
&p_{\mathrm{corrected},j}-p_{\mathrm{real},j}
=
O_p\left(
\sqrt n\,\frac{r_n}{\sqrt n}
+r_n+n^{-1/2}
+\sqrt{n/m}
\right)=
O_p\left(
r_n+n^{-1/2}
+\sqrt{n/m}
\right).
\end{align*}
This proves both stated rates.
\end{proof}

\subsection{Proof of \Cref{thm:cox_calibration}~(Local existence and stability of Cox calibration)}
\label{proof:cox_calibration}

\begin{proof}\quad
Here, $\widetilde D=(\widetilde X,\widetilde Y,\widetilde\Delta)\sim P$ denotes a generic observed-data record from the fixed baseline distribution $P$. The function $\xi_P(\widetilde D)$ is evaluated at $\beta(P)$ using the risk-set quantities under $P$ and is kept fixed as $\lambda$ varies. We first derive the calibration Jacobian and use the implicit-function theorem to construct and control the local tilt parameter. We then verify that the target score root is the unique Cox parameter under the tilted distribution. Finally, we translate the tilt-parameter bound into distributional bounds and verify that the tilt preserves the required regularity properties.

\noindent\textbf{Step 1: Derivative of the score function along the exponential tilt.}\quad
To construct a tilt attaining a target coefficient near $\beta(P)$, we first derive the calibration Jacobian required by the implicit-function theorem and then use it to control the resulting calibration parameter. Choose $\varepsilon>0$ sufficiently small that \Cref{thm:cox_wasserstein_stability,thm:cox_covariance_stability} and \eqref{eq:cox_w1_information_lower_bound} apply whenever $W_1(P,P_0)\leq\varepsilon$.

For a fixed $P$, define $F_P(\lambda,q):=U_{P_\lambda}(q)$. Since $P_{\lambda=0}=P$, $F_P\{0,\beta(P)\}=U_P\{\beta(P)\}=0$. To apply the implicit-function theorem at this root, we next identify the derivative of $F_P$ with respect to $\lambda$. First, Equation~\eqref{eq:cox_xi} gives
$$\mathbb E_P\{\xi_P(\widetilde D)\} =U_P\{\beta(P)\}
-\int_0^\tau \frac{ S_P^{(1)}\{\beta(P),t\} -\overline X_P\{\beta(P),t\} S_P^{(0)}\{\beta(P),t\} }{ S_P^{(0)}\{\beta(P),t\} } \mathbb E_P\{dN(t)\} =0.$$
The first term is zero because $\beta(P)$ solves the population score equation, and the integrand in the second term is zero by the definition of $\overline X_P$.

We now verify directly that $\xi_P$ gives the derivative of the full score function along the exponential tilt. For $j=1,\ldots,p$, the centering established above gives
\[
\left. \frac{\partial}{\partial\lambda_j} \frac{dP_\lambda}{dP}(\widetilde D) \right|_{\lambda=0}
= \xi_{P,j}(\widetilde D) -\mathbb E_P\{\xi_{P,j}(\widetilde D)\} =\xi_{P,j}(\widetilde D).
\]
Consequently, for $k=0,1$,

\[
\left. \frac{\partial}{\partial\lambda_j} S_{P_\lambda}^{(k)}(\beta,t) \right|_{\lambda=0}
=
\mathbb E_P\left[ \left. \frac{\partial}{\partial\lambda_j} \frac{dP_\lambda}{dP}(\widetilde D) \right|_{\lambda=0} e^{\beta^\top\widetilde X} \widetilde X^{\otimes k}\widetilde R(t) \right]=
\mathbb E_P\left[ \xi_{P,j}(\widetilde D) e^{\beta^\top\widetilde X} \widetilde X^{\otimes k}\widetilde R(t) \right].
\]
Applying the quotient rule to $\overline X_{P_\lambda}=S_{P_\lambda}^{(1)}/S_{P_\lambda}^{(0)}$ therefore yields
\[
\left. \frac{\partial}{\partial\lambda_j} \overline X_{P_\lambda}\{\beta(P),t\} \right|_{\lambda=0}
=
\frac{ \mathbb E_P\left[ \xi_{P,j}(\widetilde D) e^{\beta(P)^\top\widetilde X}\widetilde R(t) \{\widetilde X-\overline X_P(\beta(P),t)\} \right] }
{S_P^{(0)}\{\beta(P),t\}}.
\]
Differentiating both the outer expectation and the risk-set quantity in the full score function now gives
\[
\begin{aligned}
&~~~~\left.
\frac{\partial}{\partial\lambda_j}
U_{P_\lambda}\{\beta(P)\}
\right|_{\lambda=0}\\
&=
\mathbb E_P\left[
\xi_{P,j}(\widetilde D)
\int_0^\tau
\{\widetilde X-\overline X_P(\beta(P),t)\}
\,d\widetilde N(t)
\right]-
\int_0^\tau
\left.
\frac{\partial}{\partial\lambda_j}
\overline X_{P_\lambda}\{\beta(P),t\}
\right|_{\lambda=0}
\mathbb E_P\{dN(t)\}\\
&=
\mathbb E_P\left[
\xi_P(\widetilde D)\xi_{P,j}(\widetilde D)
\right],
\end{aligned}
\]
where the last equality follows by substituting the preceding derivative of $\overline X_{P_\lambda}$ and using Equation~\eqref{eq:cox_xi}. Combining these $p$ coordinatewise derivatives gives
\[
\left. \frac{\partial F_P(\lambda,q)}{\partial\lambda^\top} \right|_{\lambda=0,\,q=\beta(P)}
=
\mathbb E_P\left[ \xi_P(\widetilde D)\xi_P(\widetilde D)^\top \right].
\]
This identity shows that the calibration Jacobian incorporates both the direct change in the outer expectation and the induced change in the risk-set quantities. It also connects the Jacobian to the Cox sandwich covariance.

\noindent\textbf{Step 2: Local existence and bound of the calibration parameter.}\quad
By Equation~\eqref{eq:cox_if}, $\xi_P(\widetilde D)=J_P\{\beta(P)\}\psi_P(\widetilde D)$, and hence
\[
\mathbb E_P\left[ \xi_P(\widetilde D)\xi_P(\widetilde D)^\top \right]
= J_P\{\beta(P)\}\Sigma(P)J_P\{\beta(P)\}^\top.
\]

Since $\Sigma(P_0)$ is positive definite, $\lambda_{\min}\{\Sigma(P_0)\}>0$. Then for any unit vector $u,$
$$u^\top\Sigma(P)u= u^\top\Sigma(P_0)u + u^\top\{\Sigma(P)-\Sigma(P_0)\}u \geq \lambda_{\min}\{\Sigma(P_0)\} - \|\Sigma(P)-\Sigma(P_0)\|.$$
After reducing $\varepsilon$ if necessary, \Cref{thm:cox_covariance_stability} gives
$$\lambda_{\min}\{\Sigma(P)\}
\geq
\lambda_{\min}\{\Sigma(P_0)\} -\|\Sigma(P)-\Sigma(P_0)\|
\geq \frac{1}{2}\lambda_{\min}\{\Sigma(P_0)\}$$ 
whenever $W_1(P,P_0)\leq\varepsilon$. Moreover, \eqref{eq:cox_w1_information_lower_bound} gives $\lambda_{\min}[J_P\{\beta(P)\}]\geq\kappa/2$. Therefore, for every unit vector $u$, the preceding Jacobian identity gives
\[
\begin{aligned}
u^\top
\left.
\frac{\partial F_P(\lambda,q)}{\partial\lambda^\top}
\right|_{\lambda=0,\,q=\beta(P)}u
&\quad=
\left[J_P\{\beta(P)\}^\top u\right]^\top
\Sigma(P)
\left[J_P\{\beta(P)\}^\top u\right]\\
&\quad\geq
\lambda_{\min}\{\Sigma(P)\}
\left\|J_P\{\beta(P)\}^\top u\right\|^2\\
&\quad\geq
\frac{\kappa^2}{8}\lambda_{\min}\{\Sigma(P_0)\}.
\end{aligned}
\]
Thus, the calibration Jacobian is uniformly positive definite over the distributions under consideration. Writing $c_{\mathrm{Jac}} =\kappa^2\lambda_{\min}\{\Sigma(P_0)\}/8$, the assumed uniform continuity of the first derivatives permits fixed neighborhoods in which
\[
\sigma_{\min}\left\{ \frac{\partial F_P(\lambda,q)}{\partial\lambda^\top} \right\} \geq\frac{c_{\mathrm{Jac}}}{2},
\qquad
\left\| \left\{ \frac{\partial F_P(\lambda,q)}{\partial\lambda^\top} \right\}^{-1} \right\| \leq\frac{2}{c_{\mathrm{Jac}}}
\]

The root condition at $\{0,\beta(P)\}$ and this inverse-Jacobian bound verify the two conditions needed to solve the calibration equation locally for $\lambda$. The implicit-function theorem therefore gives a unique local function $q\mapsto\lambda(q)$ satisfying $\lambda\{\beta(P)\}=0$ and $F_P\{\lambda(q),q\}=U_{P_{\lambda(q)}}(q)=0$. Existence alone does not yet quantify the amount of tilting required. To obtain the bound used in the remaining steps, differentiate this identity with respect to $q$:
\[
\frac{\partial\lambda(q)}{\partial q^\top}
=
- \left\{ \frac{\partial F_P\{\lambda(q),q\}} {\partial\lambda^\top} \right\}^{-1} 
\frac{\partial F_P\{\lambda(q),q\}} {\partial q^\top}.
\]
The uniformly bounded first derivatives assumed in \Cref{thm:cox_calibration}, applied to the $q$ argument of $F_P(\lambda,q)=U_{P_\lambda}(q)$, give
\[
\left\| \frac{\partial F_P(\lambda,q)}{\partial q^\top} \right\|
\leq C_q
\]
throughout these neighborhoods for some $0<C_q<\infty$. Combining this bound with the inverse Jacobian bound gives
\[
\left\| \frac{\partial\lambda(q)}{\partial q^\top} \right\| 
\leq \frac{2C_q}{c_{\mathrm{Jac}}}. \]
Since
$\lambda\{\beta(P)\}=0$, integration along the line segment from
$\beta(P)$ to $q$ gives
\begin{equation}
\begin{aligned}
\|\lambda(q)\|
&=
\|\lambda(q)-\lambda\{\beta(P)\}\|\\
&=
\left\|
\int_0^1
\frac{\partial\lambda}{\partial q^\top}
\left[\beta(P)+t\{q-\beta(P)\}\right]
\{q-\beta(P)\}\,dt
\right\|\\
&\leq
\int_0^1
\left\|
\frac{\partial\lambda}{\partial q^\top}
\left[\beta(P)+t\{q-\beta(P)\}\right]
\right\|
\|q-\beta(P)\|\,dt\\
&\leq
\frac{2C_q}{c_{\mathrm{Jac}}}\|q-\beta(P)\|.
\end{aligned}
\label{eq:cox_calibration_lambda_bound}
\end{equation}

\noindent\textbf{Step 3: Identification and uniqueness of the calibrated Cox parameter.}\quad
Step~1 constructs $\lambda(q)$ so that $U_{P_{\lambda(q)}}(q)=0$. To conclude that $\beta\{P_{\lambda(q)}\}=q$, we must also show that this score root is unique. We do so by transferring the information lower bound from $P$ to the tilted distribution. The score derivative satisfies
\[
-\frac{\partial U_{P_{\lambda(q)}}(\beta)}{\partial\beta^\top}
= J_{P_{\lambda(q)}}(\beta). \]
After reducing $\varepsilon$ if necessary, \Cref{thm:cox_wasserstein_stability} places $\beta(P)$ in the interior of $\mathcal B_r$ uniformly over $P\in\mathcal P_0$ satisfying $W_1(P,P_0)\leq\varepsilon$. Choose a fixed $\rho>0$ within the common $\beta$-neighborhood on which the first derivatives are uniformly continuous, and choose $\delta>0$ smaller than the radius of the $q$-neighborhood obtained in Step~1 so that $\mathcal B_\rho(q) = \{\beta:\|\beta-q\|\leq\rho\} \subset\mathcal B_r$ whenever $\|q-\beta(P)\|\leq\delta.$
Since
$J_{P_\lambda}(\beta)=-\partial U_{P_\lambda}(\beta)/ \partial\beta^\top$, the assumed uniform continuity of the first derivatives, together with \eqref{eq:cox_calibration_lambda_bound}, permits $\delta$ to be chosen smaller if necessary so that
\[
\sup_{\beta\in\mathcal B_\rho(q)}
\left\| J_{P_{\lambda(q)}}(\beta)-J_P(\beta) \right\| \leq\frac{\kappa}{4}.
\]
Equation~\eqref{eq:cox_w1_information_lower_bound} then gives, for every $\beta\in\mathcal B_\rho(q)$,
\[
\lambda_{\min}\{J_{P_{\lambda(q)}}(\beta)\}
\geq
\lambda_{\min}\{J_P(\beta)\} -\|J_{P_{\lambda(q)}}(\beta)-J_P(\beta)\|
\geq
\frac{\kappa}{2}-\frac{\kappa}{4} =\frac{\kappa}{4}.
\]
Hence the population Cox partial log-likelihood under $P_{\lambda(q)}$ is strictly concave in this neighborhood, and $q$ is its unique local stationary point.

The population Cox partial log-likelihood is concave on the full parameter space. If another stationary point existed, both points would be maximizers and the line segment joining them would consist of maximizers. This would contradict strict concavity near $q$. Consequently, $\beta\{P_{\lambda(q)}\}=q$. This establishes the identification conclusion of the theorem. It remains to control how far the calibration moves the baseline distribution.

\noindent\textbf{Step 4: Distributional stability and preservation of regularity under the exponential tilt.}\quad 
Having constructed the calibrated distribution and identified its Cox parameter, we now translate the bound on $\lambda(q)$ into bounds for $d_{\mathrm{TV}}$ and $W_1$. Equation~\eqref{eq:cox_xi} and Assumptions~\ref{ass:cox_parameter}--\ref{ass:cox_risksets} give a fixed $L_\xi<\infty$ such that $\|\xi_P(\widetilde D)\|\leq L_\xi$ uniformly over $P\in\mathcal P_0$ satisfying $W_1(P,P_0)\leq\varepsilon$. Therefore,
$|\lambda^\top\xi_P(\widetilde D)|\leq L_\xi\|\lambda\|$ and
$|\zeta_P(\lambda)|\leq L_\xi\|\lambda\|$, which gives
\begin{equation}
e^{-2L_\xi\|\lambda\|}
\leq
\frac{dP_\lambda}{dP}(\widetilde D)
\leq
e^{2L_\xi\|\lambda\|}.
\label{eq:cox_calibration_density_bound}
\end{equation}

This pointwise density-ratio bound first controls the total-variation distance. Under the total-variation convention used in the paper,
\[
d_{\mathrm{TV}}(P_\lambda,P)=
\frac12\mathbb E_P \left| \frac{dP_\lambda}{dP}(\widetilde D)-1 \right| \leq \frac12\left\{e^{2L_\xi\|\lambda\|}-1\right\} \leq C_{\mathrm{TV}}\|\lambda\|, \]
where the last inequality follows from the mean-value theorem for sufficiently small $\|\lambda\|$. Because the observed-data space is bounded, this total-variation bound also controls the Wasserstein distance. Under the ground metric $\|X-X'\|+|Y-Y'|+|\Delta-\Delta'|$, the observed-data space has diameter at most $2K+\tau+1$. The coupling characterization of $W_1$ therefore gives
\[ W_1(P_\lambda,P) \leq (2K+\tau+1)d_{\mathrm{TV}}(P_\lambda,P) \leq C_W\|\lambda\|. \]
Combining these two bounds with \eqref{eq:cox_calibration_lambda_bound} gives
\[ \|\lambda(q)\| +d_{\mathrm{TV}}\{P_{\lambda(q)},P\} +W_1\{P_{\lambda(q)},P\} \leq C\|q-\beta(P)\| \]
for some $0<C<\infty$. This proves the quantitative stability bound in the theorem. We finally verify that the tilted distribution retains the regularity properties used above. Since the density ratio is strictly positive, $P_\lambda$ and $P$ have the same support. Moreover,
$S_{P_\lambda}^{(0)}(\beta,t)
=
\mathbb E_P\left[ \frac{dP_\lambda}{dP}(\widetilde D)e^{\beta^\top\widetilde X}\widetilde R(t) \right]
\geq e^{-2L_\xi\|\lambda\|}S_P^{(0)}(\beta,t) \geq a e^{-2L_\xi\|\lambda\|}.$
Thus, the risk-set denominators remain bounded away from zero, with a possibly modified lower-bound constant. Similarly, for $0\leq y<\tau$, the marginal observed-time density under the tilted distribution satisfies $f_{P_\lambda}(y) =
\mathbb E_P\left[ \left. \frac{dP_\lambda}{dP}(\widetilde D) \,\right|\,\widetilde Y=y \right]f_P(y) \leq e^{2L_\xi\|\lambda\|}M.$
Absolute continuity also preserves
$P_\lambda(Y=\tau,\Delta=1)=0$. Consequently, the exponential tilt preserves the bounded-support, positive-risk-set, and observed-time regularity properties, with possibly modified uniform constants. The local information lower bound was established separately in Step~3. Together with the construction in Step~1 and the identification in Step~3, this completes the proof.
\end{proof}

\section{Calibration Algorithm Details}
\label{app:algorithm}
The following notation defines the weighting and effective-sample-size quantities used throughout the calibration, validation, and sampling steps of \Cref{alg:cox_calibration}.
Given synthetic sample $\{D_k\}_{k=1}^K$ and a calibration parameter
$\lambda$, define the normalized exponential weights by
$$
w_k(\lambda)
=
\frac{
\exp\{\lambda^\top\widehat\xi_{\widehat P_n}(D_k)\}
}{
\sum_{r=1}^K
\exp\{\lambda^\top\widehat\xi_{\widehat P_n}(D_r)\}
},
\qquad k=1,\ldots,K.
$$
Their effective sample size and effective sample-size fraction are
$
\operatorname{ESS}(\lambda)
=
\frac{1}{\sum_{k=1}^K w_k(\lambda)^2},
\operatorname{ESSFrac}(\lambda)
=
\frac{\operatorname{ESS}(\lambda)}{K}.
$
$\delta_d$ denotes the Dirac probability measure placing unit
mass at $d$.

\begin{algorithm}[!h]
\caption{Calibration and sampling from the fitted generative distribution
$\widehat P_n$}
\label{alg:cox_calibration}
\begin{algorithmic}[1]
\Require $\mathcal D_n$, $\widehat P_n$,
 $m$, pilot sample size $L_0$, calibration sample
size $L$, validation and SIR proposal-pool sizes
$K_{\mathrm{val}}$ and $K_{\mathrm{sir}}$, a prespecified set of Cox
covariates, score-residual tolerance $\varepsilon_{\mathrm{root}}$, minimum
effective sample-size fraction $\eta_{\mathrm{ESS}}$, maximum validation
score-reduction ratio $\rho_{\mathrm{val}}$, maximum number of
calibration attempts $A_{\max}$, and sample-size growth factor $g>1$.

\State Fit the original-data Cox model and set $q=\beta(P_n)$.

\State Draw $\{D_k^{\mathrm{pilot}}\}_{k=1}^{L_0}\overset{\mathrm{iid}}{\sim} \widehat P_n$ to estimate
$\beta(\widehat P_n)$ and the risk-set quantities needed to construct
$\widehat\xi_{\widehat P_n}(D)$, which is fixed in the following steps.
All Cox quantities are calculated using the same prespecified covariates.

\State Draw
$\{D_\ell^{\mathrm{cal}}\}_{\ell=1}^L \overset{\mathrm{iid}}{\sim}  \widehat P_n$, compute
$w_\ell^{\mathrm{cal}}(\lambda)$
and define the weighted empirical distribution
$\overline P_{n,L}^{\mathrm{cal}}(\lambda)
= \sum_{\ell=1}^L w_\ell^{\mathrm{cal}}(\lambda) \delta_{D_\ell^{\mathrm{cal}}}.$ For weights $\{w_\ell^{\mathrm{cal}}(\lambda)\}_{\ell=1}^L$, calculate
$\operatorname{ESS}_{\mathrm{cal}}(\lambda)$ and $
\operatorname{ESSFrac}_{\mathrm{cal}}(\lambda).$

\State Starting from $\lambda=0$, apply a local numerical root-finding
method to solve
$U_{\overline P_{n,L}^{\mathrm{cal}}(\lambda)}(q)=0.$
Retain a solution $\widehat\lambda$ only if
$\left\|
U_{\overline P_{n,L}^{\mathrm{cal}}(\widehat \lambda)}\{q\}
\right\|_2
\leq\varepsilon_{\mathrm{root}}$
and  $\operatorname{ESSFrac}_{\mathrm{cal}}(\widehat\lambda)
\geq\eta_{\mathrm{ESS}}.$ 

\State Get $\overline P_{n,\widehat\lambda}$ by
$\frac{d\overline P_{n,\widehat\lambda}}{d\widehat P_n}(D)
\propto \exp\left\{ \widehat\lambda^\top \widehat\xi_{\widehat P_n}(D) \right\}.$

\State Draw 
$\{D_k^{\mathrm{val}}\}_{k=1}^{K_{\mathrm{val}}} \overset{\mathrm{iid}}{\sim}  \widehat P_n$ and compute
$w_k^{\mathrm{val}}(\widehat \lambda)$
and $\operatorname{ESSFrac}_{\mathrm{val}}.$

\State Define
$\widehat P_{n,K_{\mathrm{val}}}^{\mathrm{val}}
\coloneqq 1/K_{\mathrm{val}} \sum_{k=1}^{K_{\mathrm{val}}} \delta_{D_k^{\mathrm{val}}}, 
\overline P_{n,K_{\mathrm{val}}}^{\mathrm{val}}(\widehat \lambda)
\coloneqq \sum_{k=1}^{K_{\mathrm{val}}} w_k^{\mathrm{val}} (\widehat \lambda)\delta_{D_k^{\mathrm{val}}}.$ 
Retain the calibration only if
\[
\frac{
\left\|
U_{\overline P_{n,K_{\mathrm{val}}}^{\mathrm{val}}(\widehat \lambda)}\{q\}
\right\|_2
}{
\left\|
U_{\widehat P_{n,K_{\mathrm{val}}}^{\mathrm{val}}}\{q\}
\right\|_2
} \leq\rho_{\mathrm{val}}, \qquad \operatorname{ESSFrac}_{\mathrm{val}} \geq\eta_{\mathrm{ESS}}.
\] 
If either validation condition fails and fewer than $A_{\max}$ attempts have been made, multiply $L_0$, $L$, and $K_{\mathrm{val}}$ by $g$ and repeat steps 2--6. Report numerical calibration failure if the criteria remain unsatisfied after $A_{\max}$ attempts. If the denominator is zero, the ratio is defined as zero when the numerator is also zero and as infinity otherwise.

\State Draw fresh $\{D_k^{\mathrm{sir}}\}_{k=1}^{K_{\mathrm{sir}}}\overset{\mathrm{iid}}{\sim} \widehat P_n$ and compute $w_k^{\mathrm{sir}}.$

\State Define the weighted empirical SIR approximation $\overline P_{n,K_{\mathrm{sir}}}^{\mathrm{sir}}(\widehat \lambda) \coloneqq \sum_{k=1}^{K_{\mathrm{sir}}} w_k^{\mathrm{sir}}(\widehat \lambda) \delta_{D_k^{\mathrm{sir}}}.$ Draw $  \widetilde D_1,\ldots,\widetilde D_m \overset{\mathrm{iid}}{\sim} \overline P_{n,K_{\mathrm{sir}}}^{\mathrm{sir}}(\widehat \lambda)$ and define $\overline P_{n,m} \coloneqq \frac1m\sum_{j=1}^m\delta_{\widetilde D_j}.$

\Ensure A numerical calibrated distribution $\overline P_{n, \lambda}$, its weighted empirical SIR approximation $\overline P_{n,K_{\mathrm{sir}}}^{\mathrm{sir}}(\widehat \lambda)$, and a calibrated synthetic dataset $\widetilde{\mathcal D}_m =\{\widetilde D_j\}_{j=1}^m$ with empirical distribution $\overline P_{n,m}$.
\end{algorithmic}
\end{algorithm}

\section{Additional Experiment Details}
\label{app:experiment-details}

This section gives the implementation-level definitions of the empirical metrics reported in the simulation and real-data experiments. Every discrepancy compares an original sample of size $n$ with a synthetic sample of size $m$. 
The simulation tables first take the median over the 10 synthetic draws generated from each fitted generator and then the median over 100 independently generated original datasets. 

\subsection{Joint distributional metrics}

For a fair multivariate comparison, metrics are computed after placing all coordinates on a common scale. Each covariate is standardized using the mean and population standard deviation of the corresponding original sample. In both simulated and real data, the observed time is divided by the maximum observed time in the original training sample. Event status remains binary and has weight one. 
The same transformation is applied to the synthetic data.

\noindent\textbf{Sliced Wasserstein distance.}
For each comparison, let $D_1,\ldots,D_n$ denote the real  sample and let $\widetilde D_1,\ldots,\widetilde D_m$ denote the sample drawn from the fitted generative distribution $\widehat P_n$. For independently sampled directions $\theta_\ell=G_\ell/\|G_\ell\|_2$, where $G_\ell\sim N(0,I)$, the reported sliced $W_1$ distance is
\[
\widehat{\operatorname{SW}}_1(P_n,\widehat P_{n,m})
=\frac{1}{L}\sum_{\ell=1}^{L}
\int_0^1
\left|
F_{n,\ell}^{-1}(u)-
\widetilde F_{m,\ell}^{-1}(u)
\right|\,du,
\]
where $F_{n,\ell}^{-1}$ and $\widetilde F_{m,\ell}^{-1}$ denote the empirical quantile functions of the projected samples $\{\theta_\ell^{\mathsf T}D_i\}_{i=1}^n$ and $\{\theta_\ell^{\mathsf T}\widetilde D_j\}_{j=1}^m$, respectively. This formulation accommodates unequal sample sizes without downsampling. We use $L=500$ random projection directions in every reported experiment.

\noindent\textbf{MMD}
We use the Gaussian kernel $
k_\sigma(z,z')=\exp\{-\|z-z'\|_2^2/(2\sigma^2)\}$
and the biased empirical V-statistic
\[
\widehat{\operatorname{MMD}}^2
=\frac{1}{n^2}\sum_{i,j=1}^{n}k_\sigma(D_i,D_j)
 +\frac{1}{m^2}\sum_{i,j=1}^{m}k_\sigma(\widetilde D_i,\widetilde D_j)
 -\frac{2}{nm}\sum_{i=1}^{n}\sum_{j=1}^{m}
 k_\sigma(D_i,\widetilde D_j).
\]
The tables report
$\sqrt{\max(\widehat{\operatorname{MMD}}^2,0)}$. The bandwidth is selected from pooled original and synthetic records by $\sigma=\{\operatorname{median}_{a<b}\|D_a-D_b\|_2^2/2\}^{1/2}$, excluding zero pairwise distances. Kernel sums are evaluated in blocks of 1,000 records. For the downstream tables, each distribution and the pooled bandwidth sample are capped at 5,000 records. For the large-sample convergence experiment in Figure~\ref{fig:simulation-joint-convergence}, the corresponding caps are 90,000 and 9,000 records, respectively. Subsampling is seeded independently within each replicate.

\subsection{Additional experimental details in \Cref{subsec:simulation-convergence}}
\label{app:sim_conv-details}
In the Cox estimation convergence rate verification experiment, $\beta(P_0)$ and $\Sigma(P_0)$ were approximated using a fixed independent reference sample of size $200{,}000$ generated from $P_0$. The distance $W_1(\widehat P_{n,m},P_0)$ was approximated by comparing the synthetic sample with the same fixed empirical reference sample. We retain the notation $\beta(P_0)$, $\Sigma(P_0)$, and $P_0$ in \Cref{fig:simulation5.1-beta-bias-rate} for readability, with the understanding that they are evaluated with these large-sample approximations.

\subsection{Additional experimental details in \Cref{subsec:simulation-evaluation}}
\label{app:protocol-details}

S2 retains only $x_{2},x_{6},x_{15},x_{19},x_{27}$ and multiplies their S1 coefficients by 2.2525, yielding $$(-0.7317,-0.7346,0.9357,-0.5755,0.7390)$$. This construction changes sparsity while maintaining $\operatorname{Var}(X^\top\beta)=2.80$. The censoring bounds are $c=0.800$ for S1 and S2, $c=0.0896$ for S3, and $c=6.35$ for S4. These choices produce the event rates reported in Table~\ref{tab:simulation-settings}.

\subsection{Additional experimental details in
\Cref{subsec:simulation-inference}}

Each of the $1{,}000$ replications used an independent random seed.
In each replication, a new original sample of size $n=1{,}000$ was
drawn, and the Cox coefficient $\beta(P_n)$ and sandwich variance
$\Sigma(P_n)$ were computed. The same original sample was used for the
oracle and perturbed generator settings. Cox coefficients were computed
by Newton iteration, and the sandwich variance was computed from the
empirical adjusted score contributions in \Cref{eq:cox_xi}.

The numerical calibration followed \Cref{alg:cox_calibration} with
$L_0=200{,}000,
L=2{,}000{,}000,
K_{\mathrm{val}}=500{,}000,
K_{\mathrm{sir}}=500{,}000.
\varepsilon_{\mathrm{root}}=10^{-10},
\eta_{\mathrm{ESS}}=0.01,
\rho_{\mathrm{val}}=0.5,
A_{\max}=3,
g=2.$
The displayed values of $L_0$, $L$, and $K_{\mathrm{val}}$ are the
initial sample sizes. Following a failed validation attempt, they were
multiplied by $g$ before the next attempt, with at most $A_{\max}$
attempts in total. The final SIR proposal-pool size
$K_{\mathrm{sir}}$ remained fixed and was used only after validation
was successful.

Let $m_{\max}=50{,}000$. For the oracle and uncalibrated perturbed
generators, a synthetic sample of size $m_{\max}$ was drawn directly.
For the calibrated analysis, $m_{\max}$ records were generated by
SIR from the independent proposal sample of size
$K_{\mathrm{sir}}$. For each generator setting, the first $m$
records were used for $m\in
\{1{,}000,2{,}000,5{,}000,10{,}000,20{,}000,50{,}000\}.$
The main analysis used $m=10{,}000$. Because the analyses at different
values of $m$ used nested prefixes of the same size-$m_{\max}$ sample,
their rejection-rate estimates are positively correlated.

Writing an asymptotic distribution as $N(\mu,v)$, the parameters
$(\mu,v)$ are $(0,1+m/n)$ for the naive analysis under either the
oracle or calibrated generator,
$(2\sqrt{m/n},1+m/n)$ for the uncalibrated naive analysis under the
perturbed generator,
$(2/\sqrt{1+n/m},1)$ for its uncalibrated corrected analysis, and
$(0,1)$ for the corrected analyses under the oracle and calibrated
generators. Rejection rates are reported with $95\%$ Wilson intervals.

\noindent\textbf{Pilot construction and numerical implementation.}
Because this experiment contains one Cox covariate, all coefficient,
score, and calibration quantities in the following implementation are
scalar. From the pilot sample
$\{D_k^{\mathrm{pilot}}\}_{k=1}^{L_0}
=
\{(X_k^{\mathrm{pilot}},
Y_k^{\mathrm{pilot}},
\Delta_k^{\mathrm{pilot}})\}_{k=1}^{L_0},$
we computed the pilot Cox estimate
$\widehat\beta^{\mathrm{pilot}}$ of $\beta(\widehat P_n)$ and defined
$\widehat S^{(j)}(t)
=
\frac{1}{L_0}
\sum_{k=1}^{L_0}
\ind(Y_k^{\mathrm{pilot}}\geq t)
e^{\widehat\beta^{\mathrm{pilot}}X_k^{\mathrm{pilot}}}
(X_k^{\mathrm{pilot}})^j, j=0,1,$
and
$\widehat{\overline X}(t)
=
\frac{\widehat S^{(1)}(t)}
     {\widehat S^{(0)}(t)}.$
Replacing the corresponding population quantities in
\Cref{eq:cox_xi} by these pilot estimates gives
\[
\widehat\xi_{\widehat P_n}(D)
=
\Delta\{X-\widehat{\overline X}(Y)\}
-
e^{\widehat\beta^{\mathrm{pilot}}X}
\frac{1}{L_0}
\sum_{\substack{k:\,
\Delta_k^{\mathrm{pilot}}=1,\\
Y_k^{\mathrm{pilot}}\leq Y}}
\frac{
X-\widehat{\overline X}(Y_k^{\mathrm{pilot}})
}{
\widehat S^{(0)}(Y_k^{\mathrm{pilot}})
}.
\]
This pilot-estimated function was held fixed during calibration,
validation, and final SIR sampling.

For each candidate $\lambda$,
$U_{\overline P_{n,L}^{\mathrm{cal}}(\lambda)}\{q\}$ was computed in
$O(L)$ operations after sorting by follow-up time and using cumulative
risk-set sums. The local root $\widehat\lambda$ was obtained by Brent's
method on the bracket adjacent to zero containing a sign change, using
an absolute tolerance of $10^{-12}$ for $\lambda$. A solution was
retained only when the root-residual and calibration effective
sample-size requirements in \Cref{alg:cox_calibration} were satisfied.
The maximum absolute score residual across the $1{,}000$ replications
was $1.72\times10^{-12}$.

All $1{,}000$ replications also satisfied the independent validation
criteria in \Cref{alg:cox_calibration} on the first attempt. The median
and maximum validation score-reduction ratios were $0.0155$ and
$0.108$, respectively, and the median and minimum validation effective
sample-size fractions were $0.9958$ and $0.9952$. After validation, the
final SIR sample was generated from
$\overline P_{n,K_{\mathrm{sir}}}^{\mathrm{sir}}
(\widehat\lambda)$ as specified in \Cref{alg:cox_calibration}. The
median fraction of unique proposal records represented in the
size-$m_{\max}$ SIR sample was $0.951$.

\begin{figure}[!h]
    \centering
    \includegraphics[width=1\linewidth]{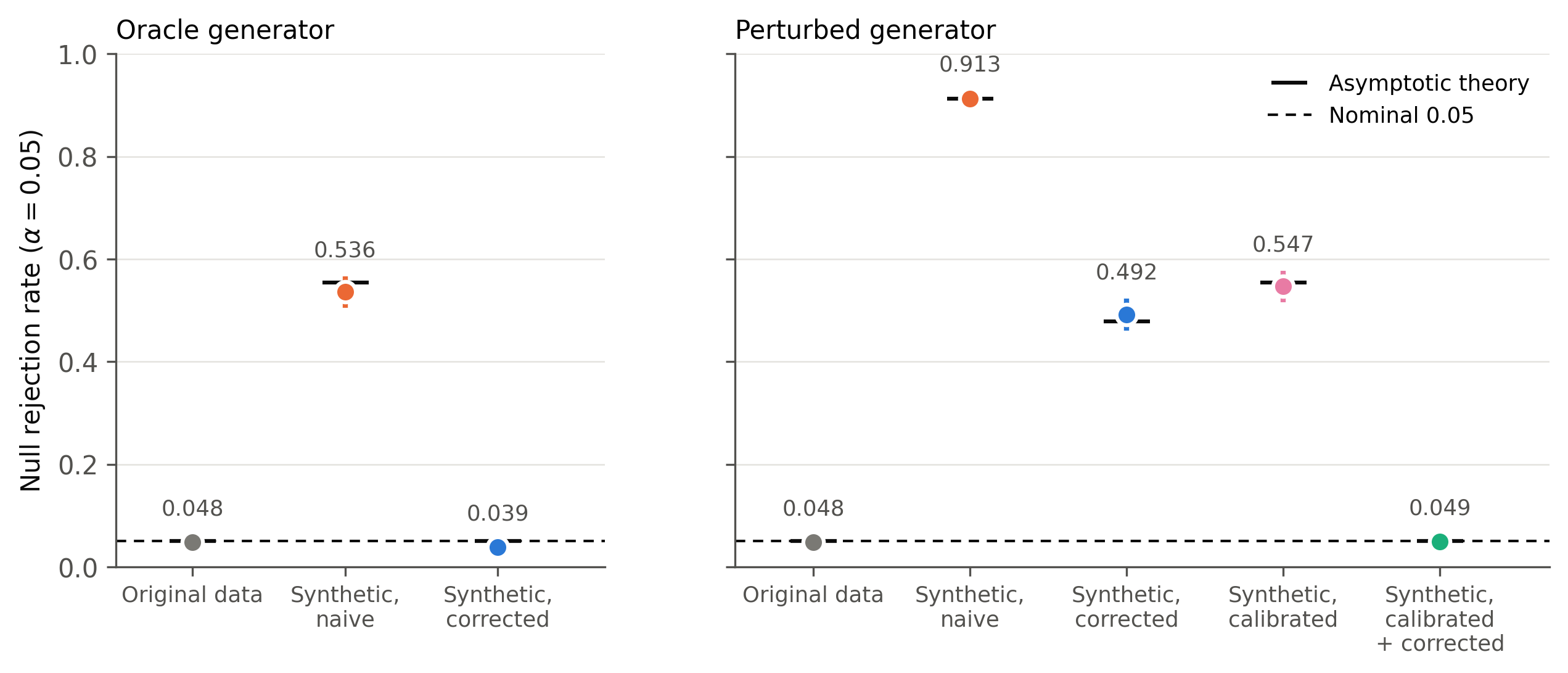}
    \caption{Null rejection rates in the corrected and calibrated Cox inference study ($n=1{,}000$, $m=10{,}000$, $R=1{,}000$). Points are empirical rejection rates with $95\%$ Wilson intervals, black ticks are asymptotic theoretical values, and the dashed line is the nominal level $0.05$.}
    \label{fig:simulation-inference-rejection}
\end{figure}

\begin{figure}[!h]
    \centering
    \includegraphics[width=1\linewidth]{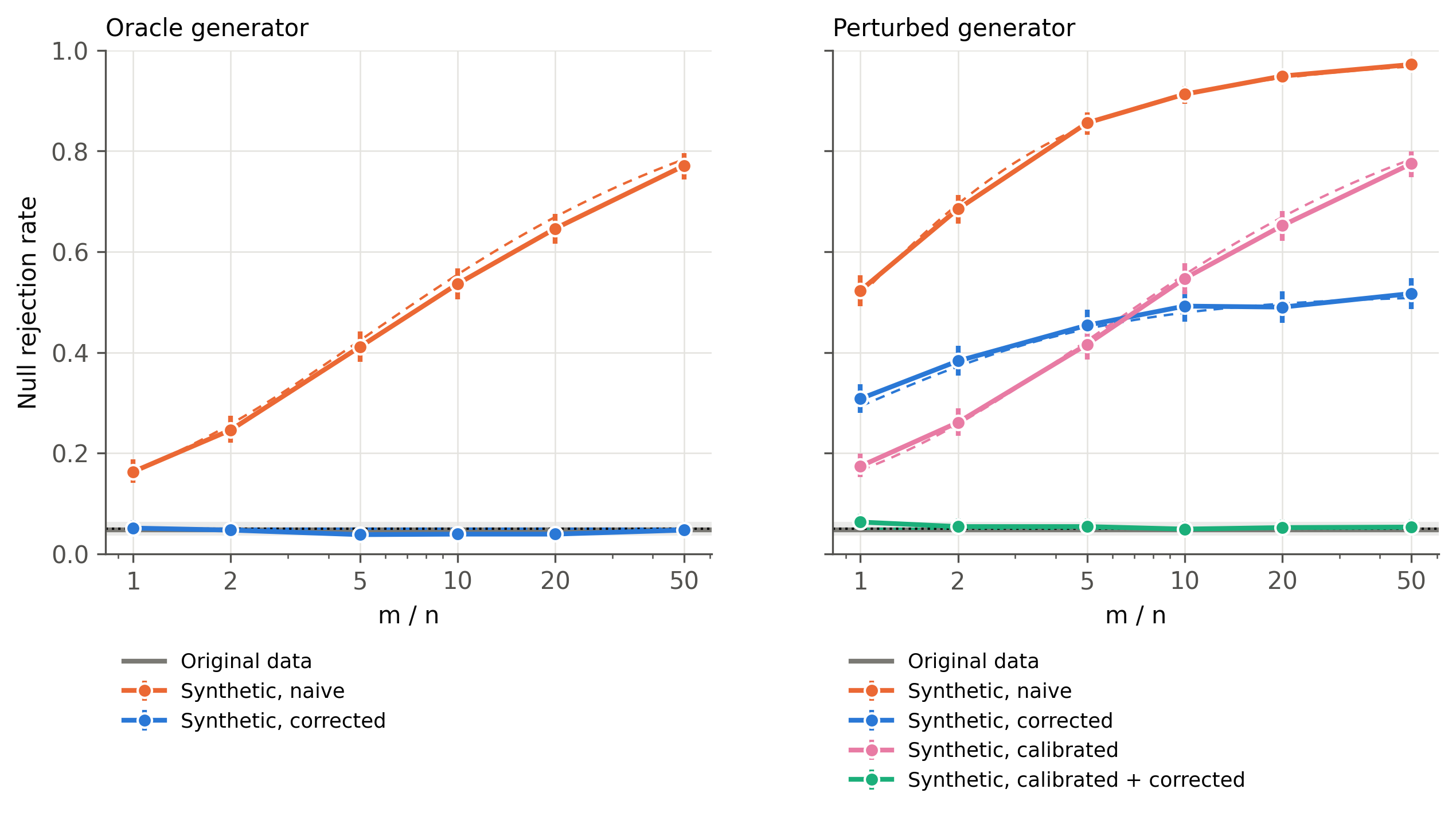}   \caption{Null rejection rates as a function of $m/n$ with $n=1{,}000$. Solid lines are simulation estimates with $95\%$ Wilson intervals, dashed lines are asymptotic theoretical values, and the grey band is the original-data rejection rate with its $95\%$ interval.}
    \label{fig:simulation-inference-sweep}
\end{figure}

\subsection{Additional experimental details in \Cref{sec:real}}

\noindent\textbf{Data splitting and generator fitting.}
\Cref{subsec:real-evaluation} used a fixed training--test split for each dataset. The NHANES split contained 20,000 training records and 12,000 test records, and the SUPPORT split contained 6,000 training records and 2,000 test records. The training and test seeds were 2026 and 2027, respectively. For each dataset, TSGSurv(diff) was fitted once to the training data and the same fitted generator was retained across the 10 synthetic replicates. Thus, the replicate variation reflects independent numerical calibrations and synthetic draws from a fixed fitted generator rather than repeated  generator fitting. The original-data split, fitted generator, preprocessing map, support bounds, target covariates, and numerical tolerances were fixed across replicates.

\noindent\textbf{Targeted calibration.}
The exponential tilt targeted four coefficient-influence directions, while the score, information matrix, and Cox risk-set  quantities were calculated using the full Cox model. Root finding and validation were applied to the corresponding four-dimensional targeted score. Tied event times were handled using the Breslow convention.

Each of the 10 replicates independently followed \Cref{alg:cox_calibration} with $L_0=200{,}000,$ $L=500{,}000,$ $K_{\mathrm{val}}=500{,}000,$ $\varepsilon_{\mathrm{root}}=10^{-8},$ $\eta_{\mathrm{ESS}}=0.01,$ $\rho_{\mathrm{val}}=0.5,$ $A_{\max}=3,$ $g=2.$ Eight of the 10 NHANES replicates succeeded on their first attempt for validation and two on their second attempt; all 10 SUPPORT  replicates succeeded on their first attempt.

\noindent\textbf{Final SIR sampling and numerical diagnostics.}  
After successful validation, a fresh independent SIR proposal pool was drawn. Its size was $K_{\mathrm{sir}}=\max(10m,100{,}000),$ giving $K_{\mathrm{sir}}=200{,}000$ for NHANES, where $m=20{,}000$, and $K_{\mathrm{sir}}=100{,}000$ for SUPPORT, where $m=6{,}000$.  $m$ records were sampled with replacement from the weighted proposal pool. For NHANES, the maximum root residual over the 10 replicates was $4.11\times10^{-13}$. The median calibration and validation ESS fractions were 0.9996 and 0.9996, respectively, and the median and maximum validation score-reduction ratios were 0.0507 and 0.128. The median final-pool ESS  fraction was 0.9996 (minimum 0.693), and the median fraction of unique  proposal records represented in the final synthetic dataset was 0.951. For SUPPORT, the maximum root residual was $4.94\times10^{-12}$; the median calibration, validation, and final-pool ESS fractions were 0.9985, 0.9985,  and 0.9985, respectively. The median and maximum validation score-reduction ratios were 0.126 and 0.178, and the median final unique fraction was  0.971. All reported replicates therefore satisfied both the score-reduction and  ESS criteria by substantial margins.

\noindent\textbf{Downstream Cox fitting and aggregation.} 
Full Cox coefficient estimates were obtained using \texttt{CoxPHSurvivalAnalysis} from \texttt{scikit-survival} version 0.23.1, with Breslow ties and no penalty. Standard errors were  calculated using the empirical sandwich estimator based on  the adjusted score contributions in \Cref{eq:cox_xi}, rather than the  inverse-information covariance. Because $m=n$, corrected synthetic standard errors used the factor $\sqrt{1+m/n}=\sqrt{2}$. Results were  summarized over the 10 synthetic replicates as specified in \Cref{app:experiment-details}.
\end{document}